\documentclass[twocolumn,tighten,twocolappendix]{aastex63}
\pdfoutput=1 
\usepackage{amsmath,amstext}
\usepackage[T1]{fontenc}
\usepackage{ae,aecompl}
\usepackage[utf8]{inputenc}
\usepackage{apjfonts} 
\usepackage[figure,figure*]{hypcap}
\usepackage{enumitem}
\usepackage{bm}
\usepackage{graphicx}
\usepackage[hang]{footmisc}
\usepackage{lineno}
\input{hyperlink-year-only-natbib-patch}

\makeatletter
\DeclareRobustCommand{\HI}{%
  \mbox{H\check@mathfonts\fontsize\sf@size\z@\selectfont I}%
}
\makeatother

\newcommand*{\http}[1]{\href{http://#1}{#1}}
\newcommand*{\https}[1]{\href{https://#1}{#1}}

\shorttitle{Dwarf Galaxy Forecasts}
\shortauthors{Nadler et al.}

\begin{document}

\title{Forecasts for Galaxy Formation and Dark Matter Constraints from Dwarf Galaxy Surveys}

\author[0000-0002-1182-3825]{Ethan O.~Nadler}
\affiliation{Carnegie Observatories, 813 Santa Barbara Street, Pasadena, CA 91101, USA}
\affiliation{Department of Physics $\&$ Astronomy, University of Southern California, Los Angeles, CA 90007, USA}

\author[0000-0002-3589-8637]{Vera Gluscevic}
\affiliation{Department of Physics $\&$ Astronomy, University of Southern California, Los Angeles, CA 90007, USA}

\author[0000-0001-9472-7179]{Trey Driskell}
\affiliation{Department of Physics $\&$ Astronomy, University of Southern California, Los Angeles, CA 90007, USA}

\author[0000-0003-2229-011X]{Risa H.~Wechsler}
\affiliation{Kavli Institute for Particle Astrophysics \& Cosmology, P. O. Box 2450, Stanford University, Stanford, CA 94305, USA}
\affiliation{Department of Physics, Stanford University, 382 Via Pueblo Mall, Stanford, CA 94305, USA}
\affiliation{SLAC National Accelerator Laboratory, Menlo Park, CA 94025, USA}

\author[0000-0003-3030-2360]{Leonidas A.~Moustakas}
\affiliation{Jet Propulsion Laboratory, California Institute of Technology, 4800 Oak Grove Drive, Pasadena, CA 91109, USA}

\author[0000-0001-5501-6008]{Andrew Benson}
\affiliation{Carnegie Observatories, 813 Santa Barbara Street, Pasadena, CA 91101, USA}

\author[0000-0002-1200-0820]{Yao-Yuan~Mao}
\affiliation{Department of Physics and Astronomy, University of Utah, Salt Lake City, UT 84112, USA}

\correspondingauthor{Ethan~O.~Nadler}
\email{enadler@carnegiescience.edu}

\label{firstpage}

\begin{abstract}
The abundance of faint dwarf galaxies is determined by the underlying population of low-mass dark matter (DM) halos and the efficiency of galaxy formation in these systems. 
Here, we quantify potential galaxy formation and DM constraints from future dwarf satellite galaxy surveys. 
We generate satellite populations using a suite of Milky Way (MW)--mass cosmological zoom-in simulations and an empirical galaxy--halo connection model, and assess sensitivity to galaxy formation and DM signals when marginalizing over galaxy--halo connection uncertainties. We find that a survey of all satellites around one MW-mass host can constrain a galaxy formation cutoff at peak virial masses of $\mathcal{M}_{50}=10^8~M_{\mathrm{\odot}}$ at the $1\sigma$ level; however, a tail toward low $\mathcal{M}_{50}$ prevents a $2\sigma$ measurement. In this scenario, combining hosts with differing bright satellite abundances significantly reduces uncertainties on $\mathcal{M}_{50}$ at the $1\sigma$ level, but the $2\sigma$ tail toward low $\mathcal{M}_{50}$ persists. 
We project that observations of one (two) complete satellite populations can constrain warm DM models with $m_{\mathrm{WDM}}\approx 10~\mathrm{keV}$ ($20~\mathrm{keV}$). Subhalo mass function (SHMF) suppression can be constrained to $\approx 70\%$, $60\%$, and $50\%$ that in cold dark matter (CDM) at peak virial masses of $10^8$, $10^9$, and $10^{10}~M_{\mathrm{\odot}}$, respectively; SHMF enhancement constraints are weaker ($\approx 20$, $4$, and $2$ times that in CDM, respectively) due to galaxy--halo connection degeneracies. 
These results motivate searches for faint dwarf galaxies beyond the MW and indicate that ongoing missions like Euclid and upcoming facilities including the Vera C.\ Rubin Observatory and Nancy Grace Roman Space Telescope will probe new galaxy formation and DM physics.
\end{abstract} 

\keywords{\href{http://astrothesaurus.org/uat/353}{Dark matter (353)}; \href{http://astrothesaurus.org/uat/416}{Dwarf galaxies (416)}; \href{http://astrothesaurus.org/uat/574}{Galaxy abundances (574)};
\href{http://astrothesaurus.org/uat/595}{Galaxy formation (595)}}

\section{Introduction}\label{sec:intro}

The faintest dwarf galaxies inhabit the smallest dark matter halos that can form galaxies. These systems are sensitive probes of galaxy formation models and simultaneously test dark matter (DM) particle physics (e.g., \citealt{Bullock170704256,2019BAAS...51c.134G,Simon190105465,Nadler:2021fkn,Sales220605295}), including DM free-streaming, interactions, and decays, wave interference of ultralight DM, and nonstandard DM production mechanisms (e.g., see \citealt{Bechtol220307354} for an overview).

To disentangle the information dwarf galaxy surveys can offer, it is crucial to understand the interplay between galaxy formation and DM physics in low-mass halos. For example, reionization inhibits star formation in halos with peak masses below $\approx 10^8$ to $10^9~M_{\mathrm{\odot}}$ because of their low virial temperatures and limited gas reservoirs (e.g., \citealt{Benitze-Llambay200406124,Munshi210105822,Kravtsov210609724,Ahvazi230813599}), suppressing the population of dwarf galaxies that occupy these systems. Observations of the faintest dwarf galaxies can shed light on these processes and can also reveal deviations in underlying halo populations from cold dark matter (CDM), which would signal new DM physics on small scales. Moreover, a measurement of the galaxy formation cutoff is necessary to infer the existence of ``dark'' (galaxy--free) halos---which remain a key, untested predicted of many DM models---in combination with gravitational probes of even smaller halos (e.g., \citealt{Gilman190806983,Banik191102663,Nadler210107810}).

Current constraints on the galaxy formation cutoff are limited by the small number of known ultrafaint dwarf galaxies, which have almost exclusively been detected as satellites of the Milky Way (MW; e.g., \citealt{Drlica-Wagner191203302}). Additionally, this measurement is plagued by large theoretical uncertainties on the faint end of the galaxy--halo connection (e.g., see \citealt{Wechsler180403097} for a review). Recently, \cite{Nadler191203303} analyzed the full-sky MW satellite population detected by the Dark Energy Survey (DES; \citealt{Abbott180103181}) and Pan-STARRS1 (PS1; \citealt{Chambers161205560}) to place a $95\%$ confidence upper limit of $8.5\times 10^7~M_{\mathrm{\odot}}$ on the peak virial mass at which $50\%$ of halos host galaxies, hereafter 
$\mathcal{M}_{50}$.\footnote{We define ``galaxies'' as dark matter-dominated systems with absolute $V$-band magnitudes $M_V<0~\rm{mag}$, which are composed of $\gtrsim 100$ stars in our fiducial models, on average, following \cite{Nadler191203303}.} This constraint accounts for observational selection effects, marginalizes over the relevant theoretical uncertainties, and implies stringent limits on a variety of DM scenarios that suppress the abundance of halos with peak masses of $\approx 10^8~M_{\mathrm{\odot}}$ (e.g., \citealt{Nadler200800022,Newton201108865,Dekker211113137}). However, current data do not permit a measurement of the galaxy formation cutoff because only tens of ultrafaints have been detected; this number is too small to rule out statistical fluctuations that mimic a cutoff or to overcome theoretical uncertainties on the predicted abundance of MW satellite galaxies.

Upcoming observational facilities will substantially improve this situation by expanding our census of faint dwarf galaxies. For example, the Vera C.\ Rubin Observatory (Rubin; \citealt{Ivezic08052366}) is expected to detect much of the remaining MW satellite population, as well as many resolved dwarf galaxies out to $\approx 5~\mathrm{Mpc}$ \citep{Drlica-Wagner190201055,Mutlu-Pakdil210501658}; comparable sensitivity has been projected for the Chinese Space Station Telescope \citep{Qu221210804}. The Nancy Grace Roman Space Telescope (Roman; \citealt{Spergel150303757}) will also be a powerful dwarf galaxy discovery machine \citep{Gezari220212311}, with the ability to detect resolved stars out to even larger distances than Rubin. The Euclid Dark Energy Mission (\citealt{EuclidMission}; launched 2023 July 1) is expected to achieve similar sensitivity \citep{Racca161005508}. Meanwhile, despite its small field of view, JWST can resolve stars out to much larger distances than Roman or Euclid and thus may play an important role in detecting faint dwarf galaxies throughout the Local Volume and beyond (e.g., \citealt{Conroy231013048}). Looking further ahead, next-generation optical/infrared space telescopes such as the Habitable Worlds Observatory, which is the top priority of the Astro2020 Decadal Survey, will revolutionize the search for dwarf galaxies at cosmic distances (e.g., see \citealt{LUVOIR191206219} for estimates of future large space telescopes' detection sensitivities).

To prepare for these advances, we study measurements of galaxy formation and DM physics in the context of future dwarf galaxy surveys. In particular, we use a suite of MW-mass cosmological zoom-in simulations \citep{Nadler220902675} to assess the sensitivity of satellite galaxy populations to a reionization-like galaxy formation cutoff, a suppression of low-mass halo abundances characteristic of warm dark matter (WDM), and the amplitude of the subhalo mass function (SHMF). We focus on subhalos and satellite galaxies (rather than isolated halos and field galaxies) because, as discussed above, future observational facilities capable of resolving ultrafaint dwarfs beyond the Local Group will have relatively small fields of view; thus, massive central galaxies will provide natural targets around which to search for dwarf satellite galaxies in deep pointings. We focus on MW-mass systems because lower-mass centrals host fewer satellites and because MW-mass systems are prevalent in the Local Volume (e.g., \citealt{Mutlu-Pakdil210501658,Carlsten220300014}) and nearby Universe (e.g., \citealt{Geha170506743,Mao200812783}).

By combining our simulations with an expanded version of the galaxy--halo connection model and probabilistic framework from \cite{Nadler180905542,Nadler191203303,Nadler200800022}, we explore degeneracies between galaxy formation and DM effects on dwarf satellite populations and quantify signals of new physics that future galaxy surveys can uncover. We forecast that observations of all satellite galaxies around two MW-mass hosts (e.g., the MW and M31 or another nearby galaxy) can constrain a galaxy formation cutoff at peak virial halo masses of $10^8~M_{\mathrm{\odot}}$ and place stringent upper limits on cutoffs at lower halo masses. We show that the same data can constrain WDM models with masses of $\approx 10$--$20~\mathrm{keV}$, significantly improving upon current constraints \citep{Bullock170704256,Drlica-Wagner190201055}. Furthermore, we find that the SHMF can be measured in a model-independent fashion at peak virial masses between $10^8~M_{\mathrm{\odot}}$ and $10^{10}~M_{\mathrm{\odot}}$, paving the way for tests of any DM or early-Universe physics that suppresses or enhances low-mass halo abundances relative to CDM.

This paper is organized as follows. In Section~\ref{sec:model}, we describe our simulations and galaxy--halo connection modeling framework; in Section~\ref{sec:forecasting}, we describe our procedure for forecasting constraints from future dwarf measurements. We then present the results of our galaxy formation (Section~\ref{sec:reion}), WDM (Section~\ref{sec:wdm}), and SHMF (Section~\ref{sec:shmf}) forecasts. We discuss our results in the context of upcoming and future observations, along with areas for theoretical development, in Section~\ref{sec:caveats}. We discuss the implications of our forecasts for the detection of ``dark'' halos in Section \ref{sec:discussion}, and we conclude in Section \ref{sec:conclusions}.

We adopt the same cosmological parameters used for our MW zoom-in simulations: $h = 0.7$, $\Omega_{\rm m} = 0.286$, $\Omega_{\Lambda} = 0.714$, $\sigma_8 = 0.82$, and $n_s=0.96$ \citep{Hinshaw_2013}. We quote (sub)halo virial masses using the \cite{Bryan_1998} virial overdensity, which corresponds to $\Delta_{\mathrm{vir}}\approx 99$ times the critical density of the Universe at $z=0$ in this cosmology. Throughout, we refer to DM halos within a host's virial radius as ``subhalos,'' and we denote galaxies that occupy a host's subhalos as ``satellites.'' Furthermore, ``log'' always refers to the base-$10$ logarithm.


\section{Modeling Dwarf Galaxy Populations}
\label{sec:model}

We begin with an overview of our forecasting framework, including the cosmological zoom-in simulations we base our forecasts on (Section \ref{sec:zoom}), our galaxy--halo connection model (Section \ref{sec:gh_model}), and our treatment of dwarf population statistics in beyond-CDM scenarios (Section \ref{sec:ncdm_model}); we then illustrate the predictions of our model for subhalo and dwarf satellite galaxy populations (Section~\ref{sec:illustration}).

\subsection{Cosmological Zoom-in Simulations}
\label{sec:zoom}

Our forecasts are based on subhalo populations from cosmological dark matter-only zoom-in simulations of MW-mass systems. Specifically, we use the ``Milky Way-mass'' suite from the Symphony zoom-in compilation \citep{Mao150302637,Nadler220902675}. This suite contains $45$ host halos in a virial mass range of~$10^{12.09 \pm 0.02}~M_{\mathrm{\odot}}$ and accurately resolves the abundance of subhalos with present-day virial masses of $M_{\mathrm{vir}}>1.2\times 10^8~M_{\mathrm{\odot}}$ \citep{Nadler220902675}. Subhalos below this convergence threshold contribute negligibly to the observable satellite populations in our main forecasts, which are restricted to dwarf galaxies with $M_V<0~\mathrm{mag}$ that occupy subhalos with $M_{\mathrm{vir}}>10^8~M_{\mathrm{\odot}}$, on average \citep{Nadler191203303}. Thus, poorly resolved subhalos do not impact our results. We use the entire subhalo population within the virial radius of each MW-mass system---with no additional constraints imposed on e.g.\ subhalo distance from the host center, infall time, or size---to forward-model the entire observable satellite population of each host.

Our simulated MW-mass hosts have a range of secondary halo properties and inhabit a variety of cosmic environments~\citep{Fielder180705180,Nadler220902675}. Thus, host-to-host scatter in subhalo populations, which results in subhalo abundance variations up to a factor of $\approx 2$ at fixed host halo mass \citep{Mao150302637}, is naturally included in our predictions. However, we do not attempt to model the specific region of the nearby Universe inhabited by the closest $40$ MW-mass systems, which provide natural observational targets for our forecasts. In particular, our simulations span a narrow range of host halo mass, whereas there is substantial variation in stellar mass and inferred host halo mass for nearby central galaxies. For example, the volume-limited sample of $31$ massive hosts within $12~\mathrm{Mpc}$ from \cite{Carlsten220300014} spans central stellar masses of $9.9\lesssim \log(M_{*}/M_{\mathrm{\odot}})\lesssim 11.1$, and \cite{Danieli221014233} models $27$ hosts from this population (excluding galaxy groups) with semianalytic realizations of host halos with $10.5\lesssim \log(M_{\mathrm{host}}/M_{\mathrm{\odot}})\lesssim 13.3$. Our forecasts should therefore be viewed as illustrative estimates of the constraining power from future dwarf satellite galaxy surveys rather than specific predictions for upcoming observations. We discuss aspects of our simulations and galaxy--halo connection model that may be generalized to capture a broader range of host halo masses, secondary properties, and environments in Section \ref{sec:theory}.

\subsection{Galaxy--Halo Connection Model}
\label{sec:gh_model}

We use the CDM galaxy--halo connection model from \cite{Nadler180905542,Nadler191203303} with a few minor modifications described below. This model associates satellite galaxies with subhalos using extrapolated abundance-matching relations to predict satellite luminosity and size. Each satellite's absolute $V$-band magnitude, $M_V$, is predicted from its subhalo's peak maximum circular velocity, $V_{\mathrm{peak}}$, using an extrapolated abundance-matching relation anchored to the GAMA luminosity function at $M_r=-13~\mathrm{mag}$ \citep{Loveday150501003}, assuming a value for the faint-end luminosity function slope $\alpha$, and for the scatter $\sigma_M$ in absolute magnitude at fixed $V_{\mathrm{peak}}$.\footnote{After performing abundance matching, we convert $r$-band to $V$-band magnitude via $M_V=M_r+0.2~\mathrm{mag}$ \citep{Geha170506743,Nadler180905542}.} Both $\sigma_M$ and $\alpha$ are free parameters in our forecasts. Stellar mass--halo mass relation constraints derived from independent galaxy samples (e.g., \citealt{Danieli221014233}) are consistent with the relation derived from GAMA, lending confidence to our approach.

Each satellite's mean azimuthally averaged projected half-light radii, $r_{1/2}$ is predicted following \cite{Kravstov12122980} via
\begin{equation}
    r_{1/2} = \mathcal{A}\left(\frac{R_{\mathrm{vir,acc}}}{R_0}\right)^n,
\label{eq:r12}\end{equation}
where $R_{\mathrm{vir,acc}}$ denotes a subhalo's virial radius, measured when it accretes into its host's virial radius; $R_0=10\ \mathrm{kpc}$ is a fixed normalization constant. The amplitude parameter $\mathcal{A}$ and power-law slope $n$ are free parameters in our analyses. After mean sizes are predicted for a given set of model parameters, each satellite's half-light radius is drawn from a lognormal distribution with scatter $\sigma_{\log R}$, which is also a free parameter. A power-law galaxy--halo size relation as in Equation~\ref{eq:r12} has been shown to fit a variety of hydrodynamic simulation results reasonably well (e.g., \citealt{Jiang180407306}), although some semianalytic models (SAMs) predict a break in the size relation for ultrafaint dwarf galaxies, which we do not attempt to model (e.g., \citealt{Kravtsov210609724}).

We model the fraction $f_{\mathrm{gal}}$ of subhalos that host galaxies as a function of peak subhalo virial mass, $M_{\mathrm{peak}}$, using a modified version of the model in \cite{Graus180803654} via
\begin{equation}
    f_{\mathrm{gal}}(M_{\mathrm{peak}}) = \frac{1}{2}\left[1+\mathrm{erf}\left(\frac{\log(M_{\mathrm{peak}}/M_{\mathrm{\odot}})-\log(\mathcal{M}_{50}/M_{\mathrm{\odot}})}{\sqrt{2}\mathcal{S}_{\mathrm{gal}}}\right)\right],
\label{eq:fgal}\end{equation}
where $\mathcal{M}_{50}$ is the peak virial mass at which $50\%$ of halos host galaxies. $\mathcal{S}_{\mathrm{gal}}$ parameterizes the occupation fraction shape; smaller $\mathcal{S}_{\mathrm{gal}}$ implies a steeper cutoff, and $f_{\mathrm{gal}}$ approaches a step function at $\mathcal{M}_{50}$ as $\mathcal{S}_{\mathrm{gal}}\rightarrow 0$.\footnote{This parameter is dimensionless, so we label it as $\mathcal{S}_{\mathrm{gal}}$ rather than $\sigma_{\mathrm{gal}}$ as in \cite{Nadler191203303}.} Both $\mathcal{M}_{50}$ and $\mathcal{S}_{\mathrm{gal}}$ are free parameters in our forecasts. Note that occupation fraction shapes predicted by hydrodynamic simulations and SAMs are similar to Equation~\ref{eq:fgal} (e.g., \citealt{Munshi210105822,Kravtsov210609724,Ahvazi230813599}).

Finally, we model subhalo disruption due to the central galaxy using the model from \cite{Nadler171204467}, which was calibrated to the hydrodynamic zoom-in simulations from \cite{Garrison-Kimmel170103792}, which used the Feedback In Realistic Environments (FIRE) code. We apply this model to the subhalos of each host in the Symphony MW suite while allowing for variations in the strength of subhalo disruption relative to FIRE by setting
\begin{equation}
p_{\mathrm{disrupt}}=\left(p_{\mathrm{disrupt},0}\right)^{1/\mathcal{B}},\label{eq:pdisrupt}
\end{equation}
where $p_{\mathrm{disrupt},0}$ is the fiducial subhalo disruption probability assigned by the \cite{Nadler171204467} model, which depends on each subhalo's mass and maximum circular velocity at infall, accretion time, and orbital properties. Note that we do not modify $p_{\mathrm{disrupt},0}$ in our beyond-CDM scenarios because it mainly depends on a given subhalo's accretion time and orbital properties \citep{Garrison-Kimmel170103792,Nadler171204467}, which do not significantly differ between CDM and models like WDM (e.g., \citealt{Knebe08021628}).

To implement the disruption model in our analysis, each satellite is weighted by $1-p_{\mathrm{disrupt}}$ when number counts are computed from realizations of our model. In Equation~\ref{eq:pdisrupt}, $\mathcal{B}$ is a free parameter that controls the strength of disruption. In particular, smaller (larger) values of $\mathcal{B}$ correspond to weaker (stronger) disruption due to baryons, and $\mathcal{B}=1$ corresponds to the fiducial amount of disruption measured in FIRE simulations. We adopt a wide prior on $\mathcal{B}$ because the efficiency of subhalo disruption due to central galaxies (e.g., \citealt{Webb200606695,Green211013044}) and the potential numerical origins of CDM subhalo disruption in general \citep{VandenBosch171105276,Errani200107077,Green210301227,Benson220601842,Mansfield230810926} are highly uncertain; our forecasts conservatively bracket these uncertainties.

For simplicity, we do not include orphan subhalos in our forecasts. \cite{Nadler191203303} showed, using the orphan model from \cite{Nadler180905542}, that orphans do not affect galaxy--halo connection constraints derived from current MW satellite observations in our modeling framework (however, see \citealt{Bose190904039}). Because our main analyses do not consider fainter satellites (and thus lower-mass subhalos) than the \cite{Nadler191203303} analysis, we do not expect orphans to impact our results. Moreover, the convergence analyses in \cite{Nadler220902675} indicate that our simulations accurately predict subhalo abundances above the fiducial resolution cuts described above. Modeling dwarf galaxies that occupy halos with $M_{\mathrm{vir}}\lesssim 10^8~M_{\mathrm{\odot}}$ would require modeling highly stripped subhalos, which is difficult for standard halo finders \citep{Mansfield230810926} but will be an important area for future work.

In summary, we model satellite galaxy populations in CDM with eight unknown parameters: the galaxy--halo luminosity parameters $\alpha$ and $\sigma_M$, the galaxy--halo size parameters $\mathcal{A}$, $n$, and $\sigma_{\log R}$, the occupation fraction parameters $\mathcal{M}_{50}$ and $\mathcal{S}_{\mathrm{gal}}$, and the disruption parameter $\mathcal{B}$.

\subsection{Modeling Subhalo Populations Beyond CDM}
\label{sec:ncdm_model}

The galaxy--halo connection model described in Section~\ref{sec:gh_model} can easily be applied to beyond-CDM scenarios. Here, we rely on subhalo populations from our CDM simulations (Section~\ref{sec:zoom}) to analyze WDM and models with a generalized SHMF. Thus, our analyses of these scenarios rely on the assumption that the following input subhalo properties do not significantly differ relative to CDM: $V_{\mathrm{peak}}$ (for the luminosity model), $R_{\mathrm{vir,acc}}$ (for the size model), $M_{\mathrm{peak}}$ (for the occupation fraction model), and (mainly) accretion time and orbital properties (for the disruption model). In addition, the radial and angular distribution of subhalos is taken directly from our CDM simulations. We justify these assumptions for the WDM and generalized SHMF cases below.

\subsubsection{Thermal-relic WDM}
\label{sec:wdm_model}

We consider thermal-relic WDM as a benchmark model with suppressed small-scale power and low-mass (sub)halo abundances \citep{Bond1983,Bode0010389}. The WDM particle mass, $m_{\mathrm{WDM}}$, sets the free-streaming scale below which the linear matter power spectrum is suppressed. We parameterize this cutoff via \citep{Nadler200800022}
\begin{equation}
    M_{\mathrm{hm}}(m_{\mathrm{WDM}}) = 5 \times 10^8 \left(\frac{m_{\mathrm{WDM}}}{3\ \mathrm{keV}}\right)^{-10/3} M_{\mathrm{\odot}},\label{eq:mhm_mwdm}
\end{equation}
where the half-mode mass, $M_{\mathrm{hm}}$, is related to the unknown WDM particle mass, $m_{\mathrm{WDM}}$. In particular, $M_{\mathrm{hm}}$ is defined by the wavenumber at which the linear matter power spectrum is suppressed by $75\%$ relative to CDM \citep{Nadler190410000}. Thus, theoretical uncertainties in Equation~\ref{eq:mhm_mwdm} are small, and we hold this relation fixed in our forecasts.\footnote{\cite{Vogel221010753} found that Equation~\ref{eq:mhm_mwdm} is slightly inaccurate, at the $\approx 10\%$ level, when compared to transfer functions output by linear Boltzmann solvers for~$m_{\mathrm{WDM}}\gtrsim 3\ \mathrm{keV}$ (also see Appendix C of \citealt{Decant211109321}). We do not correct for this discrepancy here because the (sub)halo mass function suppression we adopt is based on Equation~\ref{eq:mhm_mwdm}; this will be addressed in a forthcoming study (E.~O.~Nadler et al.\ 2024, in preparation).} 

We model the suppression of the SHMF using the fit to WDM zoom-in simulations from \cite{Lovell13081399},
\begin{equation}
    f_\mathrm{WDM}(M_{\mathrm{peak}},m_{\mathrm{WDM}}) = \left[1+\left(\frac{\alpha M_{\mathrm{hm}}(m_{\mathrm{WDM}})}{M_{\mathrm{peak}}}\right)^{\beta}\right]^{\gamma},\label{eq:wdm_shmf}
\end{equation}
where $f_\mathrm{WDM}\equiv (dn_{\mathrm{WDM}}/dM_{\mathrm{peak}})/(dn_{\mathrm{CDM}}/dM_{\mathrm{peak}})$ with $\alpha=2.7$, $\beta=1.0$, and $\gamma=-0.99$. For the WDM analyses we present below, $M_{\mathrm{hm}}$ is a free parameter and each satellite is weighted by $f_{\mathrm{WDM}} \times (1-p_{\mathrm{disrupt}})$ when number counts are computed from realizations of our model. 

We focus on models with $m_{\mathrm{WDM}}\gtrsim 5~\mathrm{keV}$ ($M_{\mathrm{hm}}\gtrsim 10^8~M_{\mathrm{\odot}}$). As a result, we primarily study halos above the half-mode mass (and well above the free-streaming mass) in our forecasts. In this regime, WDM simulations indicate that $M_{\mathrm{peak}}$ and $V_{\mathrm{peak}}$ are only reduced by $\mathcal{O}(10\%)$ relative to matched CDM halos (e.g., \citealt{Bozek180305424,Fitts181111791}). We expect $R_{\mathrm{vir,acc}}$ to be reduced by a similar amount because it is mainly determined by $M_{\mathrm{peak}}$ and the amount of pre-infall stripping a subhalo experiences, which does not differ significantly between the models (e.g., \citealt{Elahi14063413}). Meanwhile, as described in Section~\ref{sec:gh_model}, orbital properties of given WDM subhalos are not significantly altered relative to CDM.

Finally, we assume that the WDM subhalo radial distribution is unchanged relative to CDM. WDM simulations of MW-mass hosts indicate that this is a good assumption for subhalos more massive than $M_{\mathrm{hm}}$ \citep{Lovell210403322}, which we primarily study. Modeling lower-mass subhalos would benefit from the direct use of WDM simulations. Thus, our method for applying the model to WDM is well motivated, and we leave an extension of our framework to beyond-CDM simulations for future work.

\subsubsection{Model-independent SHMF}
\label{sec:shmf_model}

We also consider a purely empirical modification to the SHMF, in which subhalo abundances vary above or below those in CDM, as a function of peak subhalo mass. Specifically, we bin in decades of subhalo mass and define
\begin{equation}
    \xi_i \equiv \log\left[f_{\mathrm{NCDM}}(10^{i-0.5}<M_{\mathrm{peak}}/M_{\mathrm{\odot}}<10^{i+0.5})\right],
\end{equation}
where $f_\mathrm{NCDM}\equiv (dn_{\mathrm{NCDM}}/dM_{\mathrm{peak}})/(dn_{\mathrm{CDM}}/dM_{\mathrm{peak}})$. In this model, $\xi_8$, $\xi_9$, and $\xi_{10}$ parameterize SHMF deviations from CDM centered on peak masses of $10^8$, $10^9$, and $10^{10}~M_{\mathrm{\odot}}$, respectively. For our SHMF forecasts, $\xi_8$, $\xi_9$, and $\xi_{10}$ are free parameters, and each satellite is weighted by $f_{\mathrm{NCDM}} \times (1-p_{\mathrm{disrupt}})$ when number counts are computed from realizations of our model. 

By construction, our generalized SHMF scenarios do not alter any of the subhalo properties relevant for our galaxy--halo connection model or quantities like the radial distribution. Although these assumptions may be broken in specific beyond-CDM scenarios, this approach allows us to isolate the effects of varying the SHMF.

\subsection{Model Illustration}
\label{sec:illustration}

We will generate ``true'' satellite populations for our forecasts by evaluating the galaxy--halo connection model at fixed values of $\sigma_M$, $\mathcal{M}_{50}$, $\mathcal{S}_{\mathrm{gal}}$, and beyond-CDM parameters; the remaining parameters are always set to the best-fit values derived from the MW satellite population in \cite{Nadler191203303}, i.e., $\alpha=-1.44$, $\mathcal{B}=0.93$, $\sigma_{\log R}=0.63~\mathrm{dex}$, and $\mathcal{A}=37~\mathrm{pc}$. However, all eight galaxy--halo connection parameters and additional beyond-CDM parameters are then left free and inferred in our forecasting analyses.

\begin{figure}[t!]
\hspace{-2.5mm}
\includegraphics[trim={0 0cm 0cm 0.3cm},width=0.515\textwidth]{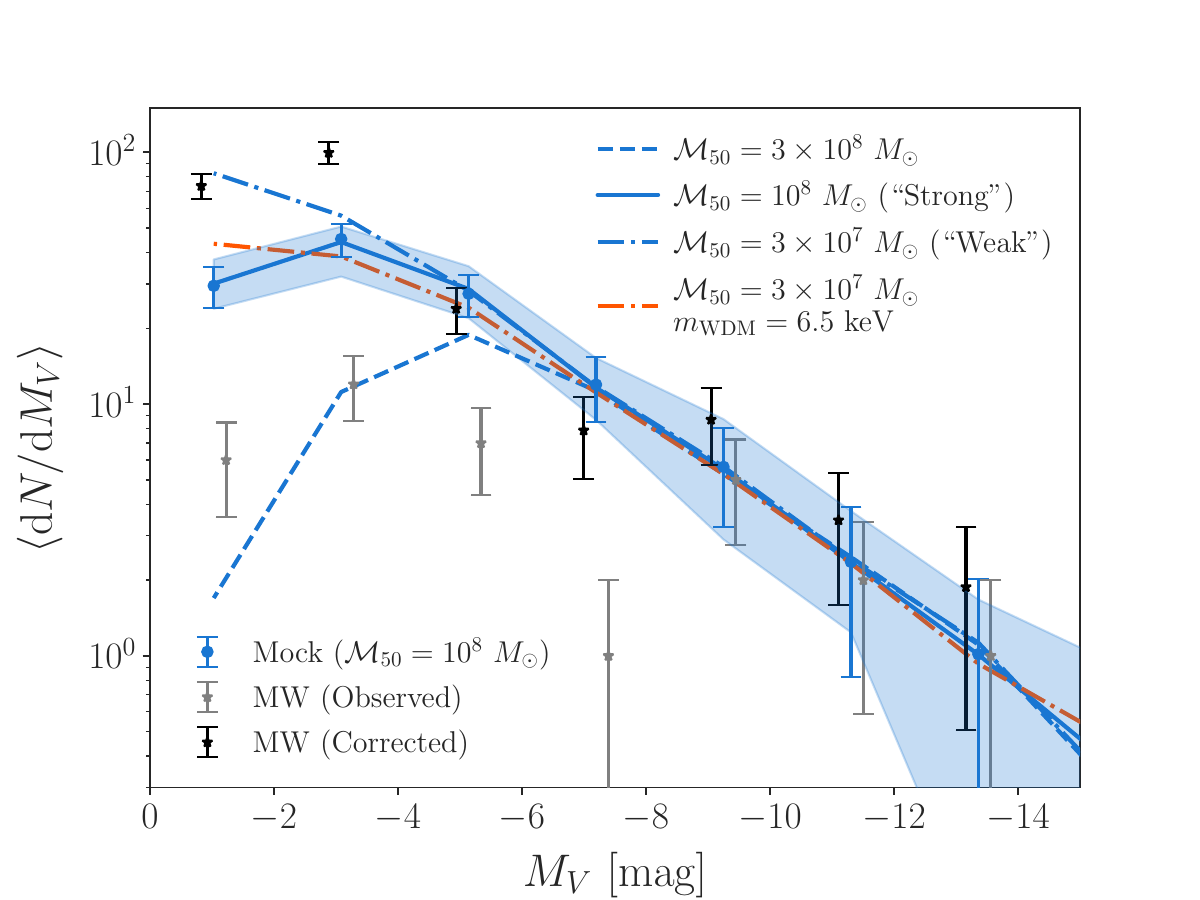}
\caption{Mean satellite luminosity function and $1\sigma$ Poisson uncertainty for the MW satellite population observed by DES and PS1 (gray), and a completeness-corrected total luminosity function from (\citealt{Drlica-Wagner191203302}; black). Blue error bars show the predicted luminosity function mean and $1\sigma$ Poisson uncertainty for $40$ complete satellite populations of MW-mass hosts generated from our galaxy--halo connection model with a ``strong'' galaxy formation cutoff ($\mathcal{M}_{50}=10^8~M_{\mathrm{\odot}}$, $\sigma_M=0.2~\mathrm{dex}$, and $\mathcal{S}_{\mathrm{gal}}=0.2$; all remaining galaxy--halo connection parameters are fixed to the input values listed in Section~\ref{sec:illustration}). Blue lines show predictions for the same $\sigma_M$ and $\mathcal{S}_{\mathrm{gal}}$, with $\mathcal{M}_{50}=3\times 10^8$ (dashed), $10^8$ (solid), and $3\times 10^7~M_{\mathrm{\odot}}$ (dotted--dashed, which we refer to as a ``weak'' cutoff throughout). The blue band shows $16$th--$84$th percentile host-to-host scatter for $\mathcal{M}_{50}=10^8~M_{\mathrm{\odot}}$. All predictions assume a satellite detectability threshold of $M_V<0~\mathrm{mag}$ and $\mu<32~\mathrm{mag\ arcsec}^{-2}$, and error bars are offset horizontally for clarity.}
\label{fig:lf}
\end{figure}

\begin{figure*}[t!]
\includegraphics[width=\textwidth]{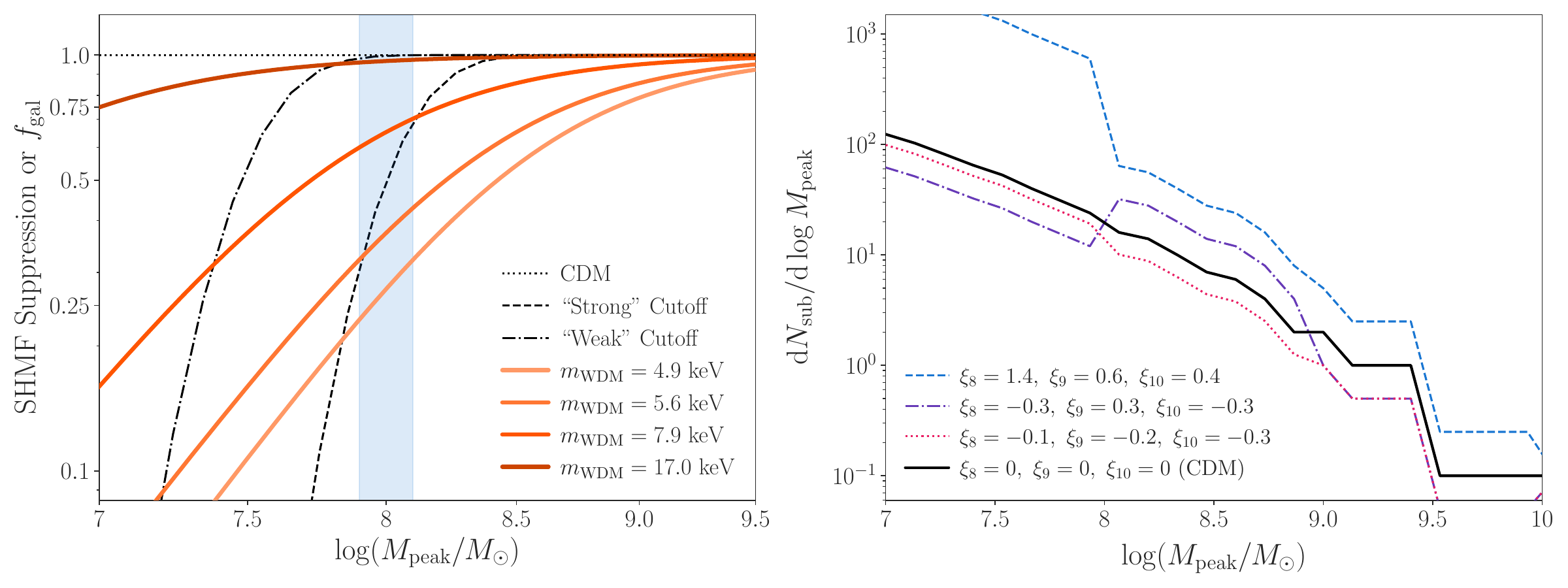}
\caption{Left: suppression of the subhalo mass function (defined as the ratio of the WDM to CDM SHMF) for WDM models with $m_{\mathrm{WDM}}=4.9$, $5.6$, $7.9$, and $17.0~\mathrm{keV}$ from lightest to darkest red. Our ``strong'' and ``weak'' galaxy formation cutoff scenarios are shown on the same axes. 
The blue band shows the range of $\mathcal{M}_{50}$ probed at $68\%$ confidence by our ``strong'' galaxy formation cutoff forecast, and thus illustrates the range of halo masses we typically probe. Right: mean SHMF (black) from our $40$ CDM MW-mass zoom-in simulations, measured using peak virial mass. Dashed blue, dotted--dashed purple, and dotted red lines illustrate the effects of our model-independent SHMF parameterization, which can enhance or suppress the SHMF relative to CDM as a function of peak halo mass, depending on the values of $\xi_8$, $\xi_9$, and $\xi_{10}$. The uniformly suppressed (dashed blue) and enhanced (dotted red) models correspond to our projections for $68\%$ confidence SHMF constraints from one complete host.}
\label{fig:wdm_pred}
\end{figure*}

Before performing these full parameter recovery tests, we illustrate the predictions of our model and the constraining power of satellite population measurements. Figure~\ref{fig:lf} shows mock observations of mean satellite luminosity functions from our $40$ MW-mass zoom-in simulations. We compare mock observations of ``true'' satellite luminosity functions in a model with $\mathcal{M}_{50}=10^8~M_{\mathrm{\odot}}$ (shown as error bars) to predictions with $\mathcal{M}_{50}=3\times 10^7$, $10^8$, and $3\times 10^8~M_{\mathrm{\odot}}$ (shown as lines). Below, we refer to $\mathcal{M}_{50}=10^8~M_{\mathrm{\odot}}$ ($3\times 10^7~M_{\mathrm{\odot}}$) as a ``strong'' (``weak'') cutoff scenario, and we set $\sigma_M=0.2~\mathrm{dex}$, and $\mathcal{S}_{\mathrm{gal}}=0.2$ in both scenarios; see Section~\ref{sec:reion} for details. Note that all remaining galaxy--halo connection parameters are fixed to the input values listed above.

For this illustration, we optimistically assume that all satellites of each host with $M_V<0~\mathrm{mag}$ and $\mu_V<32~\mathrm{mag\ arcsec}^{-2}$ are detectable; we expand on our observational assumptions in Section~\ref{sec:obsevational} and discuss them in the context of near and long-term observational capabilities in Sections~\ref{sec:obs} and \ref{sec:future}. Given this assumption, the model with $\mathcal{M}_{50}= 3\times 10^8~M_{\mathrm{\odot}}$ is detectable using the complete satellite population of a single MW-mass host. Meanwhile, models with $\mathcal{M}_{50}\lesssim 10^8~M_{\mathrm{\odot}}$ are more difficult to detect because they only affect the abundances of the faintest satellites we consider. For comparison, the gray points in Figure~\ref{fig:lf} show the luminosity function of kinematically confirmed and candidate MW satellites detected by DES and PS1 \citep{Drlica-Wagner191203302}. Because MW satellite observations are incomplete, we also show the completeness-corrected estimate from \cite{Drlica-Wagner191203302} for comparison. Note that \cite{Nadler191203303} inferred $\mathcal{M}_{50}<8.5\times 10^7~M_{\mathrm{\odot}}$ at $95\%$ confidence using these satellites, including the observational selection function derived in \cite{Drlica-Wagner191203302}. 

Figure~\ref{fig:lf} also shows a ``weak'' galaxy formation cutoff combined with an $m_{\mathrm{WDM}}=6.5~\mathrm{keV}$ WDM model, corresponding to the $95\%$ confidence WDM limit from the MW satellite population derived in \cite{Nadler200800022}. These astrophysical and DM models combine to yield a luminosity function similar to a ``strong'' galaxy formation cutoff in CDM, illustrating how degeneracies between galaxy formation and DM physics can manifest in dwarf galaxy populations.

To illustrate our beyond-CDM predictions, the left panel of Figure~\ref{fig:wdm_pred} compares the suppression of dwarf galaxy abundances in our ``strong'' and ``weak'' galaxy formation cutoff scenarios to the suppression of the SHMF in various WDM models. We show WDM masses of $m_{\mathrm{WDM}}=4.9~\mathrm{keV}$ (the fiducial value in our WDM forecasts), $5.6~\mathrm{keV}$ (the limit from one complete satellite population in our results below, assuming a ``strong'' galaxy formation cutoff), $7.9~\mathrm{keV}$ (constrained at $68\%$ confidence by the MW satellite population; \citealt{Nadler200800022}), and $17.0~\mathrm{keV}$ (constrained by two complete satellite in our results below, again assuming a ``strong'' cutoff). The shapes of the galaxy occupation fraction and SHMF cutoffs differ in detail; thus, sufficiently precise measurements of dwarf galaxy abundances can potentially disentangle these effects. However, note that the occupation fraction shape is fixed to $\mathcal{S}_{\mathrm{gal}}=0.2$ in this illustration, while our forecasts treat $\mathcal{S}_{\mathrm{gal}}$ as an unknown parameter.

Finally, the right panel of Figure~\ref{fig:wdm_pred} illustrates our model-independent SHMF parameterization. We show uniformly suppressed (dotted red) and enhanced (dashed blue) models constrained at $68\%$ confidence by one complete satellite population in our forecasts; we also show a model that mixes SHMF suppression and enhancement in different peak virial mass decades (dotted--dashed purple) at a level that is consistent with our projected dwarf galaxy sensitivity (see Section~\ref{sec:shmf_results} for details). For context, the size of these SHMF deviations relative to CDM is comparable to the SHMF sensitivity inferred from recent stellar stream and strong lensing measurements \citep{Gilman190806983,Banik191102662}. Both the shape and normalization of the SHMF can be altered in our parameterization, and only some of this parameter space can be mimicked by galaxy formation or WDM cutoffs.


\section{Forecasting Framework}
\label{sec:forecasting}

Our forecasting framework starts from realizations of the simulated satellite population following Section~\ref{sec:model}. We then analyze these mock data using probabilistic inference. Our observational assumptions are outlined in Section \ref{sec:obsevational}, our likelihood formalism is described in Section \ref{sec:likelihood}, and our evaluation of the likelihood is described in Section \ref{sec:evaluation}.

\subsection{Observational Assumptions}
\label{sec:obsevational}

We make the following observational assumptions for all forecasts presented below:
\begin{enumerate}
    \item All satellite galaxies within the virial radius of each MW-mass host that have $M_V<0\ \mathrm{mag}$ and $\mu_V<32\ \mathrm{mag\ arcsec}^{-2}$ are detectable, where $\mu_V \equiv M_V + 36.57 + 2.5\log[2\pi (r_{1/2}/\mathrm{kpc})^{2}]$ is the effective surface brightness averaged within the half-light radius.
    \item Host halo properties (and all latent variables that determine a host's subhalo population) are perfectly known.
\end{enumerate}
The second assumption is implicit in our procedure and follows because we compare predicted satellite populations, generated using the subhalo population around a given simulated host, to a ``true'' satellite population generated from the same host. Both assumptions are optimistic; we make these choices in order to assess the potential of dwarf satellite populations for constraining galaxy formation and DM physics. We place these assumptions in the context of current and future observational sensitivity and discuss related areas for modeling work in Section~\ref{sec:caveats}.

\subsection{Likelihood}
\label{sec:likelihood}

To fit the mock data and assess parameter uncertainties, we generalize the likelihood framework from \cite{Nadler180905542,Nadler191203303}, following \cite{Nadler:2021fkn}. In particular, we model realizations of satellite luminosities and sizes as a Poisson process, drawn from a distribution uniquely predicted by a set of parameter values $\boldsymbol{\theta} = \{\alpha, \sigma_M, \mathcal{A}, n, \sigma_{\log R}, \mathcal{M}_{50}, \mathcal{S}_{\mathrm{gal}}, \mathcal{B}\}$, and additionally $M_\mathrm{hm}$ or $\xi_8$, $\xi_9$, and $\xi_{10}$ in our beyond-CDM forecasts. The probability of detecting $n_{\mathrm{obs},ij}$ satellites in luminosity and size bin $j$, in host $i$, is given by
\begin{align}
&P(n_{\mathrm{obs},ij}|\boldsymbol{\theta})& \nonumber \\ &= \Big(\frac{N_{\mathrm{real.}}+1}{N_{\mathrm{real.}}}\Big)^{-({N}_{ij} + 1)}\times (N_{\mathrm{real.}}+1)^{-n_{\mathrm{obs},ij}}\frac{({N}_{ij} + n_{\mathrm{obs},ij})!}{n_{\mathrm{obs},ij}!{N}_{ij}!},&\label{eq:like}
\end{align}
where ${N_\mathrm{real.}}$ is the total number of realizations per host at a given $\boldsymbol{\theta}$, ${N}_{ij}\equiv \sum_{k=1}^{{N_\mathrm{real.}}}n_{\mathrm{pred},ijk}$, and $n_{\mathrm{pred},ijk}$ is the number of predicted satellites in host $i$, bin $j$, and realization $k$. In practice, we replace the factorials in Equation~\ref{eq:like} with appropriate Gamma functions because our model produces noninteger satellite counts.

Before moving on, we provide context for this likelihood. In particular, Equation~\ref{eq:like} is a modified Poisson likelihood derived by marginalizing over the (unknown) underlying Poisson rate in each bin of observable parameter space via \citep{Nadler180905542} 
\begin{align}
&P(n_{\mathrm{obs},ij}|n_{\mathrm{pred},ij1},\dots,n_{\mathrm{pred},ijN_{\mathrm{real.}}})&\nonumber \\ &= \int P(n_{\mathrm{obs},ij}|\lambda_{ij})P(\lambda_{ij}|n_{\mathrm{pred},ij1},\dots,n_{\mathrm{pred},ijN_{\mathrm{real.}}})\ \text{d}\lambda_{ij}&\nonumber \\
&=\frac{1}{P(n_{\mathrm{pred},ij1},\dots,n_{\mathrm{pred},ijN_{\mathrm{real.}}})}& \nonumber \\ &\times \int P(n_{\mathrm{obs},ij}|\lambda_{ij})P(n_{\mathrm{pred},ij1}|\lambda_{ij})\cdots P(n_{\mathrm{pred},ijN_{\mathrm{real.}}}|\lambda_{ij})P(\lambda_{ij})\ \text{d}\lambda_{ij} &\nonumber \\
&= \Big(\frac{N_{\mathrm{real.}}+1}{N_{\mathrm{real.}}}\Big)^{-(n_{\mathrm{pred},ij1} + \dots + n_{\mathrm{pred},ijN_{\mathrm{real.}}} + 1)}\nonumber \\ &\times (N_{\mathrm{real.}}+1)^{-n_{\mathrm{obs},ij}}\frac{(n_{\mathrm{pred},ij1} + \dots + n_{\mathrm{pred},ijN_{\mathrm{real.}}} + n_{\mathrm{obs},ij})!}{n_{\mathrm{obs},ij}!(n_{\mathrm{pred},ij1} + \dots + n_{\mathrm{pred},ijN_{\mathrm{real.}}})!},&\label{eq:like_marginalized}
\end{align}
Here, the final expression is obtained by substituting Poisson likelihoods for each conditional probability and a flat prior on positive values of $\lambda_{ij}$. The key assumption is that mock and observed satellites are drawn from the same (unknown) Poisson distribution; our likelihood converges to this distribution in the limit of many realizations \citep{Nadler180905542}.

When evaluating Equation~\ref{eq:like}, we include $N_\mathrm{real.}$ realizations at a given $\boldsymbol{\theta}$ because our galaxy--halo connection model is stochastic. In this study, we use~$N_\mathrm{real.}=10$ per host at a given~$\boldsymbol{\theta}$. We have checked that varying $N_{\mathrm{real.}}$ does not impact our results; in particular, our Markov Chain Monte Carlo (MCMC) evaluation of the posterior converges in fewer steps when $N_\mathrm{real.}$ is larger, but the posterior distribution and total run time are unchanged. We follow \cite{Nadler191203303} by using $14$ bins in $M_V$, evenly spaced from $0$ to $-20\ \mathrm{mag}$, and two bins in surface brightness that split the sample at $\mu_V=28\ \mathrm{mag\ arcsec}^{-2}$.

The joint likelihood of observing all galaxies in our mock data vector $\mathbf{s}_{\mathrm{obs}}$, at each set of theoretical parameter values $\boldsymbol{\theta}$ for all the hosts we consider in a given analysis is   
\begin{equation}
\mathcal{L}
(\mathbf{s}_{\mathrm{obs}}|\boldsymbol{\theta})=\prod_{\mathrm{hosts\ } i}\ \prod_{\mathrm{  bins\ } j} P(n_{\mathrm{obs},ij}|\boldsymbol{\theta}).\label{eq:likelihood}
\end{equation}
The mock data vector $\mathbf{s}_{\mathrm{obs}}$ is generated from one realization of our model with the same fixed galaxy--halo connection and (if applicable) beyond-CDM parameters for all hosts, with specific values described for each forecast below. 

We use Bayes' theorem to compute the joint posterior distribution over $\boldsymbol{\theta}$ for each forecast,
\begin{equation}
P(\boldsymbol{\theta}|\mathbf{s}_{\mathrm{obs}}) = \frac{\mathcal{L}(\mathbf{s}_{\mathrm{obs}}|\boldsymbol{\theta})P(\boldsymbol{\theta})}{P(\mathbf{s}_{\mathrm{obs}})},\label{eq:posterior}
\end{equation}
where $P(\boldsymbol{\theta})$ is the joint prior distribution over $\boldsymbol{\theta}$ and $P(\mathbf{s}_{\mathrm{true}})$ is the Bayesian evidence. Appendix~\ref{sec:priors} provides all prior distributions used in our forecasts.

\begin{deluxetable*}{{c@{\hspace{0.2in}}cccc}}[t!]
\centering
\tablecolumns{5}
\tablecaption{Summary of Galaxy Formation, WDM, and SHMF Forecast Results}
\tablehead{\colhead{Forecast} & \colhead{Parameter(s)} & \colhead{Input} & \colhead{1 Host} & \colhead{2 Hosts}}
\startdata
Galaxy Formation: ``weak'' cutoff (Section~\ref{sec:reion})
& $\log(\mathcal{M}_{50}/M_{\mathrm{\odot}})$ 
& $7.5$ & $7.1^{+0.3}_{-1.6^{**}}$ & $5.3^{+1.4}_{-0.2^{**}}$\\
Galaxy Formation: ``strong'' cutoff (Section~\ref{sec:reion})
& $\log(\mathcal{M}_{50}/M_{\mathrm{\odot}})$ 
& $8.0$ & $7.9^{+0.4}_{-1.6^{**}}$ & $8.0^{+0.1}_{-0.1^{**}}$\\
\hline
WDM: Scenario A (Section~\ref{sec:wdm})
& $\log(M_{\mathrm{hm}}/M_{\mathrm{\odot}})$
&  $5.0$ &  $6.1^{+0.5}_{-1.1^{**}}$ & $5.0^{+1.0}_{-0.0^{**}}$\\
WDM: Scenario B (Section~\ref{sec:wdm})
& $\log(M_{\mathrm{hm}}/M_{\mathrm{\odot}})$
&  $5.0$ &  $7.4^{+0.4}_{-1.5^{**}}$ &  $5.0^{+1.2}_{-0.0^{**}}$\\
WDM: Scenario C (Section~\ref{sec:wdm})
& $\log(M_{\mathrm{hm}}/M_{\mathrm{\odot}})$
&  $8.0$ &  $8.0^{+0.4}_{-1.5^{**}}$ &  $7.6^{+0.3}_{-1.8^{**}}$\\
\hline
SHMF: ``weak'' cutoff (Section~\ref{sec:shmf})
& $\xi_{8}$, $\xi_{9}$, $\xi_{10}$
&  $0.0$, $0.0$, $0.0$ &  $0.2^{+1.2^{**}}_{-0.3}$, $0.1^{+0.5}_{-0.3}$, $0.1^{+0.3}_{-0.4}$ & $0.5^{+0.4}_{-0.3}$, $0.3^{+0.2}_{-0.1}$, $0.1^{+0.2}_{-0.2}$ \\
SHMF: ``strong'' cutoff (Section~\ref{sec:shmf})
& $\xi_{8}$, $\xi_{9}$, $\xi_{10}$
&  $0.0$, $0.0$, $0.0$ & $0.1^{+0.9}_{-1.3^{**}}$, $0.1^{+0.7}_{-0.8}$, $0.0^{+0.3}_{-1.2^{**}}$ & $0.2^{+0.8}_{-0.3}$, $0.1^{+0.2}_{-0.2}$, $0.0^{+0.2}_{-0.2}$ \\
\hline
\hline
\enddata
{\footnotesize \tablecomments{The first column describes the forecast, the second column lists the parameter(s) of interest, the third column lists parameters' input value(s), and the fourth (fifth) columns list $68\%$ confidence interval(s) from our one (two)-host forecasts. One asterisk (two asterisks) denote limits that are consistent with our priors at $68\%$ ($95\%$) confidence (see Appendix~\ref{sec:priors}).}}
\label{tab:summary}
\end{deluxetable*}

\subsection{Posterior Sampling}
\label{sec:evaluation}

We sample the posterior in Equation~\ref{eq:posterior} using the MCMC sampler \textsc{emcee} \citep{emcee}, using $450$ walkers and a combination of the \texttt{StretchMove} and \texttt{KDEMove} sampling algorithms. For our CDM and WDM forecasts, $\sim 10^6$ samples are sufficient to accurately characterize the posterior, while $\sim 5\times 10^6$ samples are required for our SHMF forecasts due to higher dimensionality and more prominent degeneracies. We discard $\sim 100$ burn-in autocorrelation lengths for each chain, which leaves more than $100$ independent samples for every scenario we consider.

It is difficult to sample our full galaxy--halo connection (plus beyond-CDM) parameter space using more than two hosts at a time. Because our model is evaluated on high-resolution cosmological simulations, likelihood evaluations are computationally expensive when using large numbers of hosts. In addition, degeneracies among galaxy--halo connection parameters (and, in our WDM and SHMF forecasts, between galaxy--halo connection and beyond-CDM parameters) slow down posterior sampling. The most challenging parameter in this context is $\sigma_M$, which is often degenerate with $\mathcal{M}_{50}$, $\mathcal{S}_{\mathrm{gal}}$, and beyond-CDM parameters that mimic a cutoff; we discuss specific degeneracies in each forecast below. Furthermore, because the SHMF rises steeply with decreasing subhalo mass, abundant low-mass subhalos preferentially up-scatter to observable luminosities (a form of Eddington bias), resulting in non-Gaussian posteriors for individual hosts. The complexity of the fit increases with the number of hosts used to compute the joint likelihood since each host has a different underlying subhalo population and associated non-Gaussian posterior.

For each forecast, we therefore evaluate the (joint) likelihood using one and two hosts to derive our key results and characterize degeneracies in the posterior. We have checked that our posteriors for individual hosts are statistically consistent among all $45$ of our Symphony MW zoom-ins, and we describe how constraints depend on combinations of two hosts below. To determine how the uncertainty of our forecasted constraints scales with the number of hosts used in the inference, $N_{\mathrm{hosts}}$, we supplement our MCMC forecasts with projections for galaxy formation cutoff, WDM, and SHMF constraints that fix all remaining galaxy--halo connection parameters to their input values.


\section{Galaxy Formation Forecasts}
\label{sec:reion}

We first forecast sensitivity to measuring a cutoff in galaxy formation within CDM, for two fiducial values of $\mathcal{M}_{50}$:
\begin{enumerate}
    \item ``Weak'' cutoff scenario: $\mathcal{M}_{50}=3\times 10^7~M_{\mathrm{\odot}}$; 
    \item ``Strong'' cutoff scenario: $\mathcal{M}_{50}=10^8~M_{\mathrm{\odot}}$.
\end{enumerate}
In both cases, we set $\sigma_M=0.2~\mathrm{dex}$, and $\mathcal{S}_{\mathrm{gal}}=0.2$. This value of $\sigma_M$ is representative of the galaxy--halo connection scatter inferred at higher masses \citep{Wechsler180403097} and is near the upper limit derived from current MW satellite observations \citep{Nadler191203303}; thus, it is a helpful benchmark for our study. Meanwhile, our fiducial value of $\mathcal{S}_{\mathrm{gal}}$ yields a galaxy occupation fraction that agrees with the predictions of state-of-the-art galaxy formation models.

In particular, the ``weak'' cutoff scenario is broadly consistent with predictions from high-resolution simulations of galaxy formation at high redshifts (e.g., \citealt{Cote171006442}) and from various SAMs (e.g., \citealt{Kravtsov210609724,Ahvazi230813599}). We note that the ``weak'' cutoff closely matches the mean galaxy occupation fraction predicted when both atomic and molecular hydrogen cooling are included in the \cite{Ahvazi230813599} SAM. Meanwhile, the ``strong'' cutoff scenario is more consistent with predictions from hydrodynamic simulations run to lower redshifts (e.g., \citealt{Sawala14066362,Fitts161102281,Munshi210105822}) and other SAMs (e.g., \citealt{Benitze-Llambay200406124}). Furthermore, the ``strong'' cutoff is similar to (but slightly weaker than) the mean galaxy occupation fraction predicted when only atomic hydrogen cooling is included in the \cite{Ahvazi230813599} SAM. We note that galaxy occupation in hydrodynamic simulations is resolution- and definition-dependent (e.g., \citealt{Nadler191203303,Ahvazi230813599}), and that higher-resolution simulations (e.g., \citealt{Wheeler181202749}) and convergence studies~(e.g., \citealt{Munshi210105822}) indicate that progressively lower-mass halos host ``galaxies'' (i.e., simulated systems with more than a critical number of star particles) as resolution increases.

The ``strong'' cutoff scenario is informative for our forecasts because it clearly reveals degeneracies between the galaxy formation cutoff, other galaxy--halo connection parameters, and beyond-CDM parameters. Note that current observations of the MW satellite population from DES and PS1 are beginning to probe the ``strong'' cutoff and its assumed scatter of $\sigma_M=0.2~\mathrm{dex}$ \citep{Nadler191203303}. Meanwhile, the ``weak'' cutoff is not currently constrained and represents a theoretical target for the sensitivity of future dwarf galaxy surveys. Our ``strong'' and ``weak'' cutoff scenarios therefore bracket a physically and observationally motivated range of galaxy formation cutoffs.

\subsection{Cutoff Mass Scale Posteriors}
\label{sec:cutoff_result}

\begin{figure*}[t!]
\includegraphics[width=\textwidth]{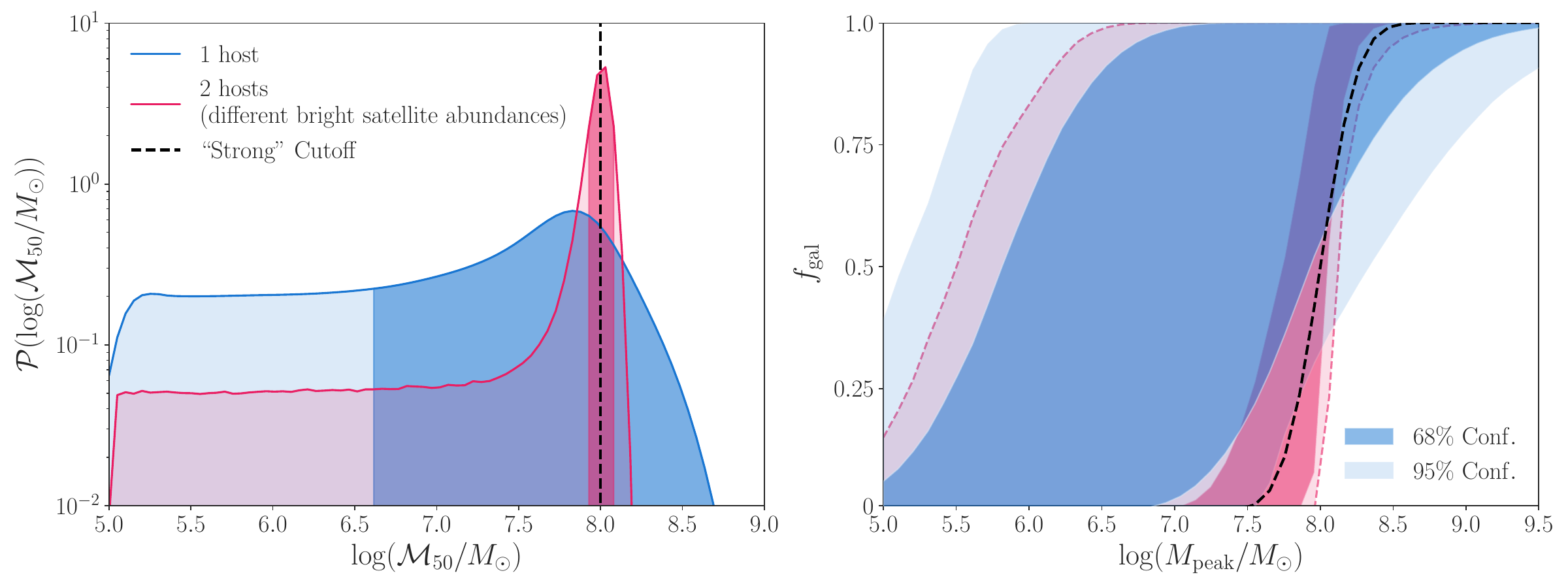}
\caption{Left: marginalized posteriors for the galaxy formation cutoff mass scale $\mathcal{M}_{50}$ (i.e., the peak virial mass at which $50\%$ of halos host galaxies with $M_V<0~\mathrm{mag}$), from our forecasted measurements of one (blue) and two (red) complete satellite populations of MW-mass hosts. Dark (light) shaded regions show $68\%$ ($95\%$) confidence intervals. Right: marginalized posteriors for the galaxy occupation fraction, defined as the fraction of halos that host galaxies with $M_V<0~\mathrm{mag}$ as a function of peak virial mass. Dark (light) bands show the $68\%$ ($95\%$) confidence intervals from our one- and two-host forecasts. In both panels, black dashed lines show the input ``strong'' galaxy formation cutoff.
}
\label{fig:reion}
\end{figure*}

Figure~\ref{fig:reion} illustrates our results for the ``strong'' cutoff scenario: the left panel shows the marginalized posterior for $\mathcal{M}_{50}$ given observations of all satellites around one (blue) and two (red) MW-mass hosts, and the right panel shows the corresponding galaxy occupation fraction posteriors. We find $\log(\mathcal{M}_{50}/M_{\mathrm{\odot}})=7.9^{+0.4}_{-1.6}$ at $68\%$ confidence for one host. Thus, $\mathcal{M}_{50}$ can be probed at the $\approx 1\sigma$ level using a single complete satellite population; however, a long tail toward low $\mathcal{M}_{50}$ prevents a measurement at the $2\sigma$ level. Note that our inferred distribution of $\mathcal{M}_{50}$ is consistent with the input value of $\log(\mathcal{M}_{50}/M_{\mathrm{\odot}})=8.0$, indicating that our galaxy--halo connection reconstruction is unbiased and that our inference framework is self-consistent.

The tail toward low $\mathcal{M}_{50}$ arises because the ``strong'' cutoff primarily suppresses satellite abundances at the lowest luminosities we consider, and this effect is not significant enough to be distinguished from models that do not suppress the abundance of satellites with $M_V<0~\mathrm{mag}$ and $\mu_V<32~\mathrm{mag\ arcsec}^{-2}$. As a result, models with $\mathcal{M}_{50}\lesssim 10^7~M_{\mathrm{\odot}}$ are not distinguishable from each other and have roughly equal probability in the marginalized posterior. Our analysis of the full one-host posterior in Appendix~\ref{sec:reion_full_posterior} shows that the low-$\mathcal{M}_{50}$ tail is enhanced by a degeneracy with the luminosity scatter $\sigma_M$. For large $\sigma_M$, galaxies hosted by abundant, low-mass halos preferentially up-scatter in luminosity, resulting in higher predicted satellite abundances at all luminosities. In this regime, galaxy--halo connection parameters adjust so that lower-mass halos explain more of the mock data; in turn, large values of $\mathcal{M}_{50}$ are disfavored because they prevent these low-mass halos from forming galaxies. In other words, higher-luminosity scatter increases the sensitivity of a fixed-depth survey to lower-mass subhalos, boosting the constraining power on large values of $\mathcal{M}_{50}$.

\begin{figure}[t!]
\includegraphics[width=0.475\textwidth]{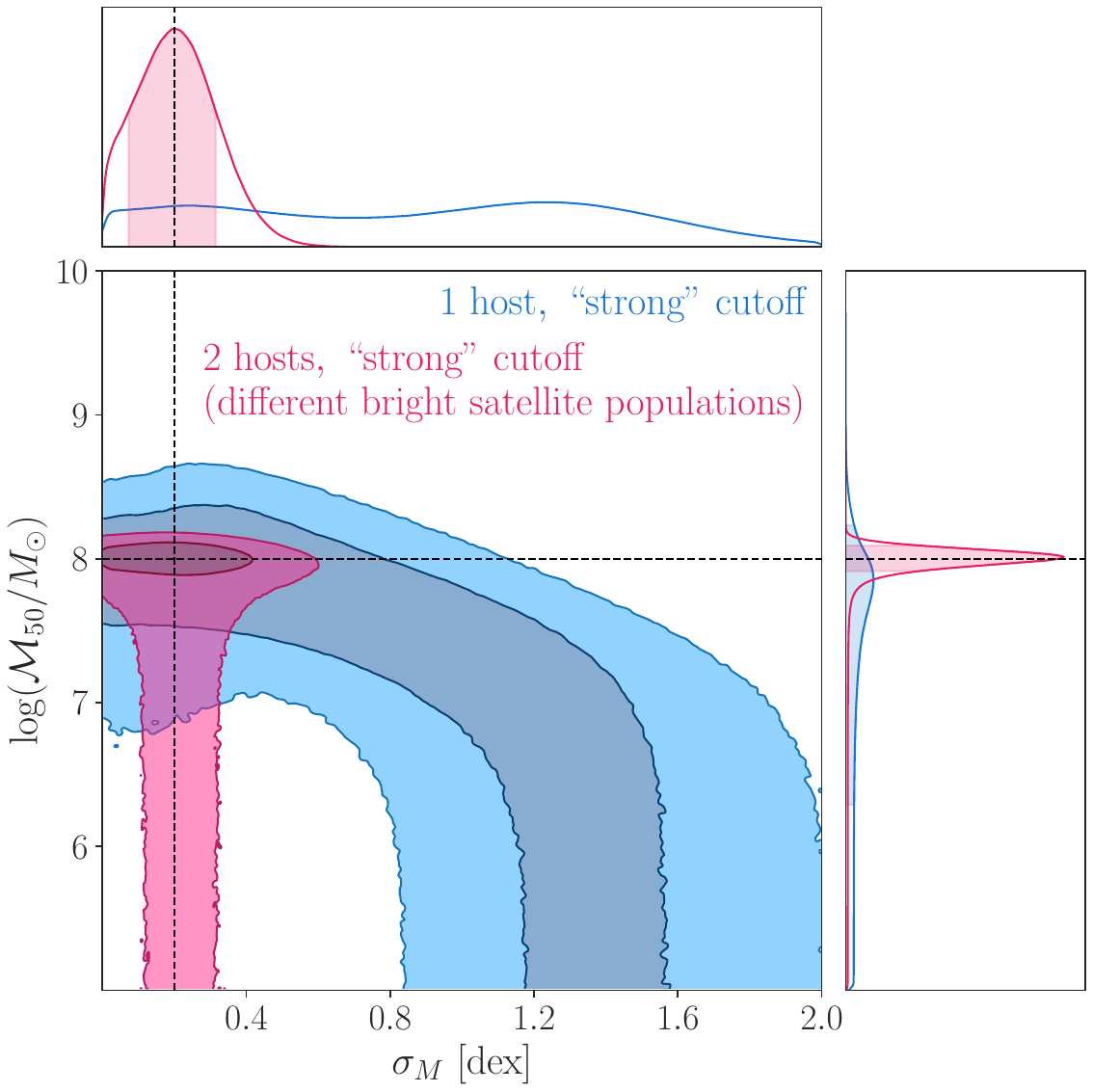}
\caption{Marginalized posterior for $\mathcal{M}_{50}$ vs.\ $\sigma_M$ from our forecast using one (blue) and two (red) complete satellite populations of MW-mass hosts in the presence of a ``strong'' galaxy formation cutoff. Dashed lines mark the input values of $\mathcal{M}_{50}$ and $\sigma_M$; two-dimensional contours represent $68\%$ and $95\%$ confidence intervals. The top and side panels show marginal posteriors, where shaded regions represent $68\%$ confidence intervals. The two hosts shown by the red contour have significantly different bright satellite abundances, which markedly improves the recovery of $\sigma_M$ and indirectly improves $\mathcal{M}_{50}$ constraints.}
\label{fig:2d_posterior}
\end{figure}

Combining the satellite populations of two hosts can improve this situation by setting a much tighter constraint on $\sigma_M$, leading to a less prominent tail toward low $\mathcal{M}_{50}$. We illustrate this effect in Figure~\ref{fig:2d_posterior} using two hosts with abundances of classical satellites ($M_V\lesssim -8~\mathrm{mag}$) that differ at a level comparable to the $1\sigma$ host-to-host luminosity function scatter in our simulation suite, shown by the blue band in Figure~\ref{fig:lf}. While $\sigma_M$ is unconstrained in our ``strong'' cutoff forecast with one host, we recover $\sigma_M=(0.2\pm 0.12)~\mathrm{dex}$ using these two hosts (recall that the input value is $\sigma_M=0.2~\mathrm{dex}$). In turn, this substantially improves our recovery of $\mathcal{M}_{50}$ to~$\log(\mathcal{M}_{50}/M_{\mathrm{\odot}})=8.0\pm 0.1$ at $68\%$ confidence. However, $\mathcal{M}_{50}$ remains difficult to measure at $95\%$ confidence due to the low-$\mathcal{M}_{50}$ tail, which persists even with two hosts. Using hosts with different bright satellite abundances effectively constrains $\sigma_M$ because the host with more classical satellites can be fit by larger values of $\sigma_M$, while the host with fewer classical satellites cannot; thus, combining such hosts breaks the $\mathcal{M}_{50}$--$\sigma_M$ degeneracy. Constraints on many other galaxy--halo connection parameters also significantly improve when such hosts are combined (see Appendix~\ref{sec:reion_full_posterior}).

In the ``weak'' cutoff scenario, our mock observations are less sensitive to $\mathcal{M}_{50}$ because galaxy formation is suppressed in a smaller fraction of halos that host observable satellites. With one complete host and our fiducial detection thresholds, $\mathcal{M}_{50}$ is only constrained at the $\approx 1\sigma$ level, with $\log(\mathcal{M}_{50}/M_{\mathrm{\odot}})=7.1^{+0.3}_{-1.6}$; the tail toward low $\mathcal{M}_{50}$ is even more prominent than in the ``strong'' cutoff scenario, which again prevents a $2\sigma$ measurement. These results do not significantly improve when combining two hosts, which only yield an upper limit on $\mathcal{M}_{50}$, with $\log(\mathcal{M}_{50}/M_{\mathrm{\odot}})=5.3^{+1.4}_{-0.2}$. 

Our two-host results in the ``weak'' cutoff scenario are less sensitive to the specific combination of hosts because the $\mathcal{M}_{50}$--$\sigma_M$ degeneracy is not the limiting factor for $\mathcal{M}_{50}$ constraints. Instead, for a ``weak'' cutoff, combining hosts with as many satellites as possible---rather than hosts with classical satellite abundances that differ substantially relative to the underlying host-to-host scatter---yields the strongest upper limits on $\mathcal{M}_{50}$. Note that current MW satellite observations mildly disfavor the ``strong'' cutoff \citep{Nadler191203303}. If future data strengthens this conclusion, our results imply that combining the Milky Way satellite population with hosts that have abundant bright satellite populations will maximize the galaxy formation constraints delivered by future surveys. Many such hosts have already been identified throughout the Local Volume \citep{Carlsten220300014} and nearby Universe \citep{Geha170506743,Mao200812783}.

\subsection{Cutoff Shape Posteriors}
\label{sec:shape_results}

With one complete satellite population (e.g., for a complete census of satellites only within the MW), the shape of the occupation fraction is difficult to recover in either galaxy formation cutoff scenario we consider. In particular, using one complete host in either the ``strong'' or ``weak'' cutoff scenario, our $f_{\mathrm{gal}}$ reconstruction falls off less slowly with decreasing halo mass than the input model over most of the allowed parameter space (see the right panel of Figure~\ref{fig:reion}). This is consistent with the tail toward low $\mathcal{M}_{50}$ and the broad $\mathcal{S}_{\mathrm{gal}}$ posterior in the one-host case (see Appendix~\ref{sec:reion_full_posterior}). Adding an additional host brings our $f_{\mathrm{gal}}$ reconstruction into agreement with the input model at the $\approx 1\sigma$ level; however, we still infer shallower occupation fractions than the input model at the $2\sigma$ level because the tail toward low $\mathcal{M}_{50}$ and large $\mathcal{S}_{\mathrm{gal}}$ (corresponding to a flat occupation fraction as a function of halo mass) remains consistent with the mock data. Our results suggest that more than two complete satellite populations are needed to probe the functional dependence of the occupation fraction on halo mass at the $\gtrsim 1\sigma$ level.

\subsection{Projections for Additional Hosts}
\label{sec:idealized_M50}

As discussed in Section~\ref{sec:likelihood}, it is computationally challenging to run our full MCMC forecasts for more than two hosts. Instead, we derive projected constraints on $\mathcal{M}_{50}$ by calculating the likelihood as a function of $\mathcal{M}_{50}$ with all remaining galaxy--halo connection parameters fixed to their input values. We perform this test as a function of $N_{\mathrm{hosts}}$, sampling realizations of our galaxy--halo connection model and different host combinations for each cutoff scenario and choice of $N_{\mathrm{hosts}}$. This procedure optimistically neglects galaxy--halo connection degeneracies to illustrate the ideal scaling of $\mathcal{M}_{50}$ constraints for statistically limited measurements; the resulting projections can be interpreted as cosmic variance-limited constraints from satellite populations.

\begin{figure}[t!]
\centering
\includegraphics[trim={0 0cm 0cm 0.5cm},width=0.5\textwidth]{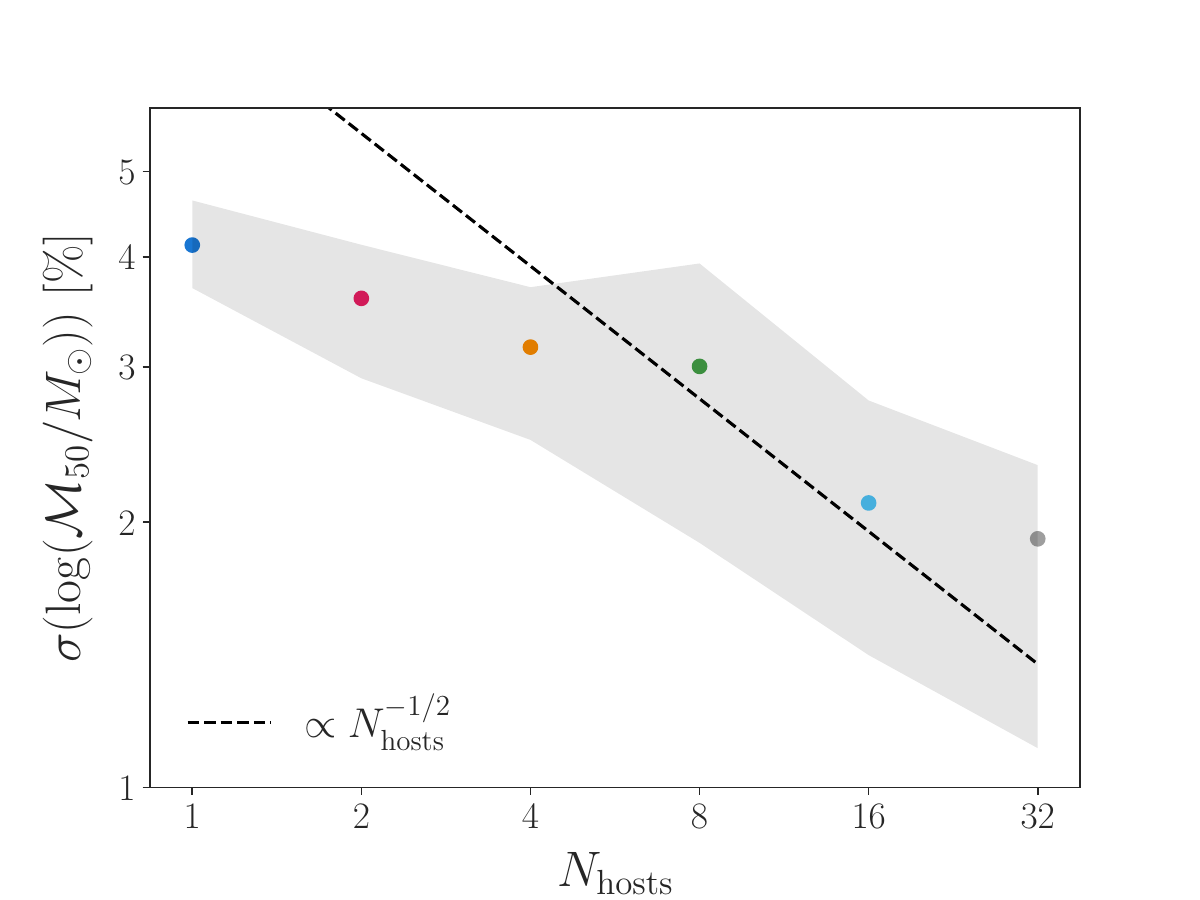}
\caption{Projected $1\sigma$ uncertainties for statistically limited measurements of $\log(\mathcal{M}_{50}/M_{\mathrm{\odot}})$ in the ``strong'' cutoff scenario, holding all remaining galaxy--halo connection parameters fixed. Points show mean measurement uncertainties as a function of the number of hosts combined in the inference for $N_{\mathrm{hosts}}=1$ (blue), $2$ (red), $4$ (orange), $8$ (green), $16$ (cyan), and $32$ (gray); the gray band shows the $68\%$ range of these uncertainties due to different host combinations. The black dashed line shows the $N_{\mathrm{hosts}}^{-1/2}$ scaling expected when combining statistically limited measurements of independent hosts.}
\label{fig:M50_proj}
\end{figure}

Figure~\ref{fig:M50_proj} shows projected $1\sigma$ measurement uncertainties on $\log(\mathcal{M}_{50}/M_{\mathrm{\odot}})$ as a function of $N_{\mathrm{hosts}}$ in the ``strong'' cutoff scenario. We derive these uncertainties from the Fisher information evaluated at the input value of $\mathcal{M}_{50}$. For large $N_{\mathrm{hosts}}$, the $1\sigma$ uncertainty on $\log(\mathcal{M}_{50}/M_{\mathrm{\odot}})$ scales as $N_{\mathrm{hosts}}^{-1/2}$, as expected when adding independent measurements. With $32$ complete hosts, we project $\approx 2\%$ uncertainties on $\log(\mathcal{M}_{50}/M_{\mathrm{\odot}})$ at the $1\sigma$ level, assuming negligible galaxy--halo connection systematics; different host combinations affect these projected constraints at the subpercent level. For reference, there are $\approx 30$ MW-mass hosts in the Local Volume with partially characterized satellite populations (e.g., \citealt{Carlsten220300014}). Our projections do not significantly change in the ``weak'' cutoff scenario, although they should only be regarded as upper limits at the $1\sigma$ level in this case.

In Appendix~\ref{sec:idealized_constraints}, we show that there is a prominent tail toward low $\mathcal{M}_{50}$ in our idealized likelihoods for the ``strong'' and ``weak'' cutoff scenarios; this tail persists for large  $N_{\mathrm{hosts}}$, particularly for the ``weak'' cutoff. Thus, $\mathcal{M}_{50}$ is difficult to measure at the $2\sigma$ level, even in the statistically limited regime, if subhalos with $M_{\mathrm{peak}}\lesssim \mathcal{M}_{50}$ are not easily detectable. As a result, we do not expect a shallower survey than our fiducial assumption to detect either the ``strong'' or ``weak'' cutoff, even if it uses more hosts. On the other hand, constraining power qualitatively improves if satellites that do not pass our fiducial detection thresholds are included in the inference. For example, using more optimistic detection thresholds of $M_V<+2~\mathrm{mag}$ (comparable to the luminosity of the recently discovered Ursa Major III/UNIONS 1 dwarf galaxy candidate; \citealt{Smith231110147}) and $\mu_V<34~\mathrm{mag\ arcsec}^{-2}$ (fainter than any known system and comparable to ``stealth galaxies'' hypothesized in previous studies; \citealt{Bullock09121873}) disfavors models with low $\mathcal{M}_{50}$ due to nondetections of extremely faint systems. Observations of satellites below our fiducial detection thresholds can therefore improve our projected galaxy formation cutoff constraints; we leave a study of these systems in the context of our modeling framework to future work.

\subsection{Interpretation}

Our forecasts indicate that complete measurements of two satellite populations (e.g., all satellites of the MW and M31) can probe a galaxy formation cutoff with $\mathcal{M}_{50}=10^8~M_{\mathrm{\odot}}$, at the $1\sigma$ level. Moreover, for statistically limited measurements, the uncertainty of this projected $1\sigma$ constraint shrinks as additional hosts are added. We emphasize that our forecasts only rely on halos that actually host galaxies. In particular, nondetections of faint, low surface brightness galaxies---which would primarily be hosted by halos with $M_{\mathrm{peak}}\lesssim \mathcal{M}_{50}$ in the absence of a galaxy formation cutoff---play a crucial role in our inference. Thus, modeling observational incompleteness (e.g., following \citealt{Drlica-Wagner191203302} and \citealt{Doliva-Dolinsky220502831} for the MW and M31 systems, respectively), which can mimic such nondetections, will be necessary to robustly extract a galaxy formation cutoff signal from future dwarf galaxy surveys. Upcoming facilities like Rubin will make substantial progress in this regard \citep{Mutlu-Pakdil210501658}; we do not specifically consider Rubin here because careful modeling of, e.g., star--galaxy separation is needed to accurately characterize its dwarf galaxy detection sensitivity (K.~Tsiane et al.\ 2024, in preparation).

Note that mock satellite populations in our analyses range from $\approx 100$ to $200$ objects per host (e.g., \citealt{Nadler191203303}). Given that subhalo and satellite statistics are approximately Poissonian (e.g., \citealt{Mao150302637}), the statistical uncertainty of our measurements is $\approx 10\%$ per host. Because faint-end galaxy--halo connection uncertainties are much larger than $10\%$, our projected measurements are limited by galaxy--halo connection systematics rather than statistical uncertainties.

Thus, using auxiliary data to constrain galaxy--halo connection parameters can reduce uncertainties in our forecasts and help realize the statistically limited measurements derived in Section~\ref{sec:idealized_M50}. For example, constraints on the faint-end luminosity function slope, $\alpha$, could be derived from observations of dwarf galaxies with luminosities comparable to the brightest satellites we consider here; forthcoming Roman data will detect such dwarf galaxies in a range of environments. Meanwhile, the galaxy--halo luminosity scatter, $\sigma_M$, can be probed more directly using dwarf galaxies' inferred dynamical masses (e.g., from stellar velocity dispersion measurements; \citealt{Simon190304743}). Finally, the efficiency of subhalo disruption due to baryons, $\mathcal{B}$, can be better constrained using large samples of bright satellite populations (e.g., from SAGA; \citealt{Geha170506743,Mao200812783}), and priors can be tightened with improved predictions for subhalo disruption in embedded-disk simulations (Y.~Wang et al.\ 2024, in preparation).

\section{Warm Dark Matter Forecasts}
\label{sec:wdm}

Next, we forecast measurements of the thermal-relic WDM particle mass, parameterized by the half-mode mass $M_{\mathrm{hm}}$ in our forecasts, using dwarf galaxy populations. We consider three scenarios, all of which include $M_{\mathrm{hm}}$ in the inference, but which differ in their assumptions regarding the underlying galaxy formation and DM physics as determined by the input values of $\mathcal{M}_{50}$ and $M_{\mathrm{hm}}$:
\begin{enumerate}
    \item Scenario A: CDM subhalo populations ($M_{\mathrm{hm}}\rightarrow 0~M_{\mathrm{\odot}}$) with a ``weak'' galaxy formation cutoff ($\mathcal{M}_{50}=3\times 10^7~M_{\mathrm{\odot}}$).
    \item Scenario B: CDM subhalo populations ($M_{\mathrm{hm}}\rightarrow 0~M_{\mathrm{\odot}}$) with a ``strong'' galaxy formation cutoff ($\mathcal{M}_{50}=10^8~M_{\mathrm{\odot}}$).
    \item Scenario C: WDM subhalo populations ($M_{\mathrm{hm}}=10^8~M_{\mathrm{\odot}}$, or $m_{\mathrm{WDM}}=4.9~\mathrm{keV}$) with a ``strong'' galaxy formation cutoff ($\mathcal{M}_{50}=10^8~M_{\mathrm{\odot}}$).
\end{enumerate}
Scenario A provides an optimistic forecast for the WDM constraining power of future dwarf galaxy surveys since it does not include an astrophysical signal that mimics the WDM cutoff. Scenarios B and C allow us to examine degeneracies between WDM and astrophysical cutoff parameters. For Scenario C, the WDM model we choose is disfavored by the MW satellite population as well as by other small-scale structure probes like strong gravitational lensing \citep{Gilman190806983,Hsueh190504182} and the Lyman-$\alpha$ forest \citep{Irsic179602,Irsic230904533}, and combinations thereof (e.g., \citealt{Nadler210107810,Enzi201013802}). Nevertheless, we chose this input model so that our predicted satellite populations are significantly affected by WDM, allowing us to study degeneracies between WDM and the galaxy--halo connection in detail.

\subsection{WDM Particle Mass Posteriors}
\label{sec:wdm_results}

\begin{figure}[t!]
\includegraphics[width=0.475\textwidth]{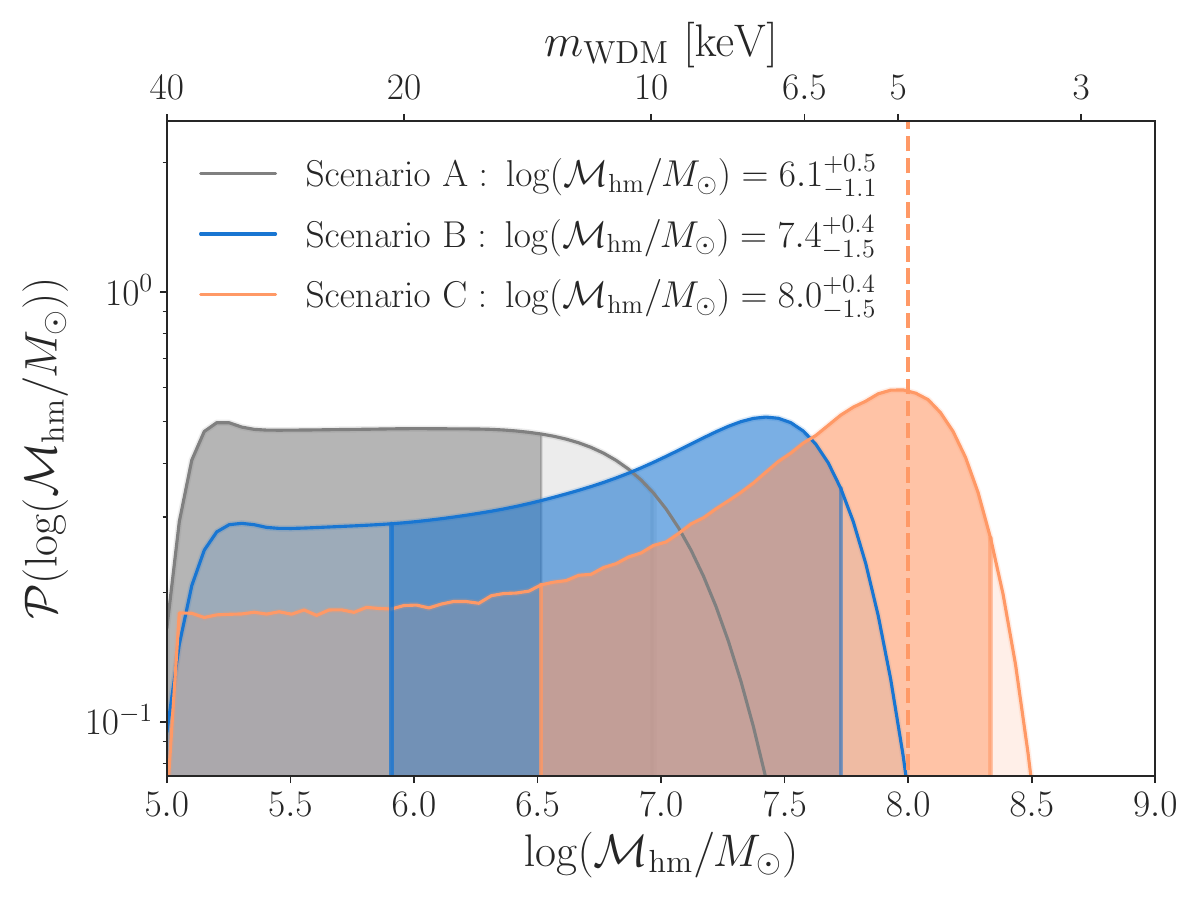}
\caption{Marginalized posterior for the WDM half-mode mass, $M_{\mathrm{hm}}$, derived from our forecasted measurements of one complete satellite population of a MW-mass system; top ticks show the corresponding values of $m_{\mathrm{WDM}}$. Results are shown for forecasts that assume CDM subhalo abundances and a ``weak'' galaxy formation cutoff (Scenario A; gray), CDM subhalo abundances and a ``strong'' galaxy formation cutoff (Scenario B; blue), and WDM subhalo abundances and a ``strong'' galaxy formation cutoff (Scenario C; light red). Dark (light) regions show $68\%$ ($95\%$) confidence intervals, and the vertical dashed red line shows the value of $M_{\mathrm{hm}}=10^8~M_{\mathrm{\odot}}$ ($m_{\mathrm{WDM}}=4.9~\mathrm{keV}$) assumed in Scenario C.}
\label{fig:wdm}
\end{figure}

Figure~\ref{fig:wdm} shows the marginalized posterior for $M_{\mathrm{hm}}$ from each of our WDM scenarios using one complete satellite population. Scenario A yields the strongest WDM constraint, with $M_{\mathrm{hm}}<10^{6.6}~M_{\mathrm{\odot}}$ ($m_{\mathrm{WDM}}>12.9~\mathrm{keV}$) from one host and $M_{\mathrm{hm}}<10^{6.0}~M_{\mathrm{\odot}}$ ($m_{\mathrm{WDM}}>19.5~\mathrm{keV}$) from two hosts at $68\%$ confidence. In Scenario B, we find $M_{\mathrm{hm}}<10^{7.8}~M_{\mathrm{\odot}}$ ($m_{\mathrm{WDM}}>5.6~\mathrm{keV}$) from one host and $M_{\mathrm{hm}}<10^{6.1}~M_{\mathrm{\odot}}$ ($m_{\mathrm{WDM}}>17.0~\mathrm{keV}$) from two hosts. These limits are consistent with our expectation that WDM constraints are most stringent in the absence of a galaxy formation cutoff that mimics the WDM signal.

To contextualize these results, note that \cite{Nadler191203303} inferred $M_{\mathrm{hm}}<10^{7.3}~M_{\mathrm{\odot}}$ ($m_{\mathrm{WDM}}>7.9~\mathrm{keV}$) at $68\%$ confidence from current DES and PS1 observations of the MW satellite population and simultaneously disfavor the ``strong'' cutoff scenario. Our forecasts are consistent with these bounds and indicate that observations of the remaining MW satellite population alone can improve WDM particle mass constraints by $\approx 60\%$ in the ``weak'' cutoff scenario. These projected one-host WDM constraints do not significantly change if we use MW-like hosts that contain Large Magellanic Cloud (LMC) analog subhalos; however, modeling the actual MW satellite population without accounting for LMC-associated satellites can bias inferred WDM constraints (e.g., see the discussion in \citealt{Nadler210107810}).

Our analysis of our full WDM forecast posteriors in Appendix~\ref{sec:wdm_full_posterior} reveals prominent degeneracies between $M_{\mathrm{hm}}$ and both $\mathcal{M}_{50}$ and $\sigma_M$. In particular, $M_{\mathrm{hm}}$ and $\mathcal{M}_{50}$ cannot simultaneously be large because they both suppress satellite abundances; meanwhile, $M_{\mathrm{hm}}$ and $\sigma_M$ cannot simultaneously be large according to our discussion of the degeneracy between $\mathcal{M}_{50}$ and $\sigma_M$ in Section~\ref{sec:cutoff_result}. Adding a second satellite population improves WDM constraints by both reducing statistical uncertainties and markedly improving the measurement of $\sigma_M$; in turn, this improves the measurement of $\mathcal{M}_{50}$ at $1\sigma$ and disfavors large values of $M_{\mathrm{hm}}$. The $m_{\mathrm{WDM}}\approx 20~\mathrm{keV}$ sensitivity of two complete hosts strongly motivates searches for remaining undiscovered satellites of the MW and M31.

In Scenario C, our $M_{\mathrm{hm}}$ posterior is consistent with the input value, indicating that the WDM extension of our inference framework is also self-consistent. We find that $M_{\mathrm{hm}}$ is only probed at the $\approx1\sigma$ level using one host, mainly due to a significant degeneracy with $\mathcal{M}_{50}$. When two complete hosts are used, the nondetection of $M_{\mathrm{hm}}$ persists while the $1\sigma$ uncertainty on $\mathcal{M}_{50}$ shrinks. In our model, $\mathcal{M}_{50}$ and $M_{\mathrm{hm}}$ both suppress satellite abundances as a function of peak halo mass, and their observable effects only differ in the shape of the halo mass-dependent suppression imprinted on satellite populations. The nondetection of $M_{\mathrm{hm}}$ is consistent with our finding in Section~\ref{sec:shape_results} that the shape of the galaxy formation cutoff is difficult to measure, even when multiple hosts are combined. Note that we vary the shape of the galaxy formation cutoff but not of the WDM cutoff, which allows $\mathcal{M}_{50}$ to be disentangled from $\mathcal{S}_{\mathrm{gal}}$ in the inference; it will be interesting to explore constraints on the functional dependence of SHMF suppression further in future work.

\subsection{Projections for Additional Hosts}

\begin{figure}[t!]
\centering
\includegraphics[trim={0 0cm 0cm 0.5cm},width=0.5\textwidth]{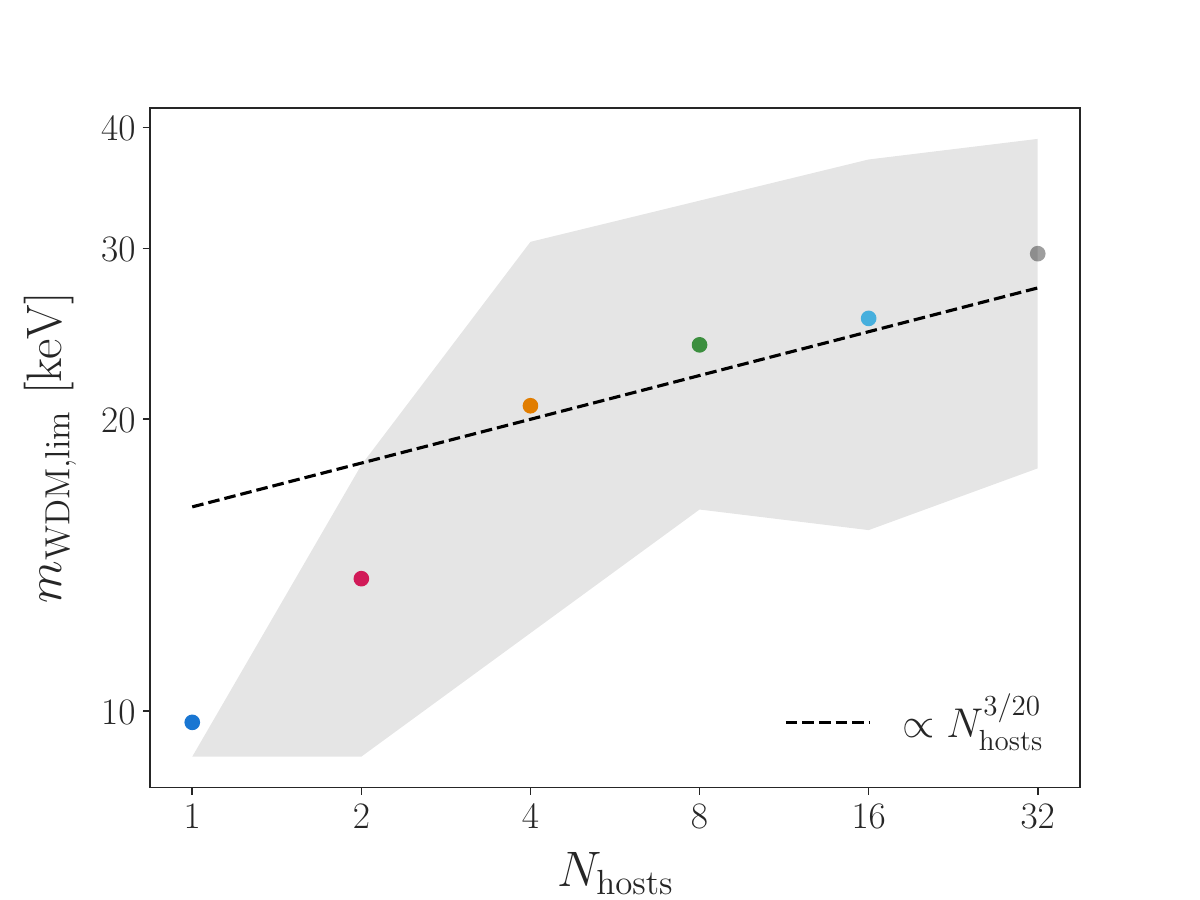}
\caption{Projected $2\sigma$ lower limits on $m_{\mathrm{WDM}}$ in the ``strong'' cutoff scenario for statistically limited measurements of dwarf satellite populations, with all galaxy--halo connection parameters held fixed. Points show mean lower limits as a function of the number of hosts combined in the inference for $N_{\mathrm{hosts}}=1$ (blue), $2$ (red), $4$ (orange), $8$ (green), $16$ (cyan), and $32$ (gray); the gray band shows the $68\%$ range of these constraints due to different host combinations. The black dashed line shows the $N_{\mathrm{hosts}}^{3/20}$ scaling expected from the functional dependence of SHMF suppression on $m_{\mathrm{WDM}}$.}
\label{fig:wdm_proj}
\end{figure}

Following Section~\ref{sec:idealized_M50}, we perform idealized tests with all galaxy--halo connection parameters held fixed to derived projections for WDM constraints with $N_{\mathrm{hosts}}>2$. Specifically, we calculate idealized likelihoods as a function of $M_{\mathrm{hm}}$ that sample over galaxy--halo connection model realizations and host combinations in the ``strong'' and ``weak'' cutoff scenarios, assuming CDM subhalo abundances. We use these to derive projected $2\sigma$ upper limits on $M_{\mathrm{hm}}$, and we convert these to lower limits on $m_{\mathrm{WDM}}$ using Equation~\ref{eq:mhm_mwdm}. Figure~\ref{fig:wdm_proj} shows the results of this test for the ``strong'' cutoff. We find that statistically limited measurements of $32$ complete hosts yield $m_{\mathrm{WDM}}\gtrsim 30~\mathrm{keV}$ at the $2\sigma$ level; different host combinations contribute a $\pm 10~\mathrm{keV}$ uncertainty to this projection. These results do not significantly change in the ``weak'' cutoff scenario because galaxy--halo connection parameters are fixed; in Section~\ref{sec:wdm_results}, our $68\%$ confidence WDM limits were more stringent for the ``weak'' versus the ``strong'' cutoff due to less prominent galaxy--halo connection degeneracies.

We derive projections for the scaling of WDM constraints with $N_{\mathrm{hosts}}$ from the dependence of the SHMF suppression on $m_{\mathrm{WDM}}$. In particular, given that our mock dwarf galaxy population measurements probe the SHMF down to $M_{\mathrm{peak}}\approx 10^8~M_{\mathrm{\odot}}$, we use Equation~\ref{eq:wdm_shmf} to predict how lower limits on $m_{\mathrm{WDM}}$, denoted $m_{\mathrm{WDM,lim}}$, depend on $N_{\mathrm{hosts}}$. Assuming that uncertainties on the SHMF amplitude shrink as $N_{\mathrm{hosts}}^{-1/2}$ and that SHMF suppression at $M_{\mathrm{peak}}\approx 10^8~M_{\mathrm{\odot}}$ determines the WDM constraint, we derive $m_{\mathrm{WDM,lim}}\propto N_{\mathrm{hosts}}^{3/20}$. Fixing the normalization of this calculation to the result of our idealized test for $N_{\mathrm{hosts}}=4$, we project $2\sigma$ lower limits of
\begin{equation}
m_{\mathrm{WDM,lim}} \gtrsim 20~\mathrm{keV} \times \left(\frac{N_{\mathrm{hosts}}}{4}\right)^{3/20},
\label{eq:wdm_scaling}
\end{equation}
which is valid for $N_{\mathrm{hosts}}\geq 4$ and statistically limited measurements. Figure~\ref{fig:wdm_proj} confirms this scaling in our idealized tests.

The relatively shallow scaling between $m_{\mathrm{WDM,lim}}$ and $N_{\mathrm{hosts}}$ in Equation~\ref{eq:wdm_scaling} follows because, as $m_{\mathrm{WDM}}$ increases, the SHMF rapidly approaches that in CDM over the range of $M_{\mathrm{peak}}$ probed by satellite galaxies. Specifically, the (sub)halos that host dwarf galaxies in our forecasts span a particular range of peak virial masses from $\approx 10^8$ to $10^{10}~M_{\mathrm{\odot}}$. According to Equation~\ref{eq:wdm_scaling}, detecting all satellites of the several hundred MW-mass systems within $\approx 40~\mathrm{Mpc}$ (e.g., \citealt{Geha170506743,Mao200812783})---which may be enabled by future optical/infrared space telescopes such as the Habitable Worlds Observatory (for potential capabilities, see the LUVOIR report; \citealt{LUVOIR191206219})---would constrain models with $m_{\mathrm{WDM}}\approx 40~\mathrm{keV}$. Additional dwarf galaxy observables and probes of lower-mass halos below the galaxy formation threshold will be necessary to qualitatively improve this WDM sensitivity.

\subsection{Interpretation}

Our results indicate that future dwarf galaxy surveys can potentially place stringent constraints on WDM, comparable to the upcoming sensitivity of other small-scale structure probes \citep{Drlica-Wagner190201055}. However, we emphasize that there are significant degeneracies between the effects of galaxy formation and DM physics on dwarf galaxy populations, which must be modeled accurately to for unbiased inference. Dwarf galaxy observables beyond those considered here, including star formation histories (which can be delayed in WDM-like models; e.g., \citealt{Bozek180305424}) and stellar velocity dispersions (e.g., \citealt{Kim210609050,Esteban230604674}) may help break these degeneracies. In parallel, combining dwarf galaxy populations with independent probes of low-mass halo abundances, such as strong lensing and stellar streams, is another promising path forward (e.g., \citealt{Banik191102662,Nadler210107810,Enzi201013802}).

Our MCMC forecasts' sensitivity to WDM masses of $\approx 10$--$20~\mathrm{keV}$ can be understood in terms of the corresponding SHMF suppression. In particular, the left panel of Figure~\ref{fig:wdm_pred} shows $\approx 20\%$--$50\%$ suppression of the SHMF at peak virial masses of $M_{\mathrm{peak}}\approx 10^8~M_{\mathrm{\odot}}$ in the most extreme WDM models that are marginally consistent with our mock CDM data. This is consistent with the fact that our CDM forecasts are sensitive to a ``strong'' galaxy formation cutoff over the range of peak halo masses indicated by the blue band. Interestingly, our most optimistic constraint of $m_{\mathrm{WDM}}>17.0~\mathrm{keV}$ in Scenario B corresponds to SHMF suppression of only $\approx 10\%$, highlighting the importance of breaking additional galaxy--halo connection degeneracies in this case.

\section{Subhalo Mass Function Forecasts}
\label{sec:shmf}

Finally, we forecast DM model-independent constraints on the amplitude of the SHMF relative to CDM. In particular, we forecast SHMF constraints for the ``weak'' and ``strong'' galaxy formation cutoff scenarios introduced above. In both scenarios, we assume underlying CDM subhalo abundances, corresponding to input values of $\xi_8=0$, $\xi_9=0$, and $\xi_{10}=0$ (recall that $\xi_i$ parameterizes the logarithmic deviation of the SHMF relative to CDM in mass decade $i$).

\subsection{SHMF Amplitude Posteriors}
\label{sec:shmf_results}

In both the ``strong'' and ``weak'' cutoff scenarios, our $\xi_8$, $\xi_9$, and $\xi_{10}$ posteriors are consistent with CDM, validating our inference framework when applied to generalized SHMFs. Furthermore, as summarized in Table~\ref{tab:summary}, we find that SHMF suppression can be constrained to $\approx 70\%$, $60\%$, and $50\%$ that in CDM at peak virial masses of $10^8~M_{\mathrm{\odot}}$, $10^9~M_{\mathrm{\odot}}$, and $10^{10}~M_{\mathrm{\odot}}$, respectively, for one complete host in the ``weak'' cutoff scenario. Meanwhile, enhanced SHMFs are only constrained relative to CDM at factors of $\approx 20$, $4$, and $3$ levels in these mass decades. Uniformly suppressed (enhanced) SHMFs corresponding to the $68\%$ confidence limits from these ``weak'' cutoff forecasts are shown by the dotted red (dashed blue) lines in the right panel of Figure~\ref{fig:wdm_pred}.

Our analysis of the full posteriors in Appendix~\ref{sec:shmf_full_posterior} reveals degeneracies between SHMF amplitude and galaxy--halo connection parameters including the faint-end luminosity function slope $\alpha$ and luminosity scatter $\sigma_M$, particularly at the low-mass end ($M_{\mathrm{peak}}\approx 10^8~M_{\mathrm{\odot}}$). These degeneracies are expected because variations in the underlying SHMF can mimic or counteract the effects of our galaxy--halo connection model for the faintest satellite we consider. At higher masses ($M_{\mathrm{peak}}\approx 10^{10}~M_{\mathrm{\odot}}$), we forecast that the SHMF can be measured reasonably precisely.

A complete census of satellites around two hosts improves the precision of these constraints. For example, at $M_{\mathrm{peak}}=10^{10}~M_{\mathrm{\odot}}$, $68\%$ confidence limits on SHMF enhancement change from three to two times that in CDM, and limits on SHMF suppression change from $50\%$ to $80\%$ that in CDM; this level of improvement is similar at lower halo masses (see Table~\ref{tab:summary}). Furthermore, with two complete hosts, both the galaxy formation cutoff and underlying SHMF are recovered more confidently at the $1\sigma$ level. Note that our projected SHMF enhancement constraints are not very sensitive to the assumed galaxy formation cutoff; however, SHMF suppression constraints are slightly weaker in our ``strong'' cutoff scenario because suppression in the underlying SHMF mimics the assumed galaxy formation cutoff. In the one-host case, these degeneracies substantially degrade the measurement of $\mathcal{M}_{50}$ relative to our galaxy formation cutoff forecasts that assume CDM subhalo abundances in Section~\ref{sec:reion}. These results underscore the importance of combining multiple satellite populations, particularly when assumptions about underlying subhalo populations are relaxed.

\begin{figure}[t!]
\hspace{-3mm}
\includegraphics[width=0.485\textwidth]{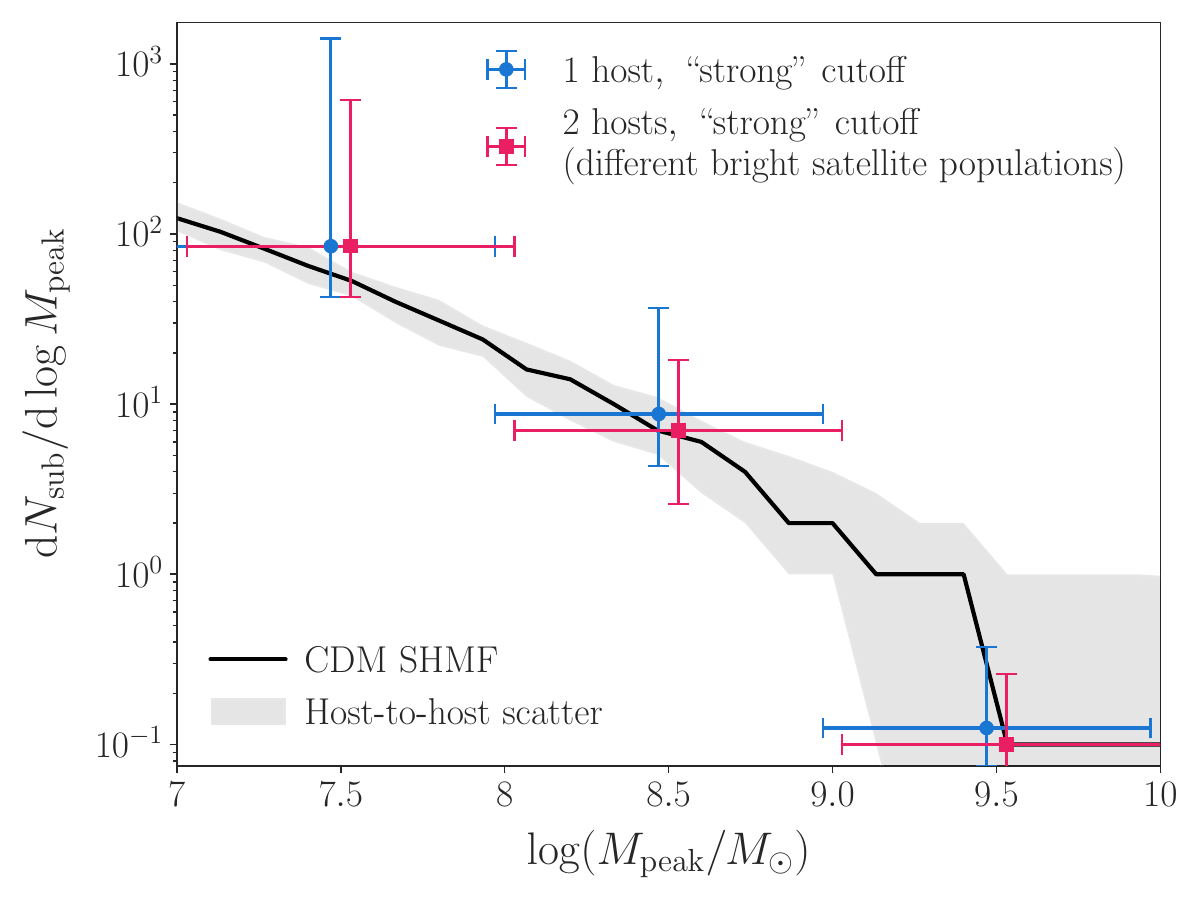}
\caption{Mean and $16$th--$84$th percentile host-to-host scatter of the SHMF (black line and band) from our $40$ CDM MW-mass zoom-in simulations. Blue circles (red squares) and error bars show the mean and $68\%$ confidence intervals from our SHMF forecast using one (two) complete satellite populations of MW-mass hosts, assuming a ``strong'' galaxy formation cutoff. For clarity, the one- and two-host error bars are offset horizontally.}
\label{fig:shmf}
\end{figure}

Figure~\ref{fig:shmf} illustrates the results of our one- and two-host SHMF forecasts in the ``strong'' cutoff scenario relative to the SHMF from our MW-mass zoom-in simulation suite. In the $M_{\mathrm{peak}}=10^9~M_{\mathrm{\odot}}$ and $10^{10}~M_{\mathrm{\odot}}$ decades, our forecasted SHMF constraints are comparable to the host-to-host scatter about the mean SHMF among our $40$ zoom-in simulations (see \citealt{Mao150302637} for a detailed discussion of the host-to-host scatter). At $M_{\mathrm{peak}}=10^8~M_{\mathrm{\odot}}$, SHMF suppression is probed at a level comparable to the host-to-host scatter, while SHMF enhancement that significantly exceeds the range of CDM predictions is compatible with the mock data.

These results imply that secondary properties beyond host halo mass that correlate with SHMF amplitude (e.g., concentration; \citealt{Mao150302637}) can be probed using multiple satellite populations. On the other hand, accurately interpreting SHMF measurements from future dwarf galaxy surveys will require modeling these secondary host halo properties, particularly for models that suppress the SHMF at low masses. We emphasize that our forecasted upper limits on the SHMF are very weak at low masses; the precision of these upper limits is mainly determined by galaxy--halo connection degeneracies for two or more hosts. Thus, we expect surveys of dwarfs in a range of environments, at the bright end of the satellite populations we consider here, to reduce galaxy--halo connection uncertainties and help determine the SHMF.

\subsection{Projections for Additional Hosts}

We derive projected uncertainties on $\xi_8$, $\xi_9$, and $\xi_{10}$ by calculating the likelihood of each parameter for fixed galaxy--halo connection parameters in the ``strong'' cutoff scenario. For large $N_{\mathrm{hosts}}$, these projected SHMF constraints shrink according to $N_{\mathrm{hosts}}^{-1/2}$, as expected from Poisson statistics; furthermore, they do not change significantly in the ``weak'' cutoff scenario. We project that statistically limited measurements of $4$ ($32$) complete hosts yield SHMF uncertainties of $\approx 5\%$ ($\approx 2\%$) relative to CDM. Unlike our results in Section~\ref{sec:shmf_results}, these projected uncertainties are roughly symmetric and do not depend strongly on peak halo mass because they are not limited by galaxy--halo connection degeneracies.

\subsection{Interpretation}

Our results indicate that future dwarf galaxy surveys can set upper limits on the SHMF at peak subhalo masses from $10^9$ to $10^{10}~M_{\mathrm{\odot}}$, while simultaneously placing a robust lower limit at $10^8~M_{\mathrm{\odot}}$. The upper limits on the SHMF amplitude at high masses result from the fact that, if high-mass subhalos are more abundant than in the input CDM model, observed luminosity functions are overpredicted. In particular, it is difficult to ``hide'' the galaxies that occupy high-mass subhalos by varying galaxy--halo connection parameters without drastically reducing the abundance of lower-mass galaxies and resulting in worse agreement with the mock CDM data (for details, see our analysis of the full SHMF forecast posteriors in Appendix~\ref{sec:shmf_full_posterior}). At lower peak halo masses, and particularly near $M_{\mathrm{peak}}\approx 10^8~M_{\mathrm{\odot}}$ where our fiducial galaxy formation cutoffs take effect, models with suppressed low-mass subhalo abundances underpredict faint-end luminosity functions; however, models with enhanced low-mass subhalo abundances can be ``hidden'' by the galaxy formation cutoff. More precise constraints on the mass scale and shape of the cutoff would therefore improve our projected SHMF enhancement measurements.

Our SHMF forecasts can be translated to any DM or early-Universe physics scenario that suppresses or enhances the SHMF relative to CDM. This is most often realized by suppressing or enhancing the formation of (sub)halos relative to CDM; it will therefore be interesting to generalize our modeling framework to directly constrain the linear matter power spectrum, which often dictates the resulting suppression or enhancement of subhalo abundances in such scenarios (e.g., \citealt{Gilman211203293}). Meanwhile, if the SHMF is altered at late times (e.g., via subhalo disruption due to new DM physics), our forecasts are applicable if the abundance of satellite galaxies is correspondingly altered. The connection between subhalo and satellite disruption is nontrivial and requires detailed study in the context of specific DM models (e.g., see \citealt{Nadler200108754,Nadler210912120} and \citealt{Mau220111740} for examples in the context of self-interacting and decaying DM); we leave studies of such scenarios to future work.

Our SHMF suppression results are relevant for a variety of warm, interacting, and ultralight DM models (e.g., \citealt{Bechtol220307354}). Meanwhile, DM models that enhance the SHMF more easily evade constraints from dwarf galaxy population statistics because their effects can be counteracted by galaxy formation physics. Such models include, e.g., ultralight axions that undergo parametric resonance in the early Universe \citep{Arvanitaki190911665,Cyncynates210909755} or DM produced during a period of early matter domination \citep{Erickcek11060536}. We expect that dwarf galaxy observables beyond luminosity, size, and distance (e.g., velocity dispersions; \citealt{Kim210609050,Esteban230604674}) will help probe DM physics that enhances (sub)halo abundances. Distinct small-scale structure probes like strong gravitational lensing and stellar streams, which are more sensitive to subhalos density profiles than dwarf galaxy population statistics, will complement this effort.


\section{Future Outlook}
\label{sec:caveats}

We now discuss our results in the context of near-future observational capabilities (Section \ref{sec:obs}), the long-term landscape of dwarf galaxy measurements (Section~\ref{sec:future}), and areas for future modeling work (Section \ref{sec:theory}).

\subsection{Near-term Observational Capabilities}
\label{sec:obs}

Our assumption that complete satellite populations of several MW-mass systems can be detected is clearly optimistic given current and near-future observational capabilities. In particular, this level of completeness will be difficult to achieve even for the MW and M31 in the next decade. Preliminary dwarf galaxy detection sensitivity estimates in \cite{Drlica-Wagner190201055} indicate that Rubin is sensitive to satellites with $M_V<0~\mathrm{mag}$ and $\mu<32~\mathrm{mag\ arcsec}^{-2}$ in the Southern Hemisphere, out to the MW virial radius. However, these projections may be optimistic due to the difficulties of star--galaxy separation; realistic mock simulations of dwarf recovery in Rubin data will be needed to accurately derive its sensitivity. Meanwhile, for M31, \cite{Doliva-Dolinsky220502831} showed that current PAndAS data is sensitive to systems with $M_V\lesssim -5.5~\mathrm{mag}$ and $\mu \lesssim 30~\mathrm{mag\ arcsec}^{-2}$; for context, \cite{Mutlu-Pakdil210501658} found that Rubin is sensitive to even fainter dwarf galaxies at comparable distances, although M31 is not within its field of view. 

Thus, upcoming observations are unlikely to achieve complete coverage of dwarf galaxy populations down to our assumed detection limits. As a result, the ``weak'' cutoff scenario will likely remain undetectable in the near future. Nonetheless, we emphasize that the MW satellite population already probes the ``strong'' cutoff scenario \citep{Nadler191203303}, and we expect these constraints to improve with additional data. Beyond observations of MW and M31 satellites, Euclid and Roman will also detect and help confirm the nature of faint dwarf galaxies throughout the Local Volume, both by discoveries with in situ data and by providing distance measurements for dwarf candidates identified in Rubin imaging data. As Euclid has already launched, a focused study of its stand-alone and combined dwarf galaxy detection and characterization capabilities is also timely. For small areas of the sky, it has recently been demonstrated that JWST can detect very faint, low surface brightness dwarf galaxies, including at cosmological distance (e.g., \citealt{Conroy231013048}), which will help characterize the dwarf galaxy population below the limit of ground-based photometric surveys.

An assumption underlying our forecasts is that properties of the MW-mass centrals that host the dwarf satellites we consider will be well characterized. Accurate and precise measurements of host halo masses are important for determining expected subhalo and satellite populations, and measurements of secondary properties like host concentration are also helpful (e.g., \citealt{Mao150302637,Fielder180705180}). At present, even the MW's halo mass remains difficult to measure precisely; although, proper motion data from Gaia have significantly reduced this uncertainty (e.g., see \citealt{Gardner210613284} for a review). For more distant systems, velocity dispersion measurements of globular clusters and of dwarf satellites themselves will likely provide the most precise host mass estimates. Jointly inferring host halo properties and the galaxy--halo connection for their satellites will be necessary to claim a detection of the astrophysical and DM signatures we have explored. Extending our model to marginalize over host halo properties is a natural avenue for future work. We do not expect marginalizing over host properties to change our qualitative conclusions (e.g., the value of combining hosts with different bright satellite populations to break galaxy--halo connection degeneracies); however, our quantitative results concerning the number of hosts needed to achieve given galaxy formation cutoff, WDM, or SHMF constraints will depend on these uncertainties in detail.

We note that our forecasts adopt broad priors on the galaxy--halo connection that are likely to be improved using measurements of brighter dwarf galaxies. As noted above, adopting tighter priors on, e.g., the luminosity function slope and galaxy--halo connection scatter shrinks our systematic error budget and therefore directly improves the precision of our galaxy formation cutoff, WDM, and SHMF constraints. Current surveys like ELVES \citep{Carlsten220300014} and SAGA \citep{Geha170506743,Mao200812783} have already detected large numbers of dwarf satellites throughout the Local Volume and out to $\approx 40~\mathrm{Mpc}$, and DESI is supplementing these efforts over a very large area \citep{Darragh-Ford221207433}. Combining our framework with data from these surveys will improve galaxy--halo connection constraints (e.g., \citealt{Danieli221014233}). Furthermore, upcoming facilities including Roman will likely detect large numbers of dwarfs in both the (partially) resolved regimes and in integrated light with distances measured via, e.g., surface brightness fluctuations \citep{Greco200407273}. In parallel, ongoing neutral hydrogen surveys like WALLABY are expected to reach \HI\ detection thresholds of $\approx 10^5~M_{\mathrm{\odot}}$ \citep{Westmeier221107094}, probing the gas content of low-mass dwarfs and providing a complementary handle on the faint-end galaxy--halo connection.

\subsection{Long-term Observational Outlook}
\label{sec:future}

In the longer term, next-generation optical/infrared space telescopes have the potential to revolutionize our ability to detect faint dwarf galaxies. For example, \cite{LUVOIR191206219} estimated that a $15~\mathrm{m}$ optical/infrared space telescope can resolve individual stars out to tens of megaparsecs and potentially detect dwarf galaxies with $M_V<0~\mathrm{mag}$ at distances of $\sim 15~\mathrm{Mpc}$. Because of the relatively small field of view of such telescope concepts, targeting satellites around known hosts at these distances will be crucial to maximize dwarf galaxy discovery. Such detection sensitivity easily satisfies the assumptions of our forecasts; for example, there are $\approx 30$ MW-mass systems to survey within $12~\mathrm{Mpc}$ (e.g., \citealt{Carlsten220300014}). As the community develops the detailed science requirements for the Astro2020-prioritized Habitable Worlds Observatory, the imaging and spectroscopic detection and characterization of faint dwarf galaxies, through both diffuse light and direct stellar-cluster or stellar identifications, is a capability that should be explored. This research will be a key step toward the galaxy formation and DM physics measurements we have forecasted.

Leveraging complementary dwarf galaxy observables will also help identify signatures of a galaxy formation cutoff and new DM physics. For example, stellar velocity dispersion measurements of nearby dwarfs probe their present-day dynamical halo masses. These measurements are necessary to disambiguate dwarf galaxies from star clusters, which is becoming increasingly challenging and may represent a limiting systematic uncertainty of satellite population measurements (e.g., \citealt{Cerny220912422}). At the same time, they provide valuable information about the galaxy--halo connection (e.g., \citealt{Simon190304743}) and probe DM physics that affects (sub)halos' inner density profiles, including WDM-like cutoffs (e.g., \citealt{Kim210609050}) or DM self-interactions (e.g., see \citealt{Tulin170502358} and \citealt{Adhikari220710638} for reviews). Excitingly, next-generation spectroscopic facilities will pursue such measurements (e.g., \citealt{Chakrabarti220306200}). In parallel, improved constraints on dwarf galaxies' star formation histories from space telescopes like Roman will help determine the detailed impact of reionization on the faint end of the galaxy--halo connection (e.g., \citealt{Weisz14047144,Weisz14094772}), providing important complementary constraints on the galaxy formation cutoff.

\subsection{Modeling Work}
\label{sec:theory}

Further modeling work will improve our ability to extract galaxy formation and DM physics from future dwarf galaxy surveys. Most importantly, including a broader range of host masses will yield mock dwarf galaxy populations that better span those probed by future surveys. For example, Symphony's LMC-mass suite contains hosts with $M_{\mathrm{host}}\approx 10^{11}~M_{\mathrm{\odot}}$, and its Group suite contains hosts with $M_{\mathrm{host}}\approx 10^{13}~M_{\mathrm{\odot}}$. Incorporating these suites will require recalibrating aspects of our galaxy--halo connection framework that are currently tuned to MW-mass systems, including our subhalo disruption model. 

In addition to the effects of host halo mass, optimally characterizing the correlated information contained in satellite population statistics and secondary host properties remains challenging. Cosmological zoom-in simulations indicate that the strength of these correlations peaks at the MW host halo mass scale \citep{Nadler220902675}. Thus, accurate measurements of the masses and secondary properties of MW-mass hosts  can significantly reduce uncertainties on their underlying SHMFs. To achieve this, the following theoretical uncertainties should be addressed:
\begin{enumerate}
    \item Even in a dark-matter-only setting, the physical origin of correlations between secondary host properties and SHMF amplitude remains unclear. \cite{Nadler220902675} showed that the mass of the largest subhalo correlates strongly with SHMF amplitude in simulations; there is a similar trend in SAGA data, which shows that the luminosity of the brightest observed satellite correlates with satellite luminosity function amplitude more strongly than central galaxy properties~\citep{Mao200812783}. For the MW, this correlation partially arises because the LMC brings its own satellites into the MW (e.g., \citealt{Kallivayalil180501448,Patel200101746}). In fact, \cite{Nadler191203303} showed that it is difficult to match the spatial anisotropy in the observed MW satellite population without modeling the LMC system. Because the LMC accreted recently and is near pericenter today, this effect is likely transient (e.g., \citealt{DSouza21041324i,Barry230305527}). Understanding the origin of the correlation between the mass of the largest subhalo and SHMF amplitude (or the luminosity of the brightest satellite and satellite luminosity function amplitude) would therefore be fruitful. This is particularly relevant because, as shown in Section~\ref{sec:reion}, combining hosts with differing bright satellite abundances can markedly improve galaxy formation constraints.
    \item Beyond dark-matter-only predictions, the presence of a central galaxy has been claimed to affect surviving subhalo and satellite abundances at the factor of $\approx 2$ level (e.g., \citealt{Garrison-Kimmel170103792,Kelley181112413,Richings181112437}). The strength of this effect is debated (e.g., \citealt{Webb200606695,Green211013044}) and may be related to artificial disruption in cosmological simulations more generally (e.g., \citealt{VandenBosch180105427,VandenBosch171105276,Mansfield230810926}). Current hydrodynamic simulations indicate that subhalo disruption is not highly sensitive to central galaxy properties beyond total mass. Because satellite populations will be measured around a variety of central galaxies in the near future, improved theoretical predictions for subhalo and satellite disruption around central galaxies with a range of properties will be helpful. Combining isolated and satellite dwarf populations will also reduce uncertainties associated with subhalo and satellite disruption, which are conservatively marginalized over in our forecasts.
\end{enumerate}

Although our simulations capture a representative range of MW-mass host halos and formation histories \citep{Mao150302637,Nadler220902675}, we have not explicitly modeled the impact of host mass and cosmic environment on the galaxy--halo connection itself. In particular, our galaxy--halo connection model populates subhalos according to the same procedure in all MW-mass hosts we consider. This assumption likely does not capture the complexity of real dwarf galaxy populations, in which galaxy formation and evolution are environmentally dependent (e.g., \citealt{Christensen231104975}; also see \citealt{Danieli221014233} for discussion in the context of Local Volume satellite populations), particularly for low-mass galaxies affected by photoionization (e.g., \citealt{Benson0210354}). Modeling environmental effects is particularly relevant because the large-scale environment of the Local Volume may be unusual (e.g., \citealt{Neuzil2020,McAlpine220204099}), thereby impacting the expected number of nearby dwarf galaxies and the effects of photoionization on these systems (e.g., \citealt{Benson0108218,Busha09013553,Li13062971,Dixon170306140}).  

We plan to incorporate host halo property and environmental dependence in our galaxy--halo connection framework in future work. In a similar spirit, applying our inference framework to dwarf galaxy populations predicted by hydrodynamic simulations and SAMs, including in beyond-CDM scenarios---rather than mock data generated by applying our empirical galaxy--halo connection model to CDM simulations---would be a useful test of our framework's flexibility.

\section{Prospects for Detecting ``Dark'' Halos}
\label{sec:discussion}

Our forecasts indicate that future dwarf galaxy population measurements can probe galaxy formation cutoffs that manifest at peak virial masses of $\approx 10^8~M_{\mathrm{\odot}}$. By sampling from our forecasts' posteriors, we confirm that subhalos with peak virial masses of $\approx 10^8~M_{\mathrm{\odot}}$ host observable satellites in our mock observations. Although slightly lower-mass halos may host galaxies in our ``strong'' and ``weak'' cutoff scenarios, it is difficult to unambiguously classify such systems as dark matter-dominated galaxies since they are predicted to form fewer than $100$ stars, on average (e.g., see \citealt{Smith231110147} for a recent example of such a system).

Thus, we assume that a peak subhalo virial mass of $3\times 10^7~M_{\mathrm{\odot}}$---corresponding to the minimum observable halo mass in our ``weak'' cutoff scenario---sets an operational limit below which halos are effectively ``dark.'' Note that the present-day masses of surviving subhalos in cosmological simulations are reduced by $\approx 50\%$ or more on average relative to their peak masses (e.g., \citealt{Nadler220902675}), although particle-tracking models \citep{Mansfield230810926} and idealized simulations \citep{Errani200107077} indicate that CDM subhalos often survive to much lower masses. Meanwhile, the present-day masses of isolated halos are roughly equal to their peak masses.

This peak mass threshold of $3\times 10^7~M_{\mathrm{\odot}}$ provides a benchmark for the detection of dark (sub)halos using gravitational DM probes. For example, \cite{Gilman190111031} predicted that subhalos with present-day masses down to $\approx 10^7~M_{\mathrm{\odot}}$ can be probed using flux ratio statistics for a sample of $50$ strong gravitational lenses. Facilities including Rubin and Roman are expected to identify thousands of strong lenses in the coming years (e.g., \citealt{Oguri10012037,Collett150702657}); it should be feasible to obtain the deeper follow-up data necessary for DM substructure studies for hundreds of these systems over the next decade, following efforts that are already underway using JWST \citep{Nierenberg230910101}. In this context, our forecasts indicate that a statistical detection of dark line-of-sight halos with masses of $\approx 3\times 10^7~M_{\mathrm{\odot}}$, may be achieved by combining upcoming dwarf galaxy and strong-lensing data. Meanwhile, an unambiguous statistical detection of dark subhalos will likely require sensitivity to substructure with stripped masses below $\approx 10^7~M_{\mathrm{\odot}}$, although it will still be necessary to model the line-of-sight contribution when interpreting lensing substructure (e.g., \citealt{Despali171005029,Sengul211200749}).

Measuring the gravitational effects of low-mass (sub)halos at an individual level is more challenging. Current gravitational imaging analyses have identified halos with masses of $\approx 10^9~M_{\mathrm{\odot}}$ \citep{Vegetti09100760,Vegetti12013643,Hezaveh160101388}; due to its excellent angular resolution, very long baseline interferometric (VLBI) observations using the MIT-Green Bank Very Large Array are beginning to probe even lower-mass (sub)halos (e.g., \citealt{Powell220703375}). It is therefore plausible that individual (sub)halos below the galaxy formation threshold will be detected in future VLBI data, e.g.\ from the Next-Generation Very Large Array \citep{SelinaVLA,KadlerVLA}. Intriguingly, recent JWST observations indicate that highly magnified and strongly lensed stars can also potentially probe extremely low-mass DM substructure \citep{Diego23071036}.

Low-mass subhalos can also be detected via their gravitational effects on stellar streams. For example, current analyses suggest that the GD-1 stream was impacted by an object with mass below $\approx 10^8~M_{\mathrm{\odot}}$ \citep{Bonaca181103631} and that power spectra of stream density fluctuations are sensitive to subhalo populations of similar masses (e.g., \citealt{Banik191102662}). Rubin is expected to increase the precision of such measurements, potentially reaching a perturber mass sensitivity of $\approx 10^6~M_{\mathrm{\odot}}$ \citep{Drlica-Wagner190201055}. Upcoming stream measurements should provide an independent constraint on the abundance and spatial distribution of low-mass MW subhalos within the next decade.

The gravitational DM probes discussed above are sensitive to both halo abundances and density profiles (e.g., \citealt{Bonaca181103631,Gilman190902573}), whereas the dwarf population statistics we have focused on mainly probe the galaxy formation cutoff and SHMF. As a result, simultaneously inferring the galaxy formation cutoff, the abundance of (sub)halos near and below the galaxy formation threshold, and the density profiles of these low-mass objects will maximize the constraining power of future searches for dark halos in the context of models beyond CDM. Note that specific DM models often affect halo abundances and density profiles in a correlated fashion; for example, suppression of the linear matter power spectrum imprinted by WDM prevents the formation of low-mass halos and simultaneously reduces the concentrations of higher-mass halos. Measurements of low-mass halo populations near and below the galaxy formation threshold will therefore contain rich, multidimensional information about galaxy formation and DM physics.


\section{Summary}
\label{sec:conclusions}

We have forecasted the galaxy formation and DM constraints that future surveys of dwarf galaxy populations around MW-mass hosts can deliver. We find the following:
\begin{itemize}
    \item A galaxy formation cutoff at peak virial subhalo masses of $\approx 10^8~M_{\mathrm{\odot}}$ can be constrained by a survey of all dwarf satellite galaxies around one MW-mass system; adding another host improves this sensitivity at the $1\sigma$ level (Figure~\ref{fig:reion}).
    \item A weaker galaxy formation cutoff at peak virial subhalo masses of $\approx 3\times 10^7~M_{\mathrm{\odot}}$ can be constrained from above, but is difficult to detect even with observations of multiple complete satellite populations. Gravitational probes of low-mass halos will be needed to help detect the cutoff in this scenario.
    \item Combining observations of multiple complete satellite populations can dramatically improve constraints on galaxy--halo connection scatter, particularly when combining hosts with differing abundances of bright satellites in the presence of a ``strong'' galaxy formation cutoff (Figure~\ref{fig:2d_posterior}).
    \item Complete measurements of one (two) satellite populations can yield lower limits on the WDM particle mass of $\approx 10~\mathrm{keV}$ ($\approx 20~\mathrm{keV}$), assuming CDM subhalo abundances. However, it is difficult to disentangle WDM and galaxy formation cutoff signals using dwarf galaxy populations alone (Figure~\ref{fig:wdm}).
    \item Future dwarf galaxy surveys can measure the SHMF from peak virial masses of $10^8$ to $10^{10}~M_{\mathrm{\odot}}$; SHMF suppression can be constrained to $\approx 50\%$ that in CDM, while enhancement can only be constrained at the factor of $\approx 3$ level times that in CDM due to galaxy--halo connection degeneracies (Figure~\ref{fig:shmf}).
\end{itemize}

These results indicate that next-generation dwarf galaxy surveys will probe unexplored galaxy formation and DM physics parameter space. Thus, searches for faint dwarf galaxies throughout the Local Volume and low-redshift Universe are timely. In parallel, developing simulation and galaxy--halo connection frameworks that accurately and flexibly model the specific dwarf galaxy populations observed by future surveys will critically enable a detection of new physics using the faintest galaxies.


\section*{Acknowledgements}

Our galaxy--halo connection framework is publicly available at \href{github.com/eonadler/subhalo\_satellite\_connection}{github.com/eonadler/subhalo\_satellite\_connection}.

We are grateful to the referee for providing constructive feedback. We thank Keith Bechtol and Alex Drlica-Wagner for comments on the manuscript and Niusha Ahvazi, Rui An, Shany Danieli, Ivan Esteban, and Josh Simon for helpful discussions related to this work.

V.G.\ acknowledges support from NASA through the Astrophysics Theory Program, award No.\ 21-ATP21-0135 and from the National Science Foundation (NSF) CAREER grant No. PHY-2239205. The work of LAM was carried out at the Jet Propulsion Laboratory, California Institute of Technology, under a contract with NASA. This work received additional support from the 
U.S. Department of Energy under contract No.\ DE-AC02-76SF00515 to SLAC National Accelerator Laboratory and from the Kavli Institute for Particle Astrophysics and Cosmology at Stanford University. This work was performed in part at the Aspen Center for Physics, which is supported by NSF grant PHY-2210452. 

The computations presented here were conducted through Carnegie's partnership with the Resnick High Performance Computing Center, a facility supported by the Resnick Sustainability Institute at the California Institute of Technology. This
work used data from the Symphony suite of simulations
(\href{http://web.stanford.edu/group/gfc/symphony/}{http://web.stanford.edu/group/gfc/symphony/}), which was supported by the Kavli Institute for Particle Astrophysics and Cosmology at Stanford University and SLAC National Accelerator Laboratory, and by the U.S. Department of Energy under contract No.\ DE-AC02-76SF00515 to SLAC National Accelerator Laboratory.

This research used \url{https://arXiv.org} and NASA's Astrophysics Data System for bibliographic information.

\software{
ChainConsumer \citep{ChainConsumer},
emcee \citep{emcee},
getdist \citep{getdist},
Jupyter (\http{jupyter.org}),
Matplotlib \citep{matplotlib},
NumPy \citep{numpy},
Scikit-Learn \citep{scikit-learn},
SciPy \citep{scipy},
Seaborn (\https{seaborn.pydata.org}).
}

\bibliographystyle{yahapj2}
\bibliography{references,software}


\appendix

\section{Galaxy--Halo Connection and Beyond-CDM Priors}
\label{sec:priors}

\begin{figure*}[t!]
\includegraphics[width=\textwidth]{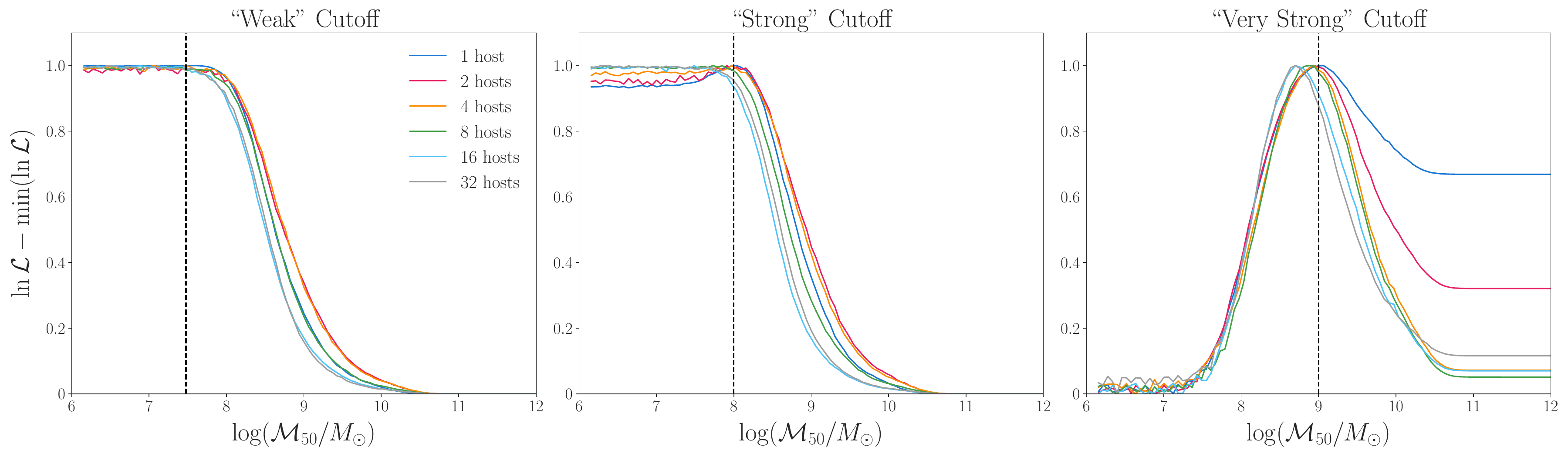}
\caption{Likelihoods as a function of $\log(\mathcal{M}_{50}/M_{\mathrm{\odot}})$, from our idealized tests that hold all remaining galaxy--halo connection parameters fixed at their input values, in the ``weak'' (left panel), ``strong'' (middle panel), and ``very strong'' (right panel) cutoff scenarios. Each panel shows the likelihood for $N_{\mathrm{hosts}}=1$ (blue), $2$ (red), $4$ (orange), and $8$ (green), $16$ (cyan), and $32$ (gray), for a fixed combination of hosts in each case. Each likelihood is normalized to its maximum value over the plotted range.}
\label{fig:M50_idealized}
\end{figure*}

Table~\ref{tab:prior} lists the priors distributions for our galaxy--halo connection and beyond-CDM forecasts. The first eight rows list our galaxy--halo connection priors (Section~\ref{sec:gh_model}), the ninth row lists our WDM prior (Section~\ref{sec:wdm_model}), and the last three rows list our SHMF amplitude priors (Section~\ref{sec:shmf_model}). Note that certain subsets of these parameters are varied in each of our forecasts. In particular, our galaxy formation cutoff forecasts vary the eight galaxy--halo connection parameters (Section~\ref{sec:reion}), our WDM forecasts vary the eight galaxy--halo connection parameters plus~$M_{\mathrm{hm}}$ (Section~\ref{sec:wdm}), and our SHMF forecasts vary the eight galaxy--halo connection parameters plus~$\xi_8$, $\xi_9$, and $\xi_{10}$ (Section~\ref{sec:shmf}).

\section{Idealized $\mathcal{M}_{50}$ Likelihoods}
\label{sec:idealized_constraints}

Here, we study $\mathcal{M}_{50}$ likelihoods in our idealized tests that fix all remaining galaxy--halo connection parameters and vary the number of hosts combined in the inference. Figure~\ref{fig:M50_idealized} shows $\mathcal{M}_{50}$ likelihoods for $N_{\mathrm{hosts}}\in[1,2,4,8,16,32]$ in ``weak'' (left), ``strong'' (middle), and ``very strong'' ($\mathcal{M}_{50}=10^9~M_{\mathrm{\odot}}$; right) cutoff scenarios. Below, we analyze the likelihoods working from the strongest to weakest signal. In all panels, we use a fixed combination of hosts for each choice of $N_{\mathrm{hosts}}$, and thus do not illustrate the combination-to-combination scatter shown in Figure~\ref{fig:M50_proj}.

In the ``very strong'' cutoff scenario, $\mathcal{M}_{50}$ is sharply constrained from below because models with $\mathcal{M}_{50}\ll 10^9~M_{\mathrm{\odot}}$ predict more satellites than the input model at a level that is significant even when using one or two hosts. $\mathcal{M}_{50}$ is less well constrained from above, particularly when using one or two hosts, because the input model only yields a handful of detectable satellites per hosts; thus, models with $\mathcal{M}_{50}\gg 10^9~M_{\mathrm{\odot}}$ that (on average) yield zero detectable satellites are only constrained when enough hosts are combined such that Poisson uncertainties are small.

In the ``strong'' cutoff scenario, $\mathcal{M}_{50}$ is sharply constrained from above because models with $\mathcal{M}_{50}\gg 10^8~M_{\mathrm{\odot}}$ predict significantly fewer satellites than the input model, even when using one or two hosts. However, the likelihood has a prominent tail toward low $\mathcal{M}_{50}$ that does not diminish as $N_{\mathrm{hosts}}$ increases. This is consistent with the results of our full MCMC forecasts in Section~\ref{sec:reion} and follows because models with $\mathcal{M}_{50}\lesssim 10^8~M_{\mathrm{\odot}}$ do not predict significantly more satellites than the ``strong'' cutoff model. As $N_{\mathrm{hosts}}$ increases and Poisson uncertainties shrink, the upper limit on $\mathcal{M}_{50}$ becomes more stringent, but the tail toward low $\mathcal{M}_{50}$ persists.

Finally, $\mathcal{M}_{50}$ likelihoods in the ``weak'' cutoff scenario are qualitatively similar to the ``strong'' cutoff results but display a weaker preference for the input model. This is again consistent with our forecasts in Section~\ref{sec:reion} and follows because the ``weak'' cutoff suppresses an even smaller fraction of the mock satellite populations in our inference, given our fiducial detection thresholds.

\section{Posterior Distributions}
\label{sec:full_posteriors}

Finally, we present the full posteriors from each forecast described in the main text. For all posteriors below, dark (light) contours represent $68\%$ ($95\%$) confidence intervals, and dashed black lines show the input value of each parameter; contours are colored blue (red) for our one-host (two-host) forecasts. Note that $\sigma_M$ and $\sigma_{\log R}$ are reported in $\rm{dex}$, $\mathcal{M}_{50}$ is reported as $\log(\mathcal{M}_{50}/M_{\rm{\odot}})$, $\mathcal{A}$ is reported in parsecs, $M_{\mathrm{hm}}$ is reported as $\log(M_{\mathrm{hm}}/M_{\rm{\odot}})$, and the remaining galaxy--halo connection and beyond-CDM parameters are dimensionless. Furthermore, the two-host posteriors below always combine hosts with different bright satellite populations (i.e., classical satellite abundances that differ at a level comparable to the $1\sigma$ host-to-host scatter in our simulation suite).

\subsection{Galaxy Formation Cutoff}
\label{sec:reion_full_posterior}

\begin{figure*}[t!]
\centering
    \includegraphics[scale=0.61]{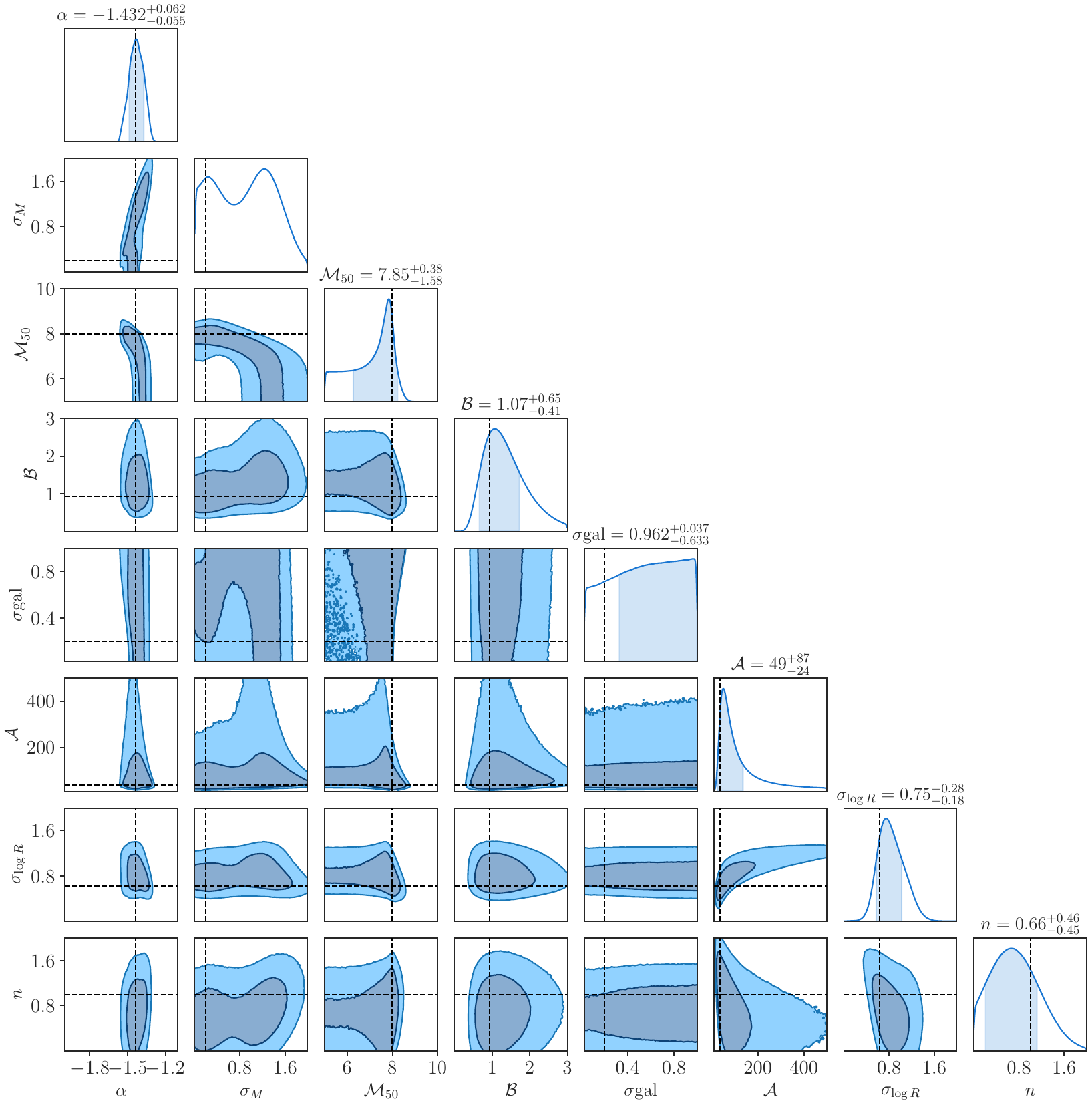}
    \caption{Posterior distribution for our ``strong'' galaxy formation cutoff forecast using one complete CDM satellite population.} \label{fig:full_posterior_reion_1}
\end{figure*}

\begin{figure*}[t!]
\centering
    \includegraphics[scale=0.61]{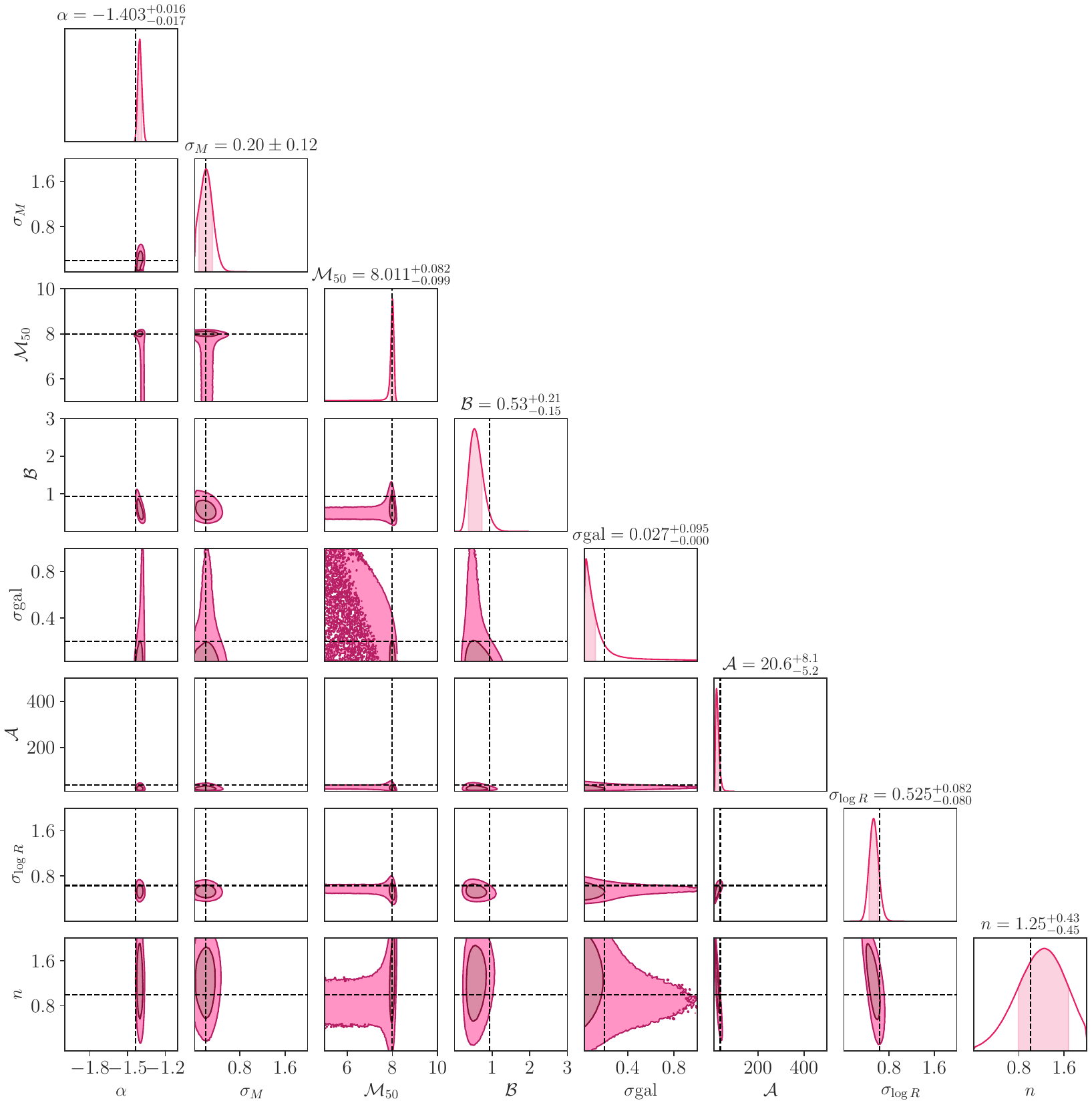}
    \caption{Posterior distribution for our ``strong'' galaxy formation cutoff forecast using two complete CDM satellite populations (compare to Figure~\ref{fig:full_posterior_reion_1}).} \label{fig:full_posterior_reion_2}
\end{figure*}

The posteriors from our one- and two-host ``strong'' galaxy formation cutoff forecasts are shown in  Figures~\ref{fig:full_posterior_reion_1} and \ref{fig:full_posterior_reion_2}, respectively. The parameters of interest for this forecast (i.e., $\sigma_M$, $\mathcal{M}_{50}$, and $\mathcal{S}_{\mathrm{gal}}$) are not strongly correlated with the other galaxy--halo connection parameters; however, there is a prominent degeneracy between $\mathcal{M}_{50}$ and $\sigma_M$ (and, to a lesser extent, between $\mathcal{S}_{\mathrm{gal}}$ and $\sigma_M$). As described in Section~\ref{sec:reion}, this follows because larger values of $\sigma_M$ preferentially cause faint galaxies to up-scatter to observable luminosities, decreasing the average masses of halos inferred to host the faintest observed galaxies. Increasing $\mathcal{M}_{50}$ cuts off galaxy formation more drastically in lower-mass halos, which is not allowed in regions of parameter space where the galaxies they host are crucial to explain the data, resulting in a negatively sloped degeneracy between $\mathcal{M}_{50}$ and $\sigma_M$.

\begin{figure*}[t!]
\centering
    \includegraphics[scale=0.61]{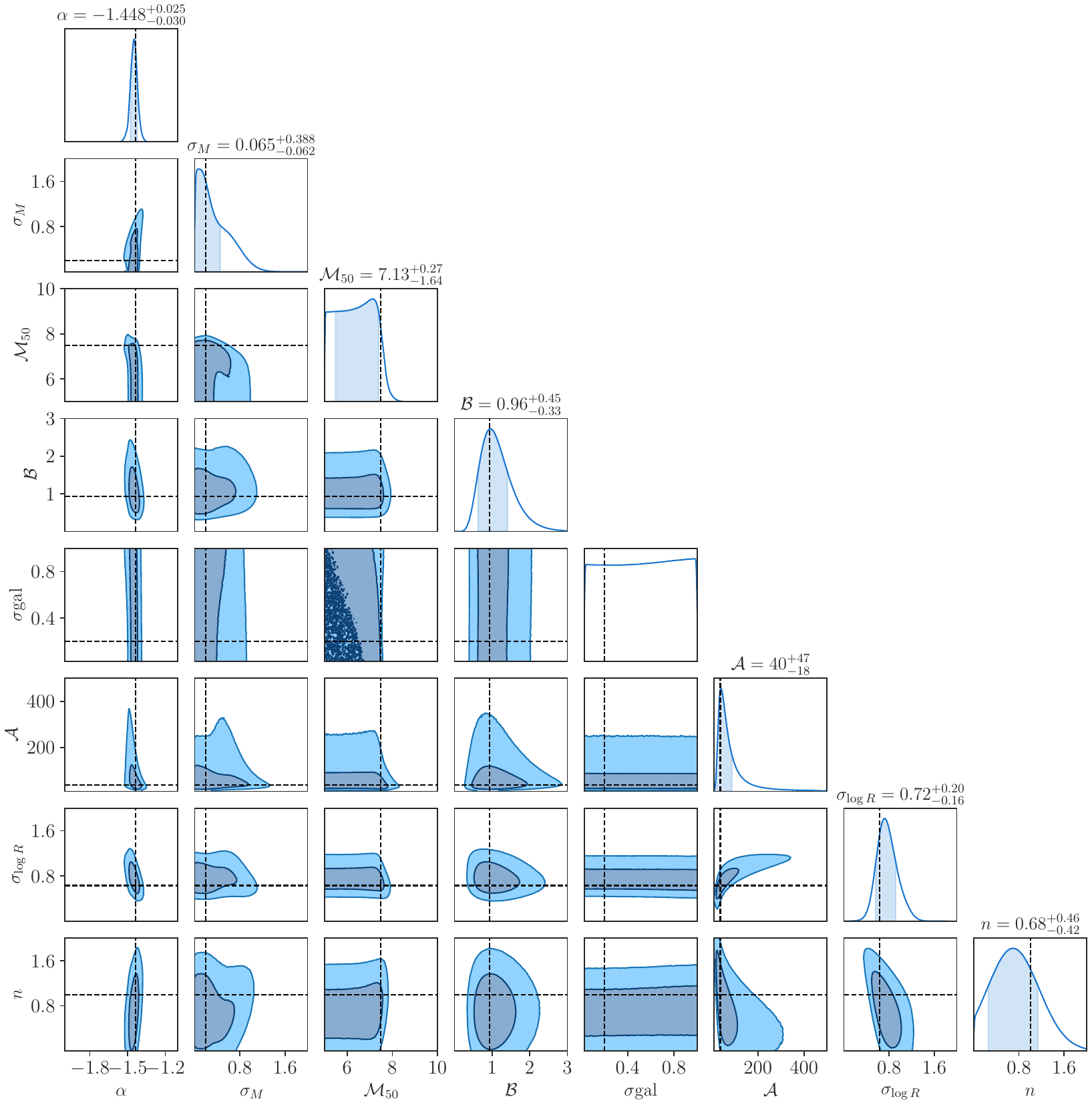}
    \caption{Posterior distribution for our ``weak'' galaxy formation cutoff forecast using one complete CDM satellite population.} \label{fig:full_posterior_weak_reion_1}
\end{figure*}

\begin{figure*}[t!]
\centering
    \includegraphics[scale=0.61]{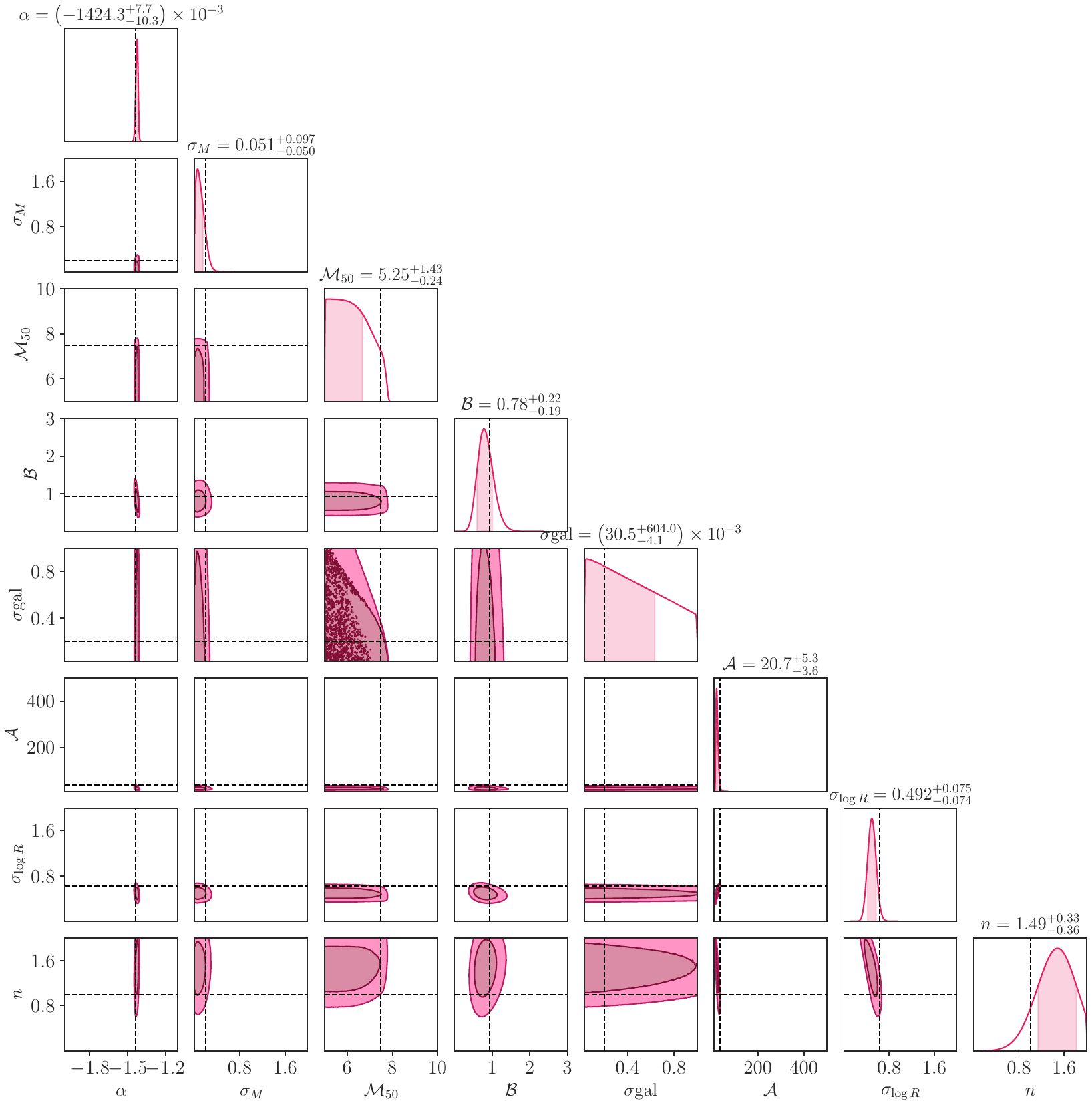}
    \caption{Posterior distribution for our ``weak'' galaxy formation cutoff forecast using two complete CDM satellite populations (compare to Figure~\ref{fig:full_posterior_weak_reion_1}).} \label{fig:full_posterior_weak_reion_2}
\end{figure*}

Next, the posteriors from our one- and two-host ``weak'' galaxy formation cutoff forecasts are shown in  Figures~\ref{fig:full_posterior_weak_reion_1} and \ref{fig:full_posterior_weak_reion_2}, respectively. In the one-host case, we only recover an upper limit on $\mathcal{M}_{50}$, and we observe weaker degeneracies between $\mathcal{M}_{50}$ and the remaining galaxy--halo connection parameters. Although uncertainties significantly shrink in the two-host forecast, $\mathcal{M}_{50}$ is still not constrained from below in this case, which is consistent with our discussion in Section~\ref{sec:reion}. Note that the upper limit on $\mathcal{M}_{50}$ in the two-host forecast is not in significant tension with its input value.

In all of these forecasts, we also observe degeneracies among the galaxy--halo connection size parameters (and particularly between $\mathcal{A}$ and $\sigma_{\log R}$) that are consistent with those reported in \cite{Nadler191203303}. For example, larger values of $\mathcal{A}$ increase the average sizes of galaxies at all luminosities, pushing some faint systems below our assumed surface brightness detectability threshold; this can be counteracted by increasing the size scatter, which preferentially causes small systems hosted by abundant, low-mass halos to up-scatter above our fiducial $r_{1/2}>10~\mathrm{pc}$ cut and become detectable.

\begin{deluxetable*}{{l@{\hspace{0.2in}}c@{\hspace{0.2in}}c}}[t!]
\centering
\tablecolumns{3}
\tablecaption{Prior Distributions for the Parameters Varied in Our Galaxy Formation, WDM, and SHMF Forecasts}
\tablehead{\colhead{Free Parameter\phantom{texttttt}} {\hspace{0.2in}}& \colhead{Prior Distribution\phantom{text}} {\hspace{0.2in}}& \colhead{Motivation}}
\startdata
Faint-end slope
& $\arctan\alpha\sim\rm{unif}(-1.1,-0.9)$ 
& Jeffreys prior over wide range of $-2<\alpha<-1.2$\\
\hline
Luminosity scatter
& $\sigma_M\sim\rm{unif}(0,2)\ \rm{dex}$
& Conservative upper limit \citep{GarrisonKimmel160304855,Lehmann151005651}\\
\hline
$50\%$ occupation mass
& $\log(\mathcal{M}_{50}/M_{\rm{\odot}})\sim\rm{unif}(5,10)$ 
& Brackets sims.\ \& SAMs \citep{Munshi210105822,Kravtsov210609724}\\
\hline
Disruption efficiency
& $\ln(\mathcal{B})\sim \mathcal{N}(\mu=1,\sigma=0.5)$ 
& Centered on simulations \citep{Garrison-Kimmel170103792,Nadler171204467}\\
\hline
Occupation shape
& $\mathcal{S}_{\rm{gal}}\sim\rm{unif}(0,1)$ 
& Brackets sims.\ \& SAMs \citep{Munshi210105822,Kravtsov210609724}\\
\hline
Size amplitude
& $\mathcal{A}\sim\rm{unif}(0,0.5)\ \rm{kpc}$ 
& Brackets empirical galaxy--halo size relation \citep{Kravstov12122980}\\
\hline
Size scatter
& $\sigma_{\log R}\sim\rm{unif}(0,2)\ \rm{dex}$ 
& Brackets empirical galaxy--halo size relation \citep{Kravstov12122980}\\
\hline
Size power-law index
& $n\sim \mathcal{N}(\mu=1,\sigma=0.5)$ 
& Centered on empirical galaxy--halo size relation \citep{Kravstov12122980}\\
\hline
WDM half-mode mass
& $\log(M_{\mathrm{hm}}/M_{\rm{\odot}})\sim\rm{unif}(5,10)$
& Brackets current constraints \citep{Gilman190806983,Banik191102663}\\
\hline
SHMF deviation at $10^8~M_{\rm{\odot}}$
& $\xi_8\sim\rm{unif}(-2,2)$
& Brackets current constraints \citep{Banik191102662,Nadler210107810}\\
\hline
SHMF deviation at $10^9~M_{\rm{\odot}}$
& $\xi_9\sim\rm{unif}(-2,2)$
& Brackets current constraints \citep{Banik191102662,Nadler210107810}\\
\hline
SHMF deviation at $10^{10}~M_{\rm{\odot}}$
& $\xi_{10}\sim\rm{unif}(-2,2)$
& Brackets current constraints \citep{Banik191102662,Nadler210107810}\\
\hline
\enddata
{\footnotesize \tablecomments{Here $\mathcal{N}(\mu,\sigma)$ denotes a normal distribution with mean $\mu$ and standard deviation $\sigma$.}}
\label{tab:prior}
\end{deluxetable*}

\subsection{Warm Dark Matter}
\label{sec:wdm_full_posterior}

\begin{figure*}[t!]
\centering
    \includegraphics[scale=0.61]{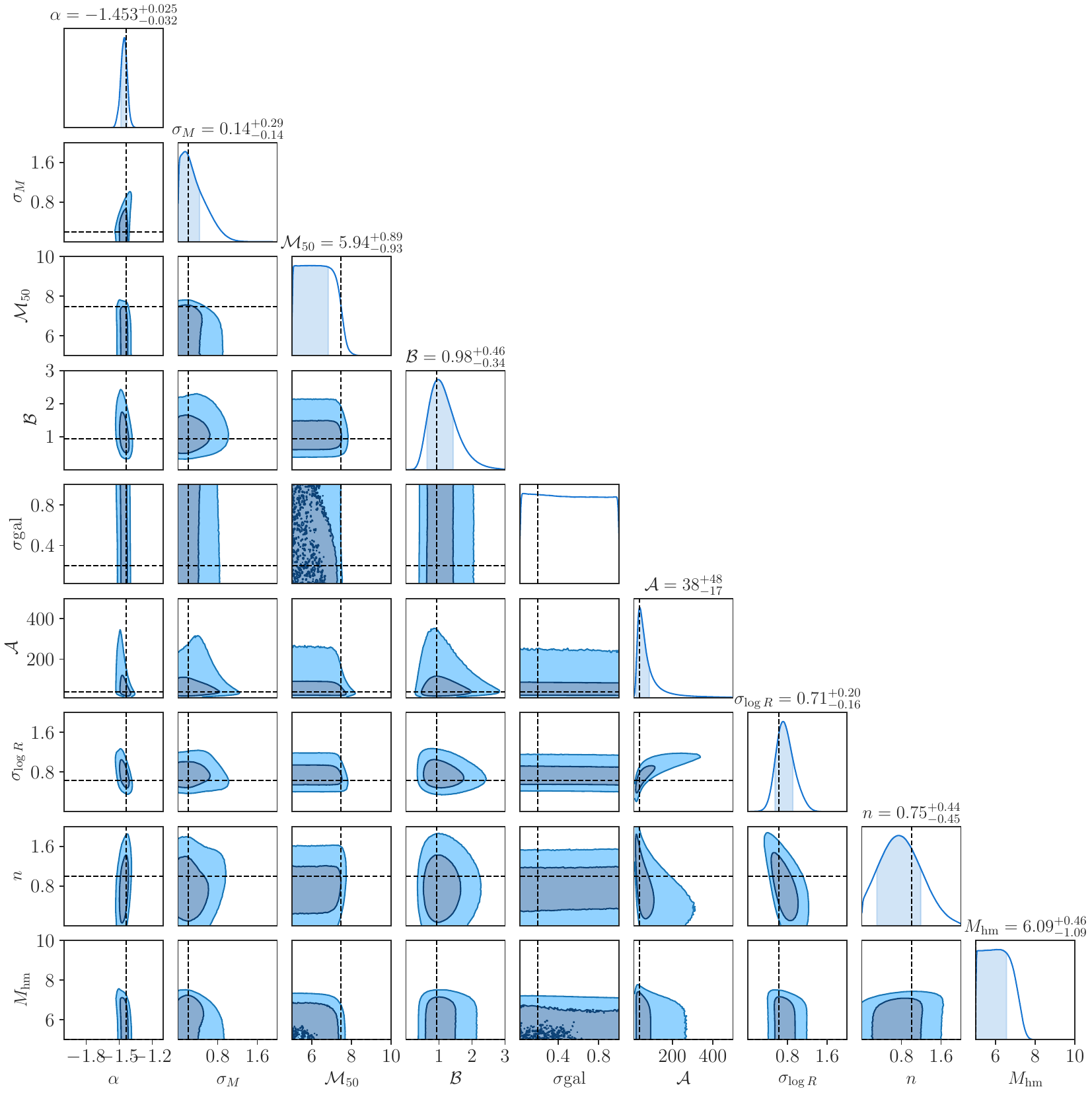}
    \caption{Posterior distribution for our WDM forecast with a ``weak'' galaxy formation cutoff assuming CDM subhalo abundances (Scenario A) using one complete satellite population.} \label{fig:full_posterior_wdm_weak_cdm_1}
\end{figure*}

\begin{figure*}[t!]
\centering
    \includegraphics[scale=0.61]{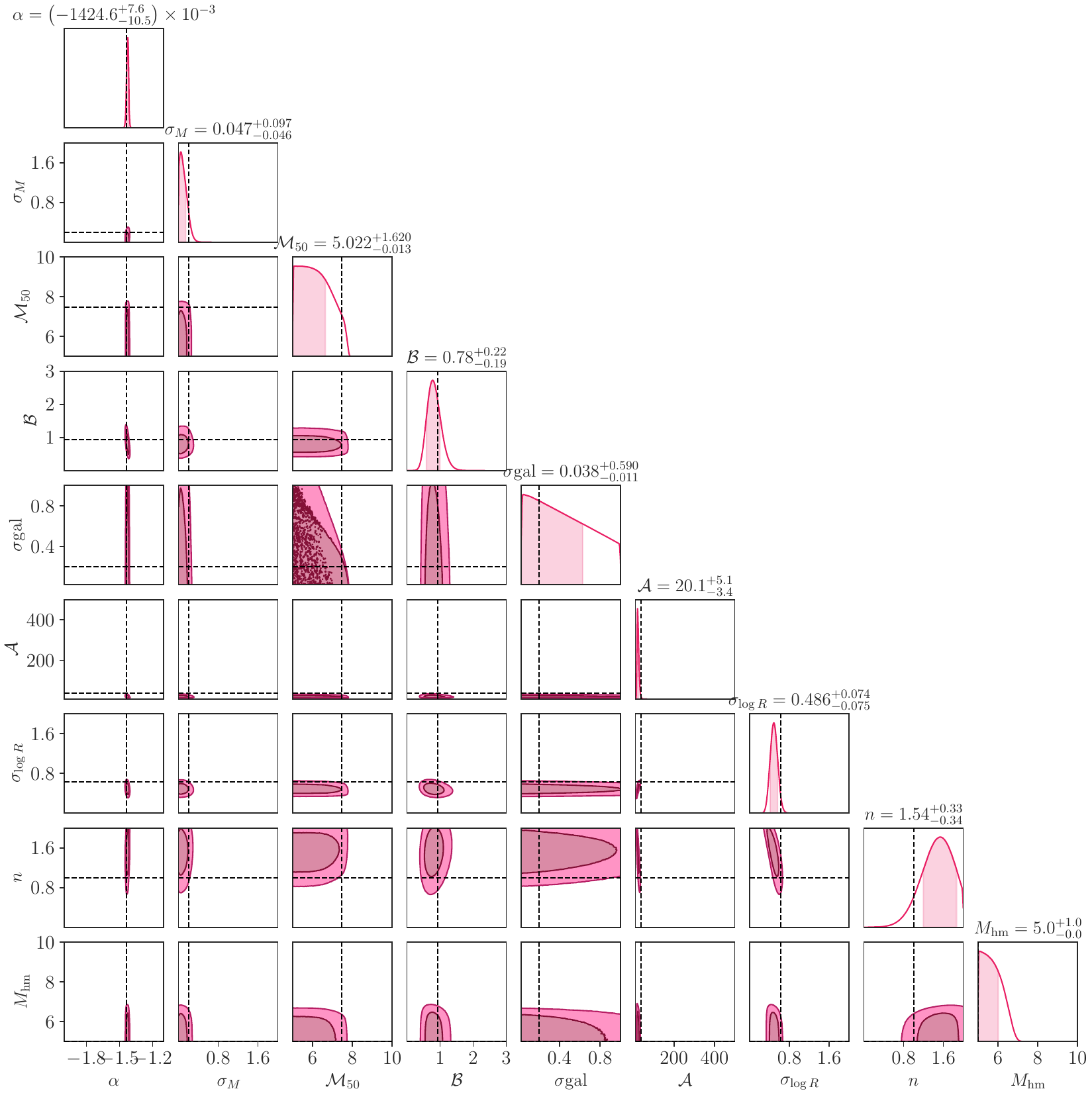}
    \caption{Posterior distribution for our WDM forecast with a ``weak'' galaxy formation cutoff assuming CDM subhalo abundances (Scenario A) using two complete satellite populations (compare to Figure~\ref{fig:full_posterior_wdm_weak_cdm_1}).} \label{fig:full_posterior_wdm_weak_cdm_2}
\end{figure*}

\begin{figure*}[t!]
\centering
    \includegraphics[scale=0.61]{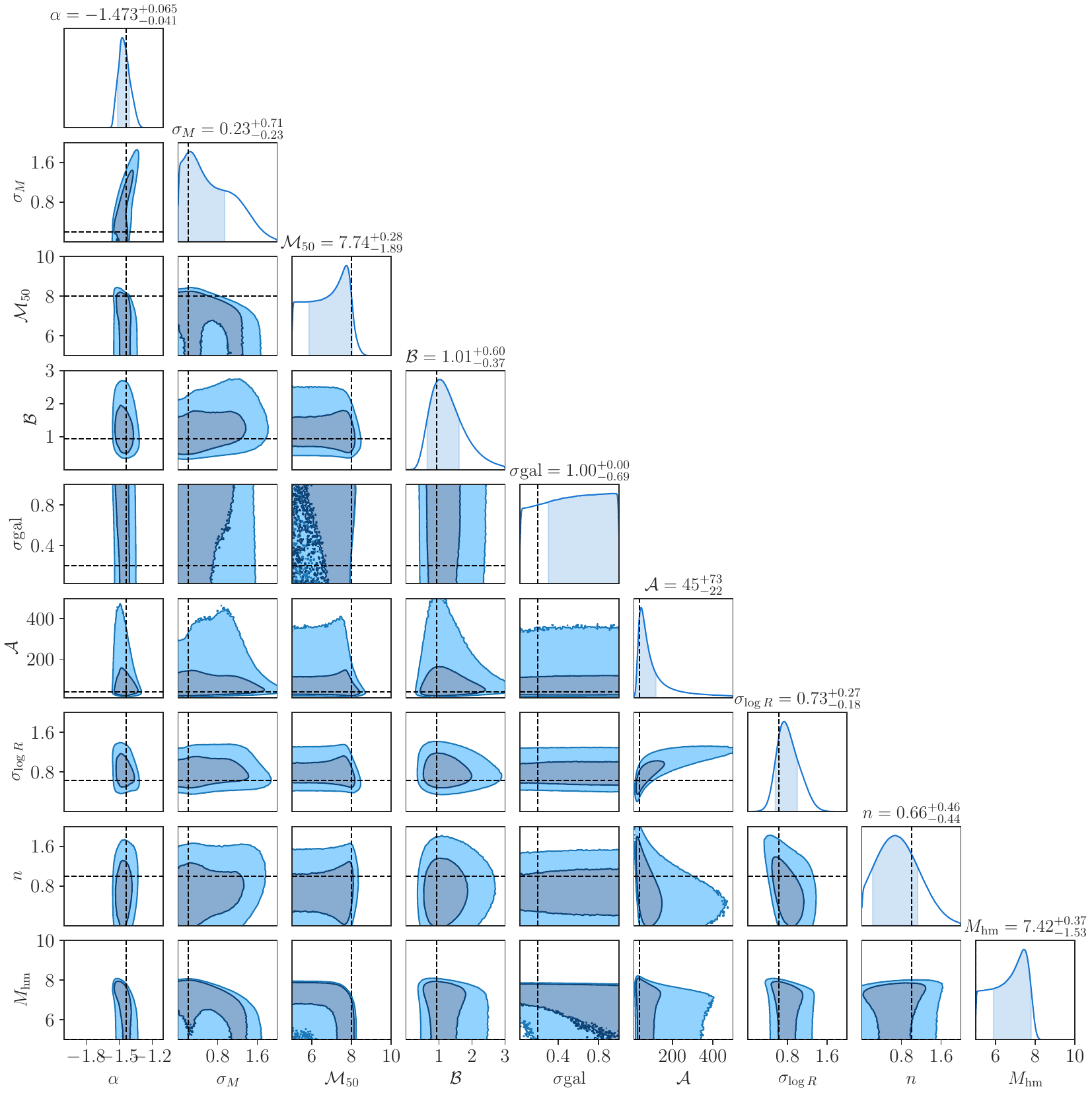}
    \caption{Posterior distribution for our WDM forecast with a ``strong'' galaxy formation cutoff assuming CDM subhalo abundances (Scenario B) using one complete satellite population.} \label{fig:full_posterior_wdm_strong_cdm_1}
\end{figure*}

\begin{figure*}[t!]
\centering
    \includegraphics[scale=0.61]{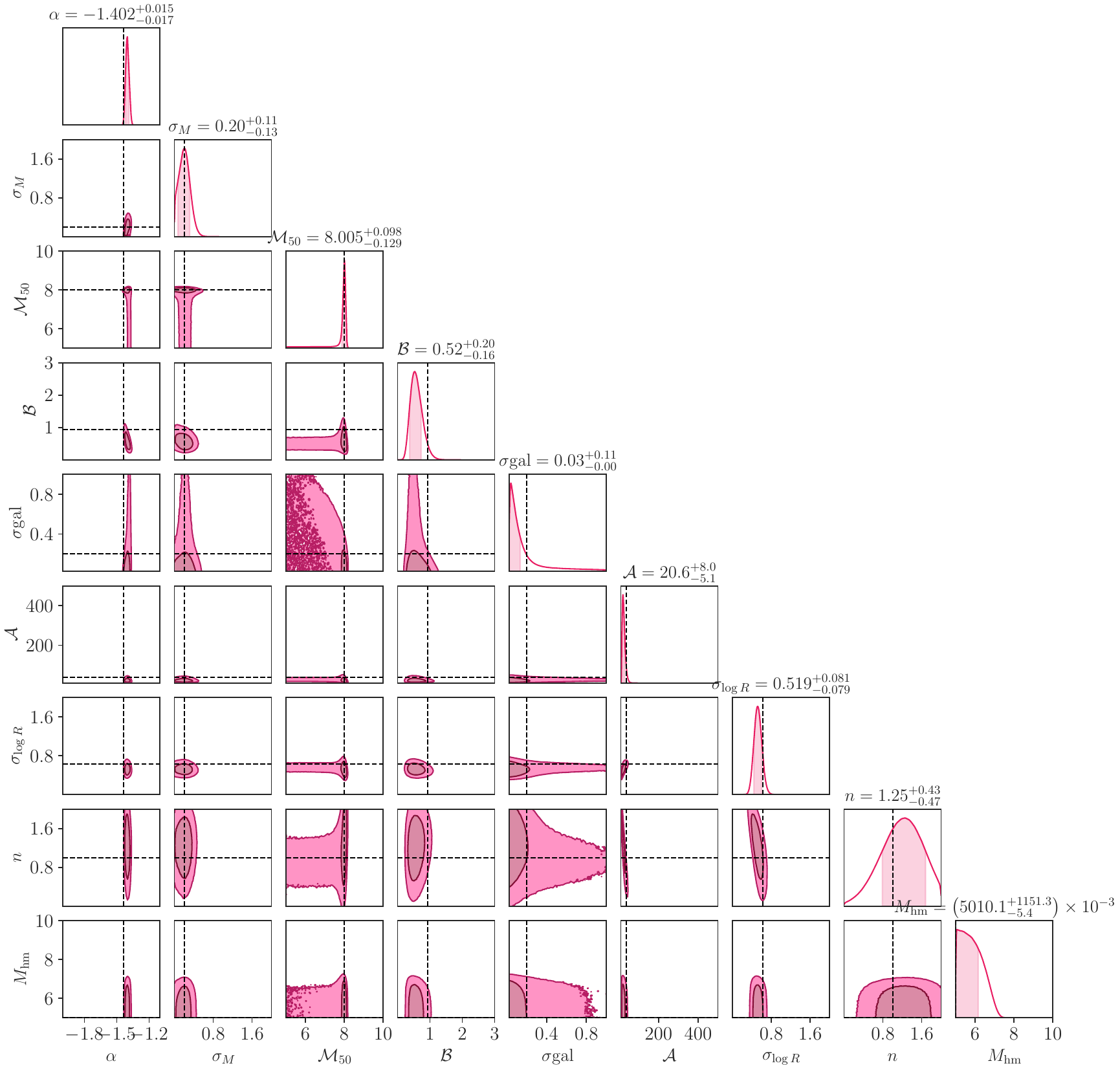}
    \caption{Posterior distribution for our WDM forecast with a ``strong'' galaxy formation cutoff assuming CDM subhalo abundances (Scenario B) using two complete satellite populations (compare to Figure~\ref{fig:full_posterior_wdm_strong_cdm_1}).} \label{fig:full_posterior_wdm_strong_cdm_2}
\end{figure*}

The posteriors from our one- and two-host WDM forecasts assuming a ``weak'' galaxy formation cutoff and CDM subhalo abundances are shown in Figures~\ref{fig:full_posterior_wdm_weak_cdm_1} and \ref{fig:full_posterior_wdm_weak_cdm_2}, and the corresponding ``strong'' galaxy formation cutoff runs are shown in Figures~\ref{fig:full_posterior_wdm_strong_cdm_1} and \ref{fig:full_posterior_wdm_strong_cdm_2}, respectively. The posteriors are qualitatively similar in these scenarios, and our description below applies to both.

All of the same degeneracies among galaxy--halo connection parameters noted in Appendix~\ref{sec:reion_full_posterior} remain when $M_{\mathrm{hm}}$ is added to the inference. In addition to the significant degeneracies between $M_{\mathrm{hm}}$ and both $\mathcal{M}_{50}$ and $\sigma_M$ discussed in Section~\ref{sec:wdm}, these posteriors also illustrate a weak degeneracy between $M_{\mathrm{hm}}$ and $\mathcal{S}_{\mathrm{gal}}$ that persists even when two complete satellite populations with different bright satellite populations are used. Figures~\ref{fig:full_posterior_wdm_strong_cdm_1} and \ref{fig:full_posterior_wdm_strong_cdm_2} clearly illustrate that the $\mathcal{M}_{50}$--$\sigma_M$ degeneracy is broken when combining two complete satellite populations with differing classical satellite abundances, which in turn yields a much stronger upper limit on $M_{\mathrm{hm}}$ in the two-host case.

\begin{figure*}[t!]
\centering
    \includegraphics[scale=0.61]{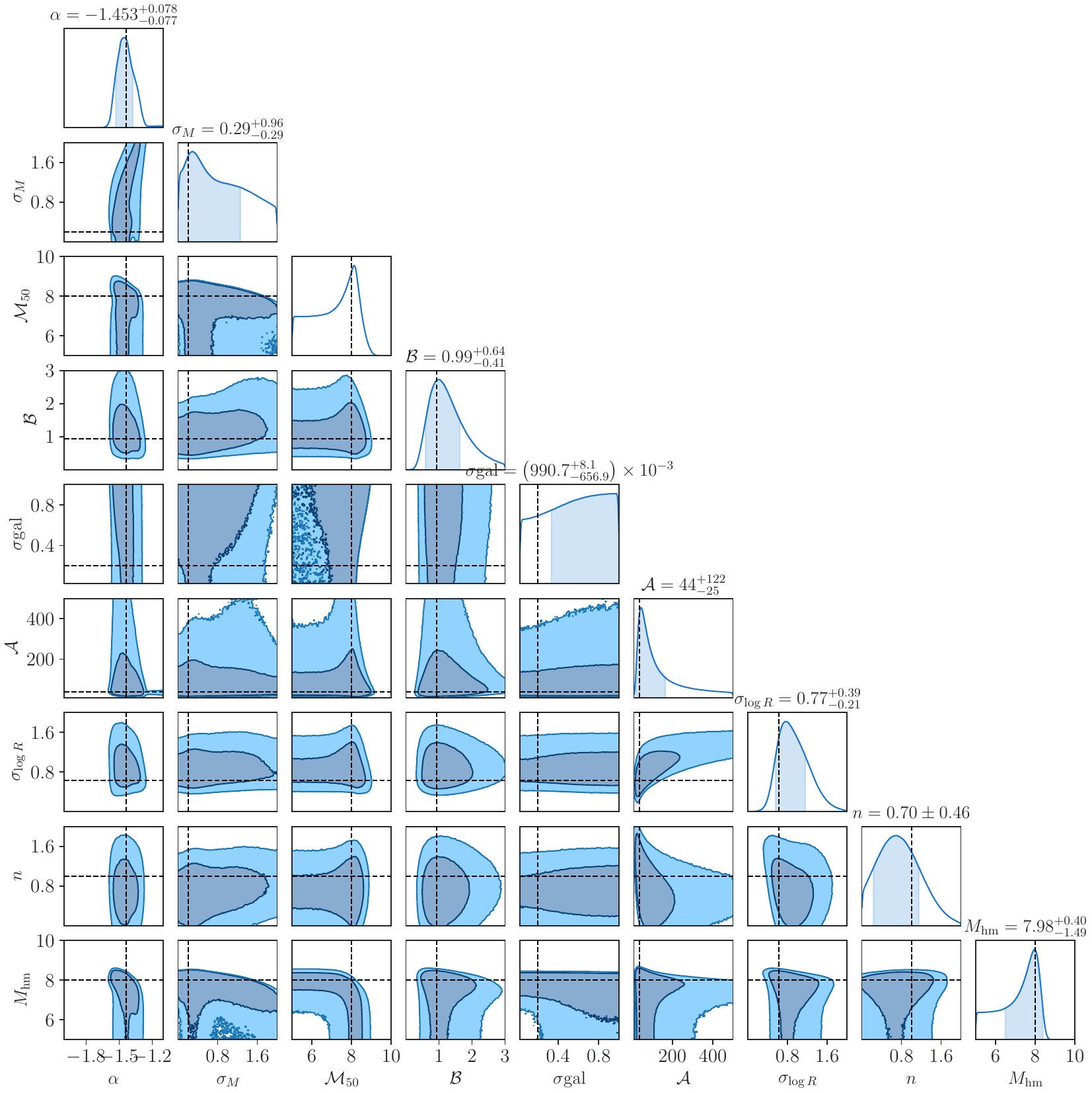}
    \caption{Posterior distribution for our WDM forecast with a ``strong'' galaxy formation cutoff assuming WDM subhalo abundances with $M_{\mathrm{hm}}=10^8~M_{\mathrm{\odot}}$, corresponding to $m_{\mathrm{WDM}}=4.9~\mathrm{keV}$ (Scenario C), using one complete satellite population.} \label{fig:full_posterior_wdm_strong_wdm_1}
\end{figure*}

\begin{figure*}[t!]
\centering
    \includegraphics[scale=0.61]{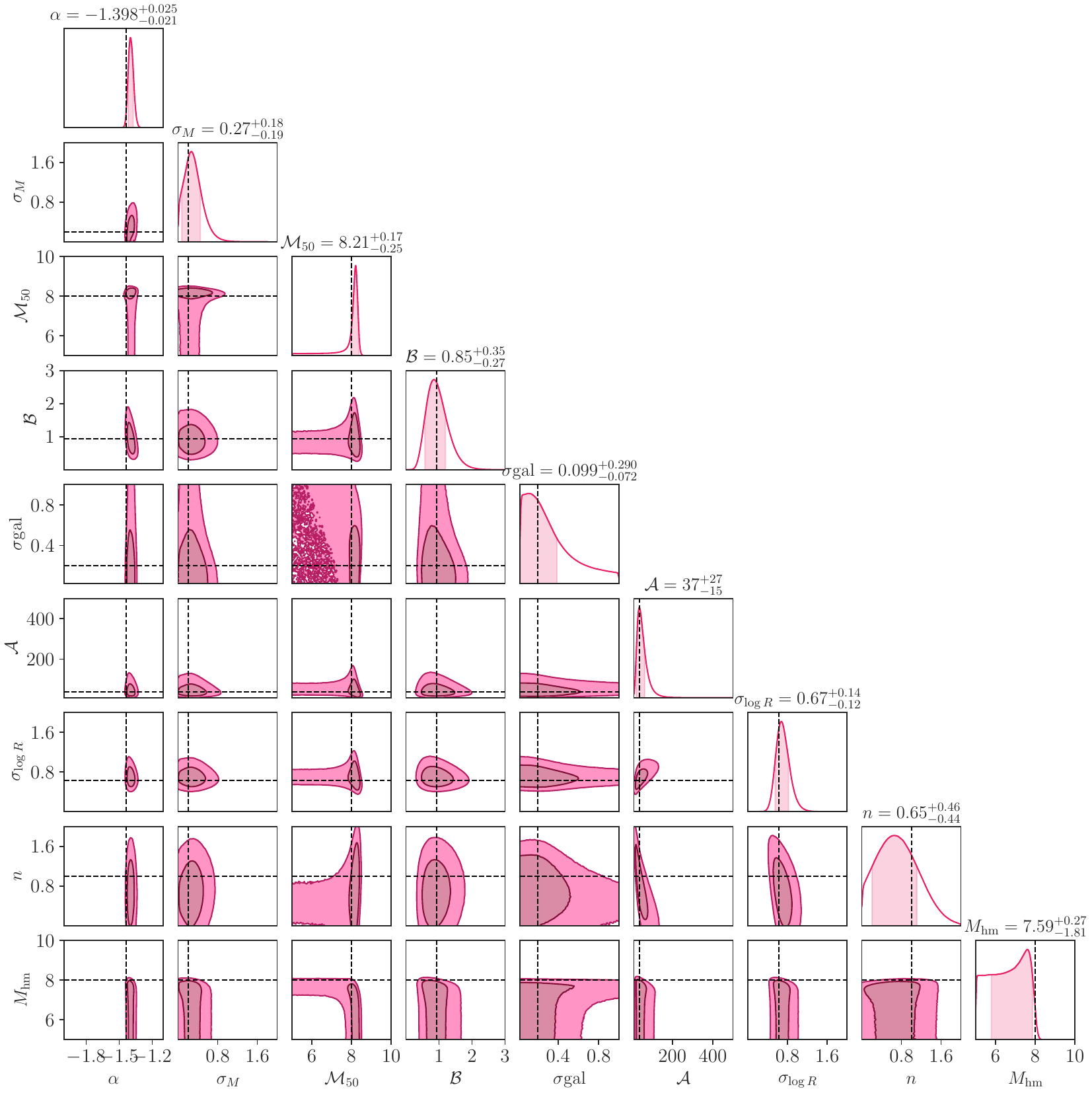}
    \caption{Posterior distribution for our WDM forecast with a ``strong'' galaxy formation cutoff, assuming WDM subhalo abundances with $M_{\mathrm{hm}}=10^8~M_{\mathrm{\odot}}$, corresponding to $m_{\mathrm{WDM}}=4.9~\mathrm{keV}$ (Scenario C), using two complete satellite populations (compare to Figure~\ref{fig:full_posterior_wdm_strong_wdm_1}).} \label{fig:full_posterior_wdm_strong_wdm_2}
\end{figure*}

Figures~\ref{fig:full_posterior_wdm_strong_wdm_1} and \ref{fig:full_posterior_wdm_strong_wdm_2}, respectively, show the posteriors from our one and two-host WDM forecasts assuming a ``strong'' galaxy formation cutoff and WDM subhalo abundances with $M_{\mathrm{hm}}=10^8~M_{\mathrm{\odot}}$ ($m_{\mathrm{WDM}}=4.9~\mathrm{keV}$). The degeneracies described in the previous paragraph all remain in this case and often become more prominent. As discussed in Section~\ref{sec:wdm}, this follows because, in our model, astrophysical and DM-induced cutoffs affect dwarf galaxy abundances in a qualitatively similar fashion. Specifically, the only difference between the effects of these cutoffs in our inference is the shape of the suppression in dwarf galaxy abundances as a function of $M_{\mathrm{peak}}$, dictated by the astrophysical and WDM cutoffs in Equations~\ref{eq:fgal} and~\ref{eq:wdm_shmf}, respectively (see Figure~\ref{fig:wdm_pred}).

The shape of the WDM SHMF cutoff has been studied in simulations (e.g., \citealt{Angulo13042406,Lovell13081399,Bose150701998,Stucker210909760}), and the shape of the astrophysical cutoff in galaxy abundances has been explored in SAMs (e.g., \citealt{Benitze-Llambay200406124,Kravtsov210609724,Ahvazi230813599}) and hydrodynamic simulations (e.g., \citealt{Sawala14066362,Fitts161102281,Munshi210105822}). Refining predictions for the shapes of both cutoffs and their dependence on halo properties beyond $M_{\mathrm{peak}}$ is an interesting area for future study and will help constrain our model. For example, halo masses evaluated during the epoch of reionization are expected to correlate more strongly with the galaxy occupation fraction than $M_{\mathrm{peak}}$ (e.g., \citealt{Benitze-Llambay200406124}), which is typically achieved after reionization. We expect that incorporating such physically motivated parameterizations will help differentiate astrophysical and DM-induced cutoffs.

In addition to degeneracies with parameters related to our satellite luminosity model, galaxy--halo size connection parameters are generally constrained less precisely than in the scenario without a WDM cutoff. For example, degeneracies between the size amplitude $\mathcal{A}$ or size scatter $\sigma_{\log R}$ and $\mathcal{M}_{50}$ are exacerbated by the larger uncertainty on $M_{\mathrm{hm}}$ in the presence of both astrophysical and DM cutoffs. Physically, degeneracies with the size model arise because our mock observations are surface-brightness-limited, such that increasing the sizes of all galaxies at fixed galaxy--halo connection parameters can mimic a cutoff by making these systems too spatially extended to be detectable at a given luminosity (see \citealt{Nadler191203303} for related discussion).

\subsection{Subhalo Mass Function}
\label{sec:shmf_full_posterior}

\begin{figure*}[t!]
\centering
    \includegraphics[scale=0.61]{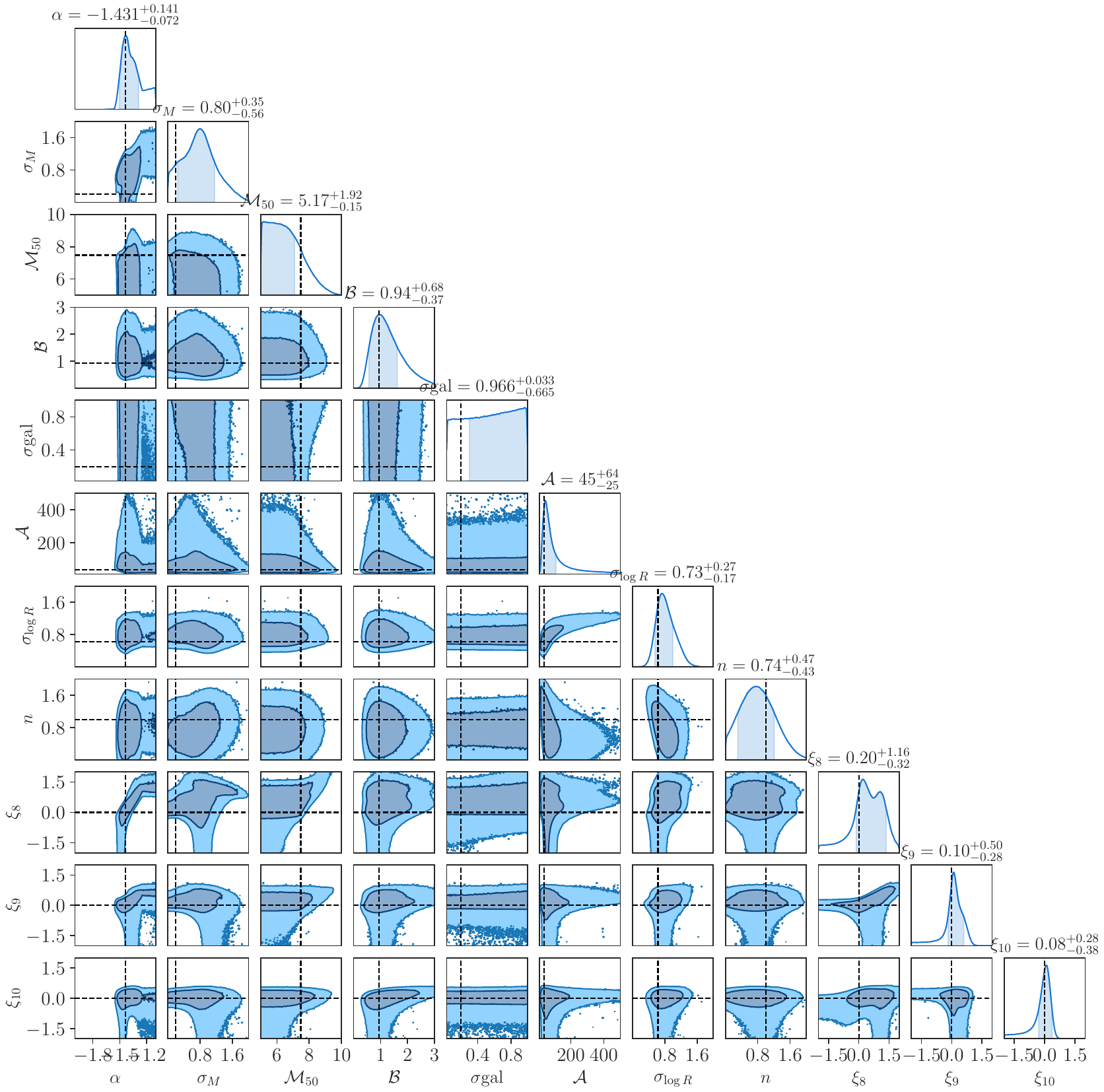}
    \caption{Posterior distribution for our SHMF forecast with a ``weak'' galaxy formation cutoff using one complete satellite population.}
    \label{fig:shmf_weak_1}
\end{figure*}

\begin{figure*}[t!]
\centering
    \includegraphics[scale=0.61]{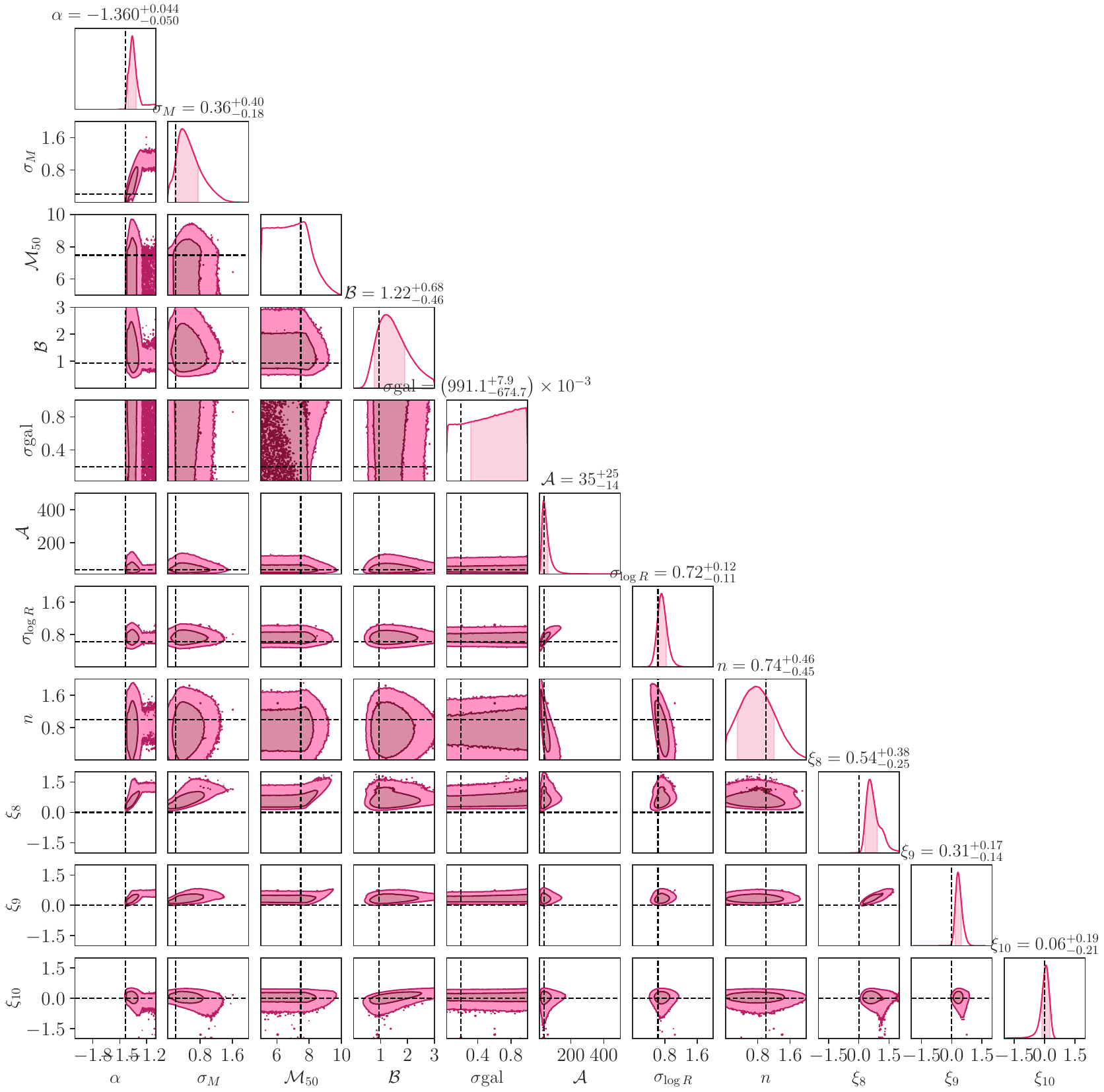}
    \caption{Posterior distribution for our SHMF forecast with a ``weak'' galaxy formation cutoff using two complete satellite populations (compare to Figure~\ref{fig:shmf_weak_1}).}
    \label{fig:shmf_weak_2}
\end{figure*}

\begin{figure*}[t!]
\centering
    \includegraphics[scale=0.61]{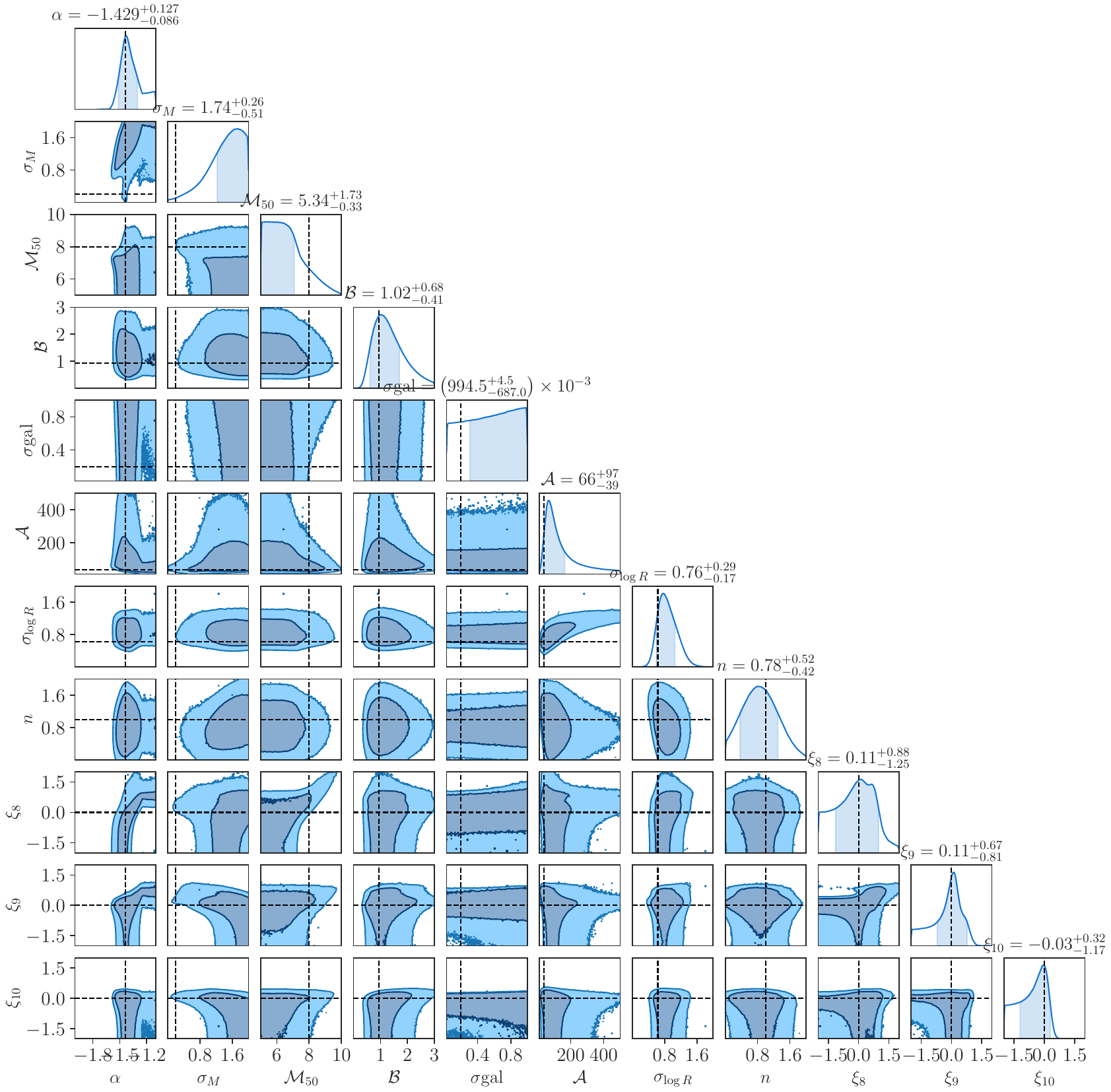}
    \caption{Posterior distribution for our SHMF forecast with a ``strong'' galaxy formation cutoff using one complete satellite population.}
    \label{fig:shmf_strong_1}
\end{figure*}

\begin{figure*}[t!]
\centering
    \includegraphics[scale=0.61]{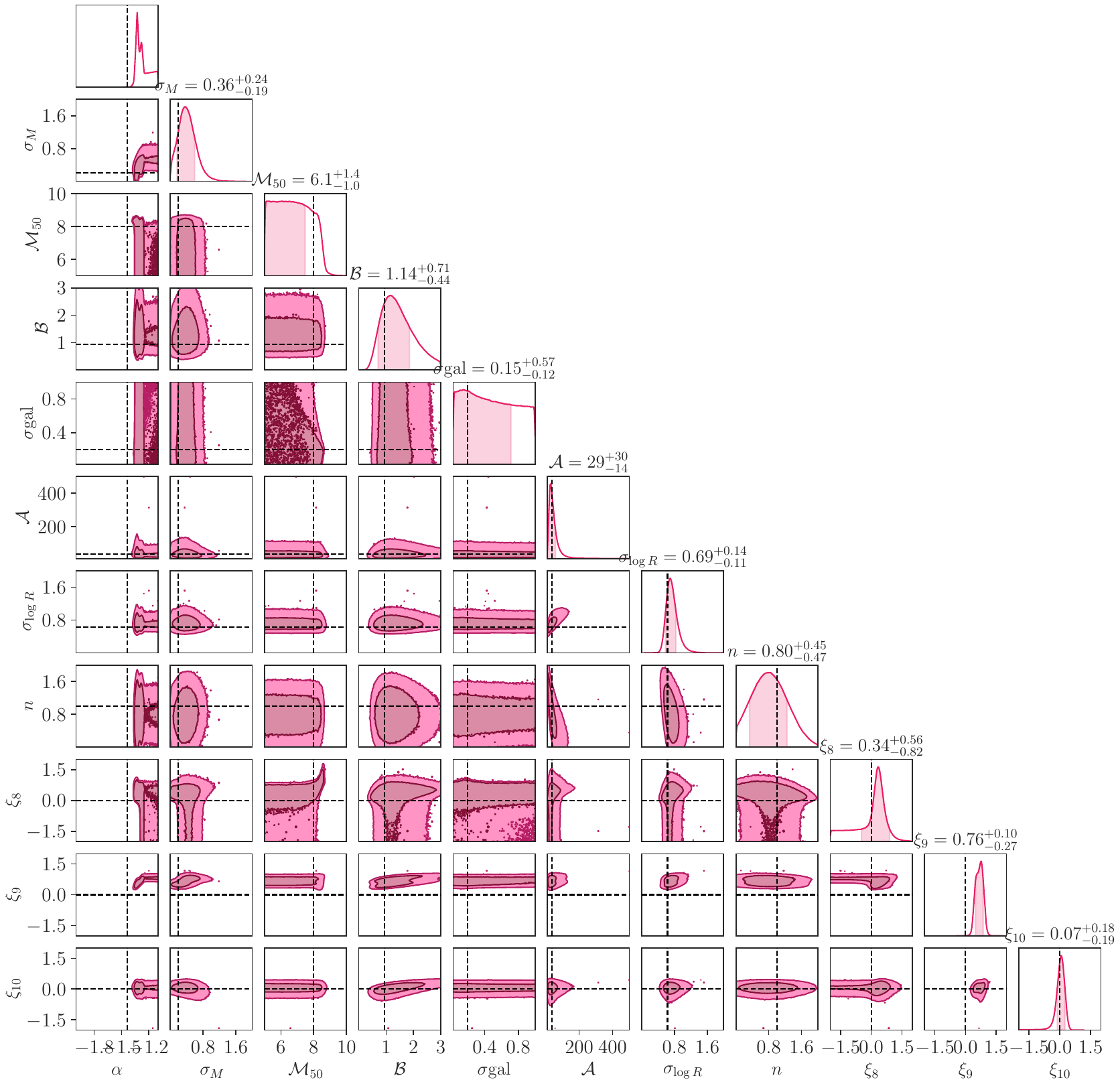}
    \caption{Posterior distribution for our SHMF forecast with a ``strong'' galaxy formation cutoff using two complete satellite populations (compare to Figure~\ref{fig:shmf_strong_1}).}
    \label{fig:shmf_strong_2}
\end{figure*}

The posteriors from our one- and two-host SHMF forecasts assuming a ``weak'' galaxy formation cutoff and CDM subhalo abundances are shown in Figures~\ref{fig:shmf_weak_1} and \ref{fig:shmf_weak_2}, and the corresponding ``strong'' galaxy formation cutoff runs are shown in Figures~\ref{fig:shmf_strong_1} and \ref{fig:shmf_strong_2}, respectively; the description below applies to both scenarios, which display similar degeneracies. We find that it is particularly challenging to sample the posterior using our MCMC method for SHMF forecasts that combine two hosts; as a result, parameters like $\alpha$ are biased at the $1\sigma$--$2\sigma$ level in our two-host forecasts (see Figures~\ref{fig:shmf_weak_2} and \ref{fig:shmf_strong_2}). However, these biases do not affect the interpretation of degeneracies presented below or the precision of our projected SHMF constraints.

The SHMF amplitude at $M_{\mathrm{peak}}=10^{10}~M_{\mathrm{\odot}}$, $\xi_{10}$, does not show noticeable degeneracies with galaxy--halo connection or other SHMF parameters. As described in Section~\ref{sec:shmf}, this follows because (i) galaxy--halo connection parameters affect satellites that occupy lower-mass halos most strongly, (ii) faint satellites are more abundant and thus contribute more to the likelihood, and (iii) the effects of varying $\xi_{10}$ cannot be hidden by nondetections, since satellites hosted by the most massive subhalos are always detectable given our observational assumptions.

Several degeneracies appear between $\xi_8$, $\xi_9$, and the galaxy--halo connection model. Both of these SHMF amplitude parameters are positively correlated with the faint-end luminosity function slope. In our model, $\alpha$ only affects the galaxy--halo connection for $M_V>-13~\mathrm{mag}$, which roughly corresponds to halos with peak virial masses between $10^9~M_{\mathrm{\odot}}$ and $10^{10}~M_{\mathrm{\odot}}$ (e.g., \citealt{Nadler191203303}). Since less negative values of $\alpha$ yield fewer faint satellites that occupy such halos, increasing the underlying SHMF amplitude counteracts this effect. In turn, the faint-end slope is measured less precisely in our SHMF forecasts relative to our galaxy formation cutoff and WDM forecasts, and also displays mild degeneracies with galaxy--halo connection parameters like $\sigma_M$ when the SHMF is allowed to vary.

The SHMF posteriors clearly show that $\xi_8$ is only constrained weakly from above, consistent with our discussion in Section~\ref{sec:shmf}. As a result, measurements of $\mathcal{M}_{50}$ degrade relative to our other forecasts when the SHMF is allowed to vary. Interestingly, $\xi_8$ and $\xi_9$ are also mildly degenerate, which follows because our galaxy--halo connection model can produce satellites with similar properties across this decade of halo mass due to, e.g., luminosity and size scatter. This tail toward large $\xi_8$ persists in the two-host posteriors, even though uncertainties on remaining SHMF and galaxy--halo connection parameters are reduced. 

\end{document}